\documentclass[twocolumn]{autart}    

\renewcommand{\Re}{\mathbb{R}}
\newcommand{\E}{\mathbb{E}}
\renewcommand{\S}{\mathcal{S}}
\renewcommand{\P}{\mathbb{P}}

\newcommand{\scriptS}{\mathcal{S}}
\newcommand{\scriptP}{\mathcal{P}}
\newcommand{\Sb}{\mathbb{S}}
\newcommand{\DeltaSet}{\mathbf{\Delta}}
\newcommand{\qedblack}{\hfill\ensuremath{\blacksquare}}

\usepackage{graphicx}
\usepackage{amsmath,amssymb,amsfonts}
\usepackage{empheq} 
\usepackage{wrapfig}
\usepackage{subcaption}
\usepackage{mathtools} 
\usepackage{textcomp}
\usepackage[table]{xcolor}
\usepackage{tabu} 
\usepackage{multirow}
\usepackage{tikz}
\usepackage{comment}
\usepackage{booktabs} 
\usepackage{bbold}
\DeclareRobustCommand{\rchi}{{\mathpalette\irchi\relax}}
\newcommand{\irchi}[2]{\raisebox{\depth}{$#1\chi$}} 
\newcommand{\dbracket}[1]{[\![ #1 ]\!]}
\usepackage[most]{tcolorbox}
\newtcolorbox{empheqboxed}{colback=white, 
    colframe=black,
    boxrule=0.25mm,
    width=\columnwidth,
    sharpish corners,
    top=-2mm, 
    left=2pt,
    bottom=5pt
}

\newcommand*{\MinNumber}{0}
\newcommand*{\MidNumber}{50}
\newcommand*{\MaxNumber}{100}


\newtheorem{prob}{Problem} 

\newcommand{\ApplyGradient}[1]{%
        \ifdim #1 pt > \MidNumber pt
            \pgfmathsetmacro{\PercentColor}{max(min(100.0*(#1 - \MidNumber)/(\MaxNumber-\MidNumber),100.0),0.00)} %
            \colorlet{feasibleColor}{green!\PercentColor!yellow}
            \colorbox{feasibleColor!20}{#1}
        \else
            \pgfmathsetmacro{\PercentColor}{max(min(100.0*(\MidNumber - #1)/(\MidNumber-\MinNumber),100.0),0.00)} %
            \colorlet{feasibleColor}{red!\PercentColor!yellow}
            \colorbox{feasibleColor!20}{#1}
        \fi
}

\graphicspath{{./Figures/}}

\newcommand{\vct}[1]{\boldsymbol{#1}}

\begin{document}

\begin{frontmatter}

\title{DUST: A Framework for \\ Data-Driven Density Steering} 

\thanks[JPnow]{Currently at RTX Technology Research Center (RTRC), East Hartford, CT (e-mail: joshua.y.pilipovsky@rtx.com)}

\author[1]{Joshua Pilipovsky\thanksref{JPnow}}
\and
\author[2]{Panagiotis Tsiotras}

\address[1]{D. Guggenheim School of Aerospace Engineering, Georgia Institute of Technology, Atlanta, GA \\
(e-mail: jpilipovsky3@gatech.edu)
}
\address[2]{D. Guggenheim School of Aerospace Engineering, Georgia Institute of Technology, Atlanta, GA \\
(e-mail: tsiotras@gatech.edu)}   

\maketitle


\begin{abstract}                          
We consider the problem of data-driven stochastic optimal control of an unknown LTI dynamical system.
Assuming the process noise is normally distributed, we pose the problem of steering the state's mean and covariance to a target normal distribution, under noisy data collected from the underlying system, a problem commonly referred to as covariance steering (CS).
A novel framework for Data-driven Uncertainty quantification and density STeering (DUST) is presented that simultaneously characterizes the noise affecting the measured data and designs an optimal affine-feedback controller to steer the density of the state to a prescribed terminal value.
We use both indirect and direct data-driven design approaches based on the notions of persistency of excitation and subspace identification to exactly represent the mean and covariance dynamics of the state in terms of the data and noise realizations.
Since both the mean and the covariance steering sub-problems are plagued with stochastic uncertainty arising from noisy data collection, we first estimate the noise realization from this dataset and subsequently compute tractable upper bounds on the estimation errors.
The first and second moment steering problems are then solved to optimality using techniques from robust control and robust optimization.
Lastly, we present an alternative control design approach based on the certainty equivalence principle and interpret the problem as one of CS under multiplicative uncertainty.
We analyze the performance and efficacy of each of these data-driven approaches using a case study and compare them with their model-based counterparts.
\end{abstract}

\end{frontmatter}

\section{Introduction}

The pursuit of safe and reliable control under uncertainty stands as a fundamental challenge in control theory.
Traditional model-based techniques, such as the linear-quadratic regulator (LQR) or model-predictive control (MPC) have been extensively explored, and have been shown to be extremely effective at controlling dynamical systems when the model accurately represents the actual physical system.
When there are model inaccuracies due to errors during system identification, robust control techniques \cite{robust_control_overview} have been used to combat these inaccuracies and ensure robust constraint satisfaction and optimality under worst-case conditions.
By the same token, exogenous disturbances affecting the state of a system have been treated in a multitude of ways; when the uncertainties are bounded, the problem 
falls under the realm of robust control \cite{Tube_MPC_OG}; when the uncertainties are probabilistic, the problem is treated using techniques from stochastic control \cite{stochastic_mpc_overview}.

Recently, there has been a paradigm shift from looking at control synthesis as an \textit{indirect} design process of first estimating a model and subsequently solving an optimal control problem using the identified model, to a \textit{direct} control design from raw data collected from the underlying physical system.
This methodological shift has been inspired by the early works on behavioral system theory by Willems et. al. \cite{WFL_OG}, which showed that one can completely characterize the trajectory space of an LTI system by solely using raw data, as long as this data is persistently exciting, a result known as the \textit{Fundamental Lemma} (FL)~\cite{WFL_OG}.
This data-driven formalism is attractive for a variety of reasons: firstly, it bypasses the technicalities and challenges of system identification methods which fail for complex models, and, instead, provides a \textit{direct} end-to-end solution from data input to control output. 
Secondly, it is, in general, a more optimal scheme for control design \cite{DD_control_overview}.
More importantly, however, it provides a \textit{non-parametric} dynamics model that is, by construction, adaptable to any linear system.
As such, many of the traditional model-based control techniques have been re-explored within this data-driven context, with notable foundational works including DeePC \cite{DeePC_Dorfler,DeePC_Allgower} and data-driven LQR \cite{DD_LQR_DePersis,DD_formulas_DePersis,DD_LQR_noisy_DePersis}.

The main issue, however, with the use of Willems' Fundamental Lemma in many real-world applications is the fact that it only holds exactly for \textit{deterministic} LTI systems.
When the system is perturbed by bounded disturbances or is subject to stochastic dynamics or measurement noise, 
these direct data-driven schemes degrade rather quickly~\cite{DeePC_CE}.
As a result, much of the recent work on this front has been tailored to exploring ways to robustify against  noisy data. 
Notable methods along this line of research include suitably regularizing the optimization problem \cite{DeePC_regularized_DR,DeePC_dist_robust}, adding slack variables to account for data infeasibility \cite{DeePC_Dorfler,DeePC_Allgower}, and low-rank approximations of the data using truncated singular value decomposition \cite{DeePC_low_rank,low_rank_Markovsky}.
In fact, certain regularizers have a one-to-one correspondence with the corresponding indirect design methods, such as certainty-equivalence and subspace predictive control \cite{FAVOREEL19994004}, providing a bridge between the two approaches~\cite{DeePC_indirect_direct_bridge,DeePC_CE}.
Extensions such as $\gamma$-DeePC \cite{gamma_DeePC} and generalized-DeePC \cite{gen_DeePC} aim to bridge the gap between subspace-predictive control and DeePC, benefiting from the positive aspects of both methods to tackle noisy data.
In the context of direct state-feedback control design, techniques from robust control have been used to robustly stabilize all possible systems consistent with the data and perturbed by energy-bounded disturbances~\cite{DD_LQR_robust_Petersen,DD_robust_invariance}.

The setting of stochastic predictive control (SPC), however, is far less studied, and the problem of designing optimal controllers for \textit{data-driven} stochastic LTI systems under noisy data is much more challenging.
Existing works on data-driven stochastic control adopt rather restrictive assumptions, such as noise-free offline data \cite{DD_SPC_noise_free} or exact polynomial chaos expansions of stochastic measurements \cite{DD_SPC_PCE1,DD_SPC_PCE2}.
A notable step towards a generalized framework for SPC  is \cite{yin2023stochastic}, where the authors explicitly quantify the statistical properties of error quantification from an indirect design based on noisy data, use a Kalman filter to estimate the initial condition, and reformulate the chance-constrained SPC problem as a semi-definite program (SDP).

In this work, we take a slightly different approach to the data-driven stochastic optimal control problem.
We pose the control problem as one of steering the probability density of the state, as opposed to just the (mean of the) state vector.
The canonical problem in this regard is to steer the state density of a linear system to some target Gaussian distribution subject to independent Gaussian disturbances and chance constraints.
This problem has been solved exactly in both discrete-time \cite{CS_discrete_time_kazu,CS_discrete_time_bakolas} and continuous-time  \cite{CS_continuous_time,OMT_stochastic_control} settings, with demonstrated success in many engineering applications, such as spacecraft rendezvous \cite{CS_IRA}, interplanetary trajectory optimization \cite{CS_interplanetary}, and autonomous driving \cite{CS_MPPI}, to name a few.
In the Gaussian setting, and since the state remains normally distributed for the entire horizon, the problem boils down to steering the first two moments of the state, hence the approach is often referred to as covariance steering (CS). 
Extensions of CS that take into account non-Gaussian disturbances \cite{CS_general_disturbance,CS_martingale_additive}, multiplicative and parametric disturbances \cite{CS_mulitplicative_knaup,CS_multiplicative_bakolas}, distributional uncertainty \cite{CS_dist_robust_wasserstein,CS_dist_robust_chebyshev}, and extensions to nonlinear systems \cite{Bakolas_NLCS,Ridderhof_NLCS} have also been reported in the literature.

\begin{figure*}
    \centering
    \includegraphics[width=\linewidth]{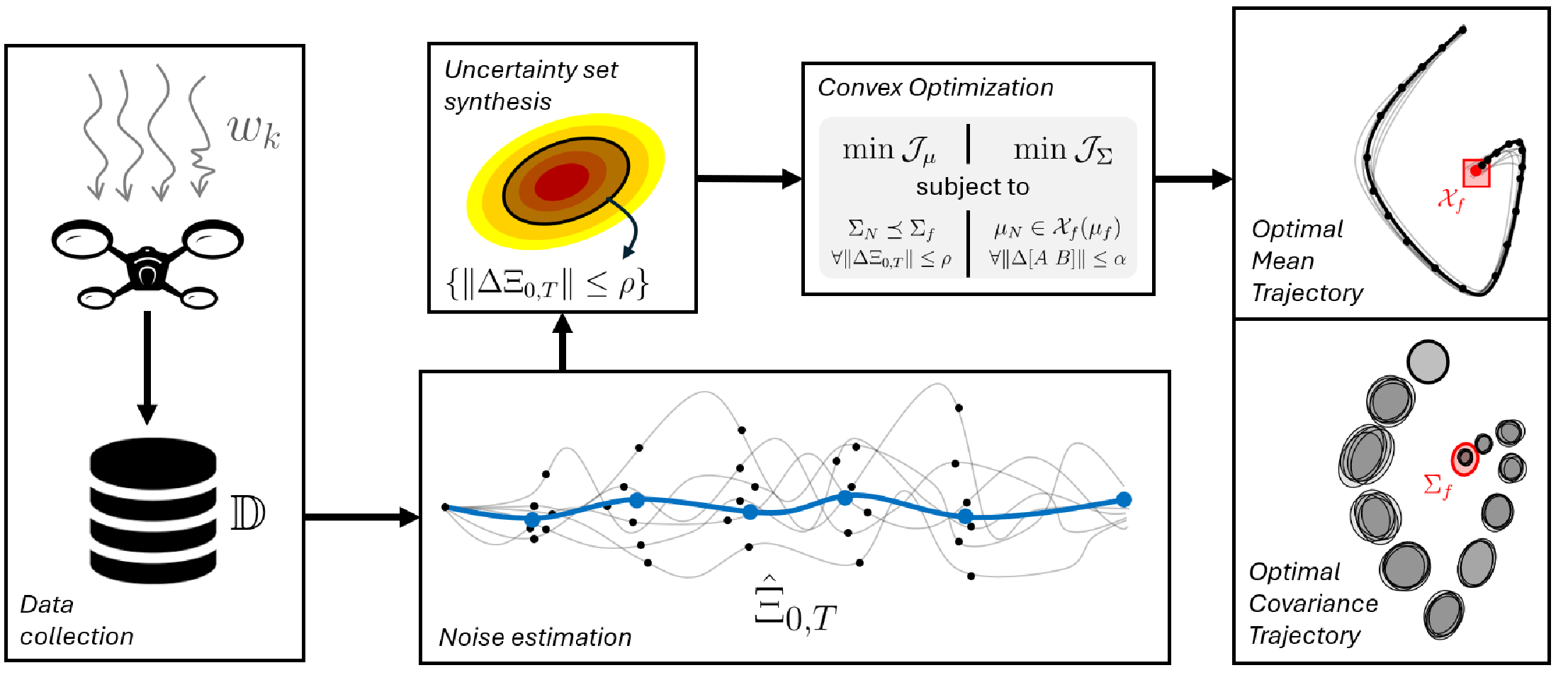}
    \caption{Breakdown of DUST framework into data-collection, noise estimation, uncertainty set construction, and robust control. The noisy dataset $\mathbb{D}$ is used to estimate the past noise realization $\hat{\Xi}_{0,T}$, which is subsequently used to generate norm-bounded uncertainty sets $\Delta_{\mathrm{model}}$ for the (indirect) DD-MS and $\Delta_{\mathrm{noise}}$ for the (direct) DD-CS problems. The end result is optimal moment trajectories that satisfy terminal distributional constraints with high probability.}
    \label{fig:schematic}
\end{figure*}

In this paper, we develop a general framework to steer the distribution of an \textit{unknown} stochastic LTI system using raw data collected offline, instead of using a known system model.
We are interested in the so-called Data-Driven Density Steering (DD-DS) control problem and
we develop a generalized framework to solve this problem, henceforth referred to as \textbf{D}ata-driven \textbf{U}ncertainty quantification and density \textbf{ST}eering (DUST). 
Since we will be dealing primarily with linear systems, we will, alternatively, investigate the Data-Driven Mean Steering (DD-MS) and Data-Driven Covariance Steering (DD-CS) problems.
By combining model-based CS theory with behavioral systems theory and statistical learning, we provide a robust framework to steer both the mean and covariance of the state distribution to a desired terminal distribution.
An illustrative flowchart of the proposed framework is shown in Figure~\ref{fig:schematic}.

We first decompose the problem into one of data-driven mean steering (DD-MS) and one of data-driven covariance steering (DD-CS), as this framework allows for a separation principle for the individual moment trajectories, excluding the presence of constraints (Section~\ref{sec:PS}).
We then follow an indirect certainty-equivalence (CE) route to exactly parameterize the mean dynamics in terms of the unknown noise realizations in the data, resulting in an uncertain quadratic program (Section~\ref{sec:DataDrivenDesign}). 
For covariance control, we parameterize the feedback gains directly using the collected data and use established model-based CS theory to arrive at an uncertain SDP.

To handle the uncertainty during data collection, Section~\ref{sec:noise_estimation} develops novel noise estimation schemes to quantify the noise using techniques from maximum likelihood estimation (MLE) and neural networks. 
We then provide connections with the corresponding indirect design techniques.
Using the known statistical properties from MLE, as well as quantitative notions of persistence of excitation, we construct high-confidence noise estimation error bounds in Section~\ref{sec:error_bounds}.
These uncertainty bounds are then used in Section~\ref{sec:RDDDS} to construct uncertainty sets to 
tractably reformulate the uncertain MS and CS problems as robust control and robust optimization problems, respectively, which can be solved efficiently to optimality using standard off-the-shelf solvers.

Since the noise enters multiplicatively in the indirect design formulation, we alternatively formulate a parametric uncertainty DD-DS (PU-DD-DS) problem and use convex relaxations to tractably solve the original problem in Section~\ref{sec:PUDDCS}.
Lastly, to illustrate the proposed framework, in Section~\ref{sec:sims} we perform an in-depth study of the proposed control design methods, analyzing their efficacy and precision, and compare them with their model-based counterparts.
We conclude with a discussion of the proposed framework, and we offer several avenues for future extensions.

\subsection*{Contributions}

This paper extends our previous work~\cite{DD_CS_no_noise,DD_CS_noise_conservative}, where we studied the limiting case of this problem, namely, the dynamics were assumed to be deterministic and the uncertainty resided solely in the boundary conditions.
In this case, Willems Fundamental Lemma holds exactly, and we can achieve an exact correspondence with model-based designs.
In addition, \cite{DD_CS_noise_conservative} studied the full data-driven density steering problem with noise quantification.
However, in that work, the uncertainty sets were unnecessarily conservative (see Section~\ref{sec:error_bounds}).
In this paper, we remove this restrictive assumption.

In summary, the contributions of this work are as follows:

\begin{enumerate}

    \item[1.] 
    We introduce a unified data-driven framework for steering the full state distribution of an unknown stochastic LTI system using offline data, by combining behavioral systems theory with model-based covariance steering ideas.

    \item[2.] 
    We develop data-driven noise estimation and uncertainty quantification tools to find the most likely noise realization, along with its covariance, and provide finite-sample/asymptotic, high-confidence ambiguity sets for the corresponding estimation errors. 

    \item[3.] 
    We provide tractable (convex optimization) formulations for both the data-driven mean steering (DD-MS) and covariance steering (DD-CS) problems.
    Each problem is solved so as 
   to guarantee terminal distribution constraint satisfaction with high probability.

    \item[4.] 
    We present a parametric-uncertainty (PU-DD-DS) formulation, interpreting the certainty equivalent (CE) problem formulation in terms of controlling a system subject to multiplicative uncertainty, and provide a tractable convex formulation for its solution, showcasing an alternative approach to the data-driven density steering problem.

    \item[5.]
    We evaluate the performance and efficacy of the proposed data-driven designs against alternative CE-based approaches and against model-based counterparts.

\end{enumerate}

%


\subsection*{Notation}

Real-valued vectors are denoted by lowercase letters, $u\in\Re^{m}$, matrices are denoted by upper-case letters, $V\in\Re^{n\times m}$, and random vectors are denoted by boldface letters, $\vct w\in\Re^{p}$.
$\vct\rchi^2_{p}$ denotes the chi-square distribution with $p$ degrees of freedom, and $Q_{\vct\rchi_p^2}(q)$ denotes its associated quantile function (i.e., the inverse of the cumulative distribution function) at the quantile $q\in (0, 1)$.
The Kronecker product is denoted as $\otimes$ and the vectorization of a matrix $A$ is denoted as $\mathrm{vec}(A) = [a_1^\intercal, \ldots, a_{M}^\intercal]^\intercal$, where $a_i$ is the $i$th column of $A$.
We use the shorthand notation $[A; B]$ to denote the vertical stacking of two matrices or vectors of compatible dimension.
We define the set $\dbracket{T} \triangleq \{1,\ldots,T\}$ and similarly $\dbracket{T}_{0} \triangleq \{0\} \cup \dbracket{T}$, for any natural number $T\in\mathbb{N}$.
Given two matrices $A$ and $B$ of compatible dimensions, the inner product between $A$ and $B$ is defined as $\langle A, B \rangle \triangleq \mathrm{tr} (A^\intercal B)$.
We denote the minimum non-zero eigenvalue of a matrix by $\lambda_{\min}^{+}$.
For a matrix $A$, we denote its Moore-Penrose pseudoinverse by $A^{\dagger}$.
%
We denote the induced matrix two-norm (or spectral norm) by $\|\cdot\|$ 
and the matrix Frobenius norm by $\|\cdot\|_{\mathrm{F}}$.
We denote the identity matrix of size $n\times n$ as $I_{n}$ and the zero matrix of size $m\times n$ as $0_{m\times n}$.
The set of $n \times n$ symmetric matrices is denoted by $\Sb^n$, the set of positive 
definite matrices by $\Sb^n_{++}$, and the set of positive semi-definite matrices by $\Sb^n_{+}$.
For simplicity, we denote the $m$-long vector of zeros as $0_{m}$.
Often, we will drop the subscript in these matrices if the dimension is clear from the context.
Lastly, we succinctly denote a discrete-time signal $z_{0},z_{1},\ldots,z_{T}$ by $\{z_k\}_{k=0}^{T}$.


\section{Problem Formulation}~\label{sec:PS}
\subsection{Problem Statement}~\label{subsec:DDCS_problem}

We consider the following linear, discrete-time stochastic time-invariant system
\begin{equation}~\label{eq:dynamics}
	\vct x_{k+1} = A \vct x_k + B \vct u_k + D \vct w_k,
\end{equation}
where $\vct x_k\in\mathbb{R}^{n}$ is the state, $\vct u_k\in\mathbb{R}^{m}$ is the control input, and $\vct w_k\sim \mathcal{N}(0, I_{d})$ are i.i.d. Gaussian disturbances, with time steps $k \in \dbracket{N-1}_0$, where $N$ represents the horizon length.
Alternatively, we may re-write \eqref{eq:dynamics} as 
\begin{equation}~\label{eq:dynamics2}
	\vct x_{k+1} = A \vct x_k + B \vct u_k + \vct \xi_k,
\end{equation}
where $\vct \xi_k\sim \mathcal{N}(0, \Sigma_{\vct\xi})$ are i.i.d. random vectors for all $k\in\dbracket{T-1}_{0}$, and where $\Sigma_{\vct\xi} \triangleq DD^\intercal$ is the covariance of the noise.
%
%
The constant system matrices $A, B, D$ are assumed to be \textit{unknown}.
The initial state $\vct x_0$ is a random $n$-dimensional vector drawn from the normal distribution
\begin{equation}~\label{eq:init_condition}
	\vct x_0 \sim \mathcal{N}(\mu_i, \Sigma_i),
\end{equation}
where $\mu_i\in\mathbb{R}^n$ is the initial state mean and $\Sigma_i\in\mathbb{S}^n_{++}$
is the initial state covariance.
It is assumed that the initial state is independent of the noise sequence, that is, $\E[\vct x_0 \vct w_k^\intercal] = 0$, for all $k\in\dbracket{N-1}_{0}$.

The objective is to steer the trajectories of (\ref{eq:dynamics}) from the initial distribution (\ref{eq:init_condition}) to the terminal distribution
\begin{equation}~\label{eq:term_condition}
	\vct x_N = \vct x_f \sim \mathcal{N}(\mu_f, \Sigma_f),
\end{equation}
where $\mu_f\in\mathbb{R}^n$ and $\Sigma_f\in\Sb^n_{++}$ are the desired state mean and covariance at time step $N$, respectively. 
The cost function to be minimized is 
\begin{equation}~\label{eq:cost_function}
	J(\vct u) \triangleq \mathbb{E}\bigg[\sum_{k=0}^{N-1} (\vct x_k - x_k^{r})^\intercal Q_k (\vct x_k - x_k^r) + \vct u_k^\intercal R_k \vct u_k\bigg],
\end{equation}
where $\{x_k^{r}\}_{k=0}^{N-1}$ is a given reference trajectory, and $Q_k\succeq 0$ and $R_k \succ 0$ for all $k \in \dbracket{N-1}_{0}$. 
\begin{rem}~\label{rem:controllability}
	We assume that the system (\ref{eq:dynamics}) is controllable, that is, for any $x_0,x_f\in\mathbb{R}^n$, and no noise ($w_k\equiv 0,~ k\in\dbracket{N-1}_{0}$), there exists a sequence of control inputs $\{u_k\}_{k=0}^{N-1}$ that steer the system from $x_0$ to $x_f$.
\end{rem}
It is assumed that we are given a $T$-long trajectory dataset $\mathbb{D} \triangleq \{x_k^{(d)}, u_k^{(d)}, x_{T}^{(d)}\}_{k=0}^{T-1}$ for control design.
Although 
it is also possible to synthesize the controllers using multiple episodic datasets $\mathbb{D}_{\ell}$ ($\ell=1,\ldots,M$), for simplicity, in the present work we assume only a single dataset.
Borrowing from the work in \cite{CS_george_journal},
we assume affine state feedback policies of the form, 
\begin{equation}~\label{eq:controlLaw}
	\vct u_k = K_k(\vct x_k - \mu_k) + v_k,
\end{equation}
parameterized by an open-loop control sequence $v = \{v_k\}_{k=0}^{N-1}$ and a feedback gain sequence $K = \{K_k\}_{k=0}^{N-1}$,
where 
$\mu_k \triangleq{} \mathbb{E}[\vct x_k]$ is the mean state.

%
In summary, the data-driven distribution steering (DD-DS) problem is stated below.

\begin{prob}[DD-DS]~\label{problem:DDCS_full}
	Given the dataset $\mathbb{D}$, find
    the feedforward $\{v_k\}_{k=0}^{N-1}$ and feedback gain 
 $\{K_k\}_{k=0}^{N-1}$ sequences
 that minimize the objective function (\ref{eq:cost_function}), subject to the initial state (\ref{eq:init_condition}) and terminal state (\ref{eq:term_condition}) boundary conditions.
\end{prob}

We note that under the control law \eqref{eq:controlLaw}, and with complete system knowledge, it is possible to re-write Problem~\ref{problem:DDCS_full} as a convex program~\cite{CS_george_journal}, which can be solved to optimality using off-the-shelf solvers \cite{MOSEK}.
Next, we proceed to solve Problem~\ref{problem:DDCS_full} in the case where the system matrices are not known.

\subsection{Problem Reformulation}~\label{subsec:PS_reformulation}

Since the state distribution remains Gaussian at all time steps, we decompose the system dynamics \eqref{eq:dynamics} into the mean dynamics and covariance dynamics.
Plugging in the control law \eqref{eq:controlLaw} into the dynamics \eqref{eq:dynamics} yields the decoupled dynamics
\begin{subequations}~\label{eq:dynamics_decoupled}
	\begin{align}
		\mu_{k+1} &= A\mu_k + B v_k, \label{eq:meanDynamics} \\
		\Sigma_{k+1} &= (A + B K_k)\Sigma_k(A + B K_k)^\intercal + DD^\intercal, \label{eq:covDynamics}
	\end{align}
\end{subequations}
where the state covariance $\Sigma_k \triangleq \mathbb{E} [ (\vct x_k - \mu_k) (\vct x_k - \mu_k)^\intercal]$.
Note that $v_k\in\Re^{m}$ controls the state mean and
$K_k\in\Re^{m\times n}$ controls the state covariance.

Contrary to the approaches in \cite{CS_discrete_time_bakolas,CS_discrete_time_kazu} that formulate a convex program in the lifted space of state and control trajectories, in this work, and similarly to \cite{CS_george_journal}, we treat the moments of the intermediate states $\{\mu_k,\Sigma_k \}_{k=0}^{N}$ over the steering horizon as decision variables subject to the dynamic constraints~\eqref{eq:dynamics_decoupled}.

First, and similar to the decoupled dynamics~\eqref{eq:dynamics_decoupled}, the cost function \eqref{eq:cost_function} can be decoupled in terms of the mean and covariance as follows
\begin{equation}
		J = J_{\mu}(\mu_k, v_k) + J_{\Sigma}(\Sigma_k, K_k),
\end{equation}
where,
\begin{subequations}
	\begin{align}
		J_{\mu} &\triangleq \sum_{k=0}^{N-1}\Big((\mu_k - x_k^r)^\intercal Q_k (\mu_k - x_k^r) + v_k^\intercal R_k v_k\Big), \label{eq:meanCost} \\
		J_{\Sigma} &\triangleq \sum_{k=0}^{N-1}\Big(\textrm{tr}(Q_k\Sigma_k) + \textrm{tr}(R_kK_k\Sigma_k K_k^\intercal) \Big). \label{eq:covCost}
	\end{align}
\end{subequations}
The two boundary conditions \eqref{eq:init_condition} and \eqref{eq:term_condition} are written as 
\begin{subequations}
	\begin{align}
		&\mu_0 = \mu_i, \quad \mu_N = \mu_f, \label{eq:meanBCs} \\
		&\Sigma_0 = \Sigma_i, \quad \Sigma_N = \Sigma_f, \label{eq:covBCs}
	\end{align}
\end{subequations}
where $\Sigma_i, \Sigma_f \succ 0$.
Problem~\ref{problem:DDCS_full} is now recast as the following two sub-problems.
\begin{prob}[DD-MS]~\label{problem:DDMS}
	Given the (unknown) mean dynamics \eqref{eq:meanDynamics}, find the
    optimal feed-forward control sequence $\{v_k\}_{k=0}^{N-1}$ and the corresponding
    optimal mean trajectory $\{\mu_k\}_{k=0}^{N}$ that minimize the mean cost \eqref{eq:meanCost} subject to the boundary conditions \eqref{eq:meanBCs}.
\end{prob} 
\begin{prob}[DD-CS]~\label{problem:DDCS}
	Given the (unknown) covariance dynamics \eqref{eq:covDynamics}, find the optimal 
    feedback gain sequence $\{K_k\}_{k=0}^{N-1}$
    and the corresponding
    covariance trajectory $\{\Sigma_k\}_{k=0}^{N}$  that minimize the covariance cost \eqref{eq:covCost} subject to the boundary conditions \eqref{eq:covBCs}.
\end{prob} 
%
%
\begin{rem}~\label{rem:convexity}
	Under complete model knowledge, Problem~\ref{problem:DDMS} is a standard quadratic program with linear constraints that can be solved analytically given knowledge of the system matrices.
	Problem~\ref{problem:DDCS}, on the other hand, is a non-linear and non-convex program due to the term $K_k \Sigma_k K_k^\intercal$ arising both in the cost function \eqref{eq:covCost} and the covariance dynamics \eqref{eq:covDynamics}.
    However, 
    as it was shown in~\cite{CS_george_journal} this problem can be relaxed to a convex optimization problem by the introduction of additional slack variables.
    Moreover, this relaxation is lossless. 
    For further details, we refer the interested reader to~\cite{CS_george_journal}.
\end{rem}


\section{Data-Driven Parameterization}~\label{sec:DataDrivenDesign}

In this section, we review the main concepts from behavioral systems theory~\cite{behavioral_systems_theory} that will allow us to parametrize the decision variables in Problems~\ref{problem:DDMS} and \ref{problem:DDCS} in terms of the collected input and output data streams.

First, recall the following definitions.
\begin{defn}~\label{def:1}
	Given a signal $\{z_k\}$ where $z_k\in\Re^{\sigma}$, we denote the Hankel matrix of depth $\ell$ by
	\begin{equation}
		Z_{i,\ell, j} \triangleq 
		\begin{bmatrix}
			z_i & z_{i+1} & \ldots & z_{i+j-1} \\
			z_{i+1} & z_{i+2} & \ldots & z_{i + j} \\
			\vdots & \vdots & \ddots & \vdots \\
			z_{i+\ell-1} & z_{i+\ell} & \ldots & z_{i+\ell+j-2}
		\end{bmatrix} \in \Re^{\sigma\ell \times j},
	\end{equation}
	where $i\in\mathbb{Z}$ and $\ell,j\in\mathbb{N}$.
	For shorthand of notation, if $\ell=1$, we denote the Hankel matrix by
	\begin{equation}
		Z_{i,1,j} \equiv Z_{i,j} = [z_i \ z_{i+1} \ \ldots \ z_{i+j-1}].
	\end{equation}
\end{defn}

\begin{defn}~\label{def:2}
	The signal $\{z_k\}_{k=0}^{T-1}: \dbracket{T-1}_{0}\rightarrow\Re^{\sigma T}$
    is \emph{persistently exciting} of order $\ell$ if the matrix $Z_{0,\ell,j}$ with $j = T - \ell + 1$ has rank $\sigma\ell$.
\end{defn}
%
%
Suppose we carry out an experiment of duration $T\in\mathbb{N}$, where we collect input and noisy state data $\{u_k^{(d)}\}_{k=0}^{T-1}$ and $\{x_k^{(d)}\}_{k=0}^{T}$, respectively.
Let the corresponding Hankel matrices for the input sequence, state sequence, and shifted state sequence (with $\ell = 1$) be 
\begin{subequations}~\label{eqUX}
    \begin{align}
        U_{0,T} &\triangleq [u_0^{(d)} \ u_1^{(d)} \ \ldots \ u_{T-1}^{(d)}], \\
        X_{0,T} &\triangleq [x_0^{(d)} \ x_1^{(d)} \ \ldots \ x_{T-1}^{(d)}], \\
        X_{1,T} &\triangleq [x_1^{(d)} \ x_2^{(d)} \ \ldots \ x_{T}^{(d)}].
    \end{align}
\end{subequations}
Using~\eqref{eqUX}, the system dynamics~\eqref{eq:dynamics2} take the form
\begin{equation} \label{eq:dynamics_realization}
X_{1,T} = A X_{0,T} + B U_{0,T} + \Xi_{0,T},
\end{equation}
where $\Xi_{0,T} \triangleq [\xi_{0}^{(d)}, \ldots, \xi_{T-1}^{(d)}]\in\Re^{n\times T}$ is the Hankel matrix of the (unknown) disturbances, and $\vct\xi_k^{(d)}\sim\mathcal{N}(0,\Sigma_{\vct\xi})$.

Assuming that the data is \textit{persistently exciting} (PE), the block Hankel matrix of the input and state data has full row rank
\begin{equation}~\label{eq:PE}
	\mathrm{rank}
	\begin{bmatrix}
		U_{0,T} \\
		X_{0,T}
	\end{bmatrix} = n + m.
\end{equation}
This PE assumption is crucial for direct data-driven control design, and is generally a mild assumption in practice, especially when noisy data is used \cite{DD_LQR_noisy_DePersis}.
%
%
The condition in \eqref{eq:PE} implies that \textit{any} arbitrary input-state sequence of \eqref{eq:dynamics} can be expressed as a linear combination of the collected input-state data. 
Furthermore, 
this idea can be used  to parameterize any arbitrary feedback interconnection~\cite{DD_LQR_DePersis}.
In the following section, we parameterize the feedback gains in terms of the input-state data and reformulate the covariance steering problem as a semi-definite program (SDP).
First, we focus on the mean steering problem,  by first identifying the system matrices using the collected input-state data.
Because of this extra system identification step, the data-driven mean steering problem is solved using an indirect approach.

\subsection{Indirect Data-Driven Mean Steering (DD-MS)}

Given the mean dynamics \eqref{eq:meanDynamics} in terms of the open-loop control $v_k$, the PE condition \eqref{eq:PE} provides a system identification type of result using the following theorem.

\begin{thm}~\label{theorem:sysID}
    Suppose $\mathbb{D} = \{x_k^{(d)}, u_k^{(d)}, x_{T}^{(d)}\}_{k=0}^{T-1}$ is a dataset collected from the underlying system \eqref{eq:dynamics} such that the rank condition \eqref{eq:PE} holds.
    Then, system \eqref{eq:meanDynamics} has the following equivalent representation
    \begin{equation} ~\label{eq:mut+1}
        \mu_{k+1} = (X_{1,T} - \Xi_{0,T})
        \begin{bmatrix}
            U_{0,T} \\
            X_{0,T}
        \end{bmatrix}^{\dagger}
        \begin{bmatrix}
            v_k \\
            \mu_k
        \end{bmatrix}.
    \end{equation}
\end{thm}
\begin{pf}
	See \cite{DD_formulas_DePersis} for details.
\end{pf}
%
	Theorem~\ref{theorem:sysID} provides a data-based representation of a linear system.
	Assuming exact knowledge of the noise realization $\Xi_{0,T}$, one may equivalently interpret equation~\eqref{eq:mut+1}
	as the solution to the least-squares problem
	\begin{equation}~\label{eq:sysID_least_squares}
		\min_{B,A} \ \left\|X_{1,T} - \Xi_{0,T} - [B \ \ A]
		\begin{bmatrix}
			U_{0,T} \\
			X_{0,T}
		\end{bmatrix}
		\right\|_{\mathrm{F}},
	\end{equation}
	where $\|\cdot\|_{\mathrm{F}}$ is the Frobenius norm.
	Data-driven indirect designs based on the certainty-equivalence (CE) principle compute an approximate system description by solving \eqref{eq:sysID_least_squares} assuming no noise (i.e., $\Xi_{0,T} = 0$).
    %
%
Using Theorem~\ref{theorem:sysID}, we can express the mean steering problem as the following convex optimization problem
\begin{empheqboxed}
    \begin{subequations}~\label{eq:meanProblem}
    	\begin{equation}
    		\min_{\mu_k, v_k} J_{\mu} = \sum_{k=0}^{N-1} \Big((\mu_k-x_k^{r})^\intercal Q_k (\mu_k-x_k^{r}) + v_k^\intercal R_k v_k\Big), \label{eq:meanProblem_cost}
    	\end{equation}
    	such that, for all $k\in\dbracket{N-1}_{0}$,
    	\begin{equation}
    	F_{v}(\Xi_{0,T}) v_k +	F_{\mu}(\Xi_{0,T}) \mu_k  - \mu_{k+1} = 0, \label{eq:meanProblem_eqConstraint1}
    	\end{equation}
    with the boundary conditions \eqref{eq:meanBCs}, where $F_{\mu}\in\Re^{n\times n}$ and $F_{v}\in\Re^{n\times m}$ result from the partition of the matrix $F$ as follows
    \begin{equation}~\label{eq:mean_dynamics_partition}
    	F \triangleq (X_{1,T} - \Xi_{0,T})
    	\begin{bmatrix}
    		U_{0,T} \\
    		X_{0,T}
    	\end{bmatrix}^{\dagger} = 
    	\begin{bmatrix}
    		F_{v}(\Xi_{0,T}) & F_{\mu}(\Xi_{0,T})
    	\end{bmatrix}.
    \end{equation}
 \end{subequations}
\end{empheqboxed}

\subsection{Direct Data-Driven Covariance Steering (DD-CS)}~\label{sec:directCS}

To solve the DD-CS problem, first notice that 
from the rank condition \eqref{eq:PE}, we can express the feedback gains as follows
\begin{equation}~\label{eq:WFL_gains}
	\begin{bmatrix}
		K_k \\
		I_n
	\end{bmatrix}
	= 
	\begin{bmatrix}
		U_{0,T} \\
		X_{0,T}
	\end{bmatrix}
	G_k,
\end{equation}
where $G_k \in \Re^{T\times n}$ are newly defined decision variables that provide the link between the feedback gains and the input-state data.
Furthermore, using this parameterization, we can re-write the covariance dynamics \eqref{eq:covDynamics} as
\begin{align}  \label{eq:covDynamics_Gvars}
	\Sigma_{k+1} &= [B \ \ A]
	\begin{bmatrix}
		K_k \\
		I_n
	\end{bmatrix}
	\Sigma_k 
	\begin{bmatrix}
		K_k \\
		I_n
	\end{bmatrix}^\intercal
	[B \ \ A]^\intercal + DD^\intercal \nonumber \\
	&= (X_{1,T} - \Xi_{0,T}) G_k \Sigma_k G_k^\intercal (X_{1,T} - \Xi_{0,T})^\intercal + \Sigma_{\vct\xi}.
\end{align}

Similarly, the covariance cost \eqref{eq:covCost} can be re-written as 
\begin{equation}~\label{eq:covCost_Gvars}
	J_{\Sigma, k} = \mathrm{tr}(Q_k\Sigma_k) + \mathrm{tr}(R_k U_{0,T} G_k \Sigma_k G_k^\intercal U_{0,T}^\intercal).
\end{equation}
To remedy the nonlinearity $G_k\Sigma_k G_k^\intercal$ in the covariance dynamics and the cost, define the new decision variables $S_k \triangleq G_k \Sigma_k\in\Re^{T\times n}$, which yields
\begin{equation}~\label{eq:covDynamics_Svars}
	\Sigma_{k+1} = (X_{1,T} - \Xi_{0,T}) S_k \Sigma_k^{-1} S_k^\intercal (X_{1,T} - \Xi_{0,T})^\intercal + \Sigma_{\vct\xi},
\end{equation}
and 
\begin{equation}~\label{eq:covCost_Svars}
	J_{\Sigma, k} = \mathrm{tr}(Q_k \Sigma_k) + \mathrm{tr}(R_k U_{0,T} S_k \Sigma_k^{-1} S_k^\intercal U_{0,T}^\intercal).
\end{equation}
This problem is still non-convex due to the nonlinear term $S_k\Sigma_k^{-1}S_k^\intercal$.
To this end, we define a new set of decision variables $Y_k \succeq S_k \Sigma_k^{-1} S_k^\intercal$ and relax the objective function with these new variables.
Similarly, we relax the covariance propagation constraints to soft (inequality) constraints, which yields the relaxed optimization problem
\begin{empheqboxed}
    \begin{subequations}~\label{eq:convexProblem}
        \begin{equation}
            \min_{\Sigma_k, S_k, Y_k} \bar{J}_{\Sigma} = \sum_{k=0}^{N-1}\left(\mathrm{tr}(Q_k\Sigma_k) + \mathrm{tr}(R_kU_{0,T}Y_kU_{0,T}^\intercal)\right),  \label{eq:convexProblem_cost}
        \end{equation}
        such that, for all $k\in\dbracket{N-1}_{0}$,
        \begin{align}
            &S_k \Sigma_k^{-1} S_k^\intercal - Y_k \preceq 0, \label{eq:convexProblem_ineqConstraint} \\
            &(X_{1,T} - \Xi_{0,T}) S_k \Sigma_k^{-1} S_k^\intercal (X_{1,T} - \Xi_{0,T})^\intercal \nonumber \\
            &\hspace{3.5cm}+\Sigma_{\vct{\xi}} - \Sigma_{k+1} \preceq 0, \label{eq:convexProblem_ineqConstraint2} \\
            &\Sigma_k - X_{0,T} S_k = 0, \label{eq:convexProblem_eqConstraint1} \\
            &\Sigma_{N} - \Sigma_{f} = 0. \label{eq:convexProblem_eqConstraint2}
        \end{align}
    \end{subequations}
\end{empheqboxed}
The equality constraint \eqref{eq:convexProblem_eqConstraint1} comes from the second block in \eqref{eq:WFL_gains} by multiplying $\Sigma_k$ on the right.
The relaxed optimization problem \eqref{eq:convexProblem} is convex, since the constraints \eqref{eq:convexProblem_ineqConstraint} and \eqref{eq:convexProblem_ineqConstraint2} can be written using the Schur complement in terms of the linear matrix inequalities (LMI)
\begin{subequations}
    \begin{align}
        & \hspace*{6mm} \begin{bmatrix}
            \Sigma_k & S_k^\intercal \\
            S_k & Y_k
        \end{bmatrix} \succeq 0, \label{eq:convex_program_LMI_1} \\
    G_\Sigma^k &\triangleq	
    \begin{bmatrix}
        \hspace*{-3mm}	\Sigma_{k+1} - \Sigma_{\vct\xi} & (X_{1,T} - \Xi_{0,T})S_k \\
            S_k^\intercal (X_{1,T} - \Xi_{0,T})^\intercal & \Sigma_{k}
        \end{bmatrix} \succeq 0. \label{eq:convex_program_LMI_2}
    \end{align}
\end{subequations}
The cost \eqref{eq:convexProblem_cost} and the equality constraints \eqref{eq:convexProblem_eqConstraint1}-\eqref{eq:convexProblem_eqConstraint2}, on the other hand, are linear in all the decision variables, and hence are trivially convex.

In Appendix~\ref{AppA} we show that 
the optimal $\{\Sigma_k^\star\}$ from the solution of \eqref{eq:convexProblem} satisfy $P_{k} \preceq \Sigma_k^\star$ with respect to the closed-loop covariance evolution under $K_k^\star$, where $P_k$ denotes the true covariance at time step $k$.
Hence, the solution from the relaxed DD-CS problem \eqref{eq:convexProblem} will satisfy the terminal covariance constraints $P_{N} \preceq \Sigma_f$. %
\color{black}
It should be mentioned, however, that, in general, the relaxed convex program \eqref{eq:convexProblem} is \textit{not} lossless, contrary to the model-based case~\cite{CS_george_journal}.

\color{black}


\section{Noise Estimation Algorithms}~\label{sec:noise_estimation}

The optimization problems \eqref{eq:meanProblem} and \eqref{eq:convexProblem}, as they stand, albeit convex, are still intractable because we know neither the disturbance covariance $\Sigma_{\vct\xi}$ nor the noise realization history $\Xi_{0,T}$.
To this end, in this section we propose a method to estimate the noise realization matrix $\Xi_{0,T}$ and the disturbance matrix $\Sigma_{\vct \xi}$ using the collected data $\mathbb{D}$.
We propose two methods 
to find these matrices: the first method trains a feed-forward neural network (NN) to estimate both the disturbance and noise realization matrices; the second method estimates $\Xi_{0,T}$ and $\Sigma_{\vct\xi}$ using maximum likelihood estimation (MLE). 
We illustrate these noise estimation techniques next.

\subsection{ML Noise Estimation}~\label{subsec:MLE}

To encode the stochastic linear system dynamics~\eqref{eq:dynamics2} as a constraint that we can use in the MLE scheme, we need to enforce \textit{consistency} of the realization data.
To this end, any realization of the dynamics \eqref{eq:dynamics2} must satisfy
\eqref{eq:dynamics_realization}.
For notational convenience, henceforth, we denote the augmented Hankel matrix in \eqref{eqUX} as
\begin{equation} \label{eq:Smat:def}
	\S \triangleq
	\begin{bmatrix}
		U_{0,T} \\
		X_{0,T}
	\end{bmatrix} \in \Re^{(m + n) \times T},
\end{equation}
from which we may re-write the dynamics realization \eqref{eq:dynamics_realization} as 
\begin{equation} \label{eq:dynamics_realizationN}
X_{1,T} = [B \ A]\S + \Xi_{0,T}.
\end{equation}
Additionally, noting that the matrix pseudoinverse satisfies the property $\S \S^\dagger \S = \S$, equation~\eqref{eq:dynamics_realizationN} can be written, equivalently, as
\begin{equation} \label{eq:dynamics_realization_rewrite}
	X_{1,T} - \Xi_{0,T} = [B \ A] \S \S^\dagger \S.
\end{equation}
Inserting the relation $X_{1,T} - \Xi_{0,T} = [B \ A]\S$ into the right-hand side of \eqref{eq:dynamics_realization_rewrite} yields
\begin{equation}~\label{eq:consistency_equation}
	(X_{1,T} - \Xi_{0,T})(I_{T} - \S^\dagger \S) = 0.
\end{equation}
Equation \eqref{eq:consistency_equation} is a \textit{model-free} type of condition that must be satisfied for all noisy linear system data realizations, and hence is a consistency relation for any feasible set $\S$ of data.

Given the constraint \eqref{eq:consistency_equation}, the MLE problem then becomes
\begin{empheqboxed}
\begin{subequations}~\label{eq:ML_problem}
	\begin{align}
		&\max_{\Xi_{0,T}, \Sigma_{\vct\xi}} \ &&\mathcal{J}_{\mathrm{ML}}(\Xi_{0,T}, \Sigma_{\vct\xi}) = \sum_{k=0}^{T - 1} \log \rho_{\vct\xi}(\xi_k) \label{eq:ML_problem_cost} \\
		&\quad  \ &&(X_{1,T} - \Xi_{0,T})(I_{T} - \S^\dagger \S) = 0,
	\end{align}
\end{subequations}
\end{empheqboxed}
where $\rho_{\vct\xi}(\xi)$ is the probability density function (PDF) of the random vector $\vct\xi$, given as
\begin{equation}~\label{eq:Gaussian}
	\rho_{\vct\xi}(\xi) = \frac{1}{(2\pi)^{n/2}}(\det\Sigma_{\vct\xi})^{-1/2}\exp\left(-\frac{1}{2}\xi^\intercal\Sigma_{\vct\xi}^{-1} \xi\right).
\end{equation}
\begin{rem}
Since the dynamics \eqref{eq:dynamics} are uncertain, it follows that there may be \textit{multiple} noise realizations $\Xi_{0,T}$ that satisfy the linear dynamics constraints~\eqref{eq:consistency_equation}.
Consequently, the purpose of the constrained MLE problem \eqref{eq:ML_problem} is to find the \textit{most likely}
noise realization sequence $\hat{\Xi}_{0,T}$ given that the noise is normally distributed according to \eqref{eq:Gaussian}.
\end{rem}
The next theorem provides the optimal solution to the MLE problem \eqref{eq:ML_problem}.

\begin{thm}~\label{thm:MLE_solution}
    The solution to the MLE problem for the most probable noise realization $\Xi_{0,T}$ and disturbance covariance matrix $\Sigma_{\vct\xi}$ is given by
    \begin{subequations}~\label{eq:MLE_soln}
        \begin{align}
            \Xi_{0,T}^\star &= X_{1,T}(I_{T} - \mathcal{S}^\dagger \mathcal{S}), \label{eq:MLE_noise_soln} \\
            \Sigma_{\vct\xi}^{\star} &= \frac{1}{T}X_{1,T}(I_{T} - \mathcal{S}^\dagger\mathcal{S})X_{1,T}^\intercal. \label{eq:MLE_cov_soln}
        \end{align}
    \end{subequations}
\end{thm}

\begin{pf}
The proof is given in Appendix~\ref{AppB}.
\end{pf}

\begin{rem}
    The solution \eqref{eq:MLE_soln} of the MLE program in \eqref{eq:ML_problem} is contingent on $\Sigma_{\vct\xi} \succ 0$. 
    In fact, the optimal covariance estimate is simply the sample covariance of the dataset with respect to the estimated noise realizations, i.e., $\Sigma_{\vct\xi}^{\star} = \frac{1}{T}\Xi_{0,T}^\star (\Xi_{0,T}^\star)^\intercal$.
    If $\Sigma_{\vct\xi}$ is singular, however, then $\log\det\Sigma_{\vct\xi}$ is undefined, hence the problem is infeasible.
    As a result, other methods, such as NN estimation (Section~\ref{subsec:NN} below), or regularization techniques (e.g., GLASSO \cite{GLASSO}, distributionally-robust estimation \cite{Kuhn_inverse_covariance}) should be used in these cases, instead.
    It should also be noted that such degenerate cases arise when the number of disturbance channels is \textit{less} than the number of states channels, i.e., $D\in\Re^{n\times d}$, with $d < n$.
\end{rem}

\subsection{NN Noise Estimation}~\label{subsec:NN}

An alternative way to estimate the realization noise given the data set $\mathbb{D}$ 
is by training a feed-forward neural network.
To this end, let $f: \Re^{n(T + 1) + mT} \rightarrow \Re^{nT}$ denote the NN mapping, where the input is $x \triangleq [x_0^{(d)^\intercal}, x_1^{(d)^\intercal},\ldots, x_T^{(d)^\intercal}, u_0^{(d)^\intercal},\ldots, u_{T-1}^{(d)^\intercal}]^\intercal$, and the output is $y \triangleq 
[\xi_0^{(d)^\intercal},\ldots, \xi_{T-1}^{(d)^\intercal}]^\intercal
= \mathrm{vec}(\Xi_{0,T})$.
Without loss of generality, we may consider a NN with ReLU activation functions.
A ReLU NN transforms, at each layer $k$, the input as
\begin{equation*}
	x_{k} = f_k(x_{k-1}) = \max(W_k x_{k-1} + b_k, 0),
\end{equation*}
where $W_k \in \mathbb{R}^{\ell_{k}\times \ell_{k-1}}$ is the weight and $b_k\in\mathbb{R}^{\ell_{k}}$ is the bias.
Once an estimate, $\hat{\Xi}_{0,T}$, of the noise realization history is obtained, the noise covariance is computed simply as the sample covariance of the estimated data via \eqref{eq:MLE_cov_soln}.
Of course, it is also possible to construct more elaborate networks to estimate both matrices of interest simultaneously.

Learning-based disturbance estimators (e.g., NN above) can achieve strong empirical performance on specific problem instances; however, in this work we choose to adopt an MLE-based approach because it yields closed-form, finite-sample characterizations of the estimation errors (see Appendix~\ref{AppF}), which we leverage to construct tractable convex uncertainty sets that certify terminal constraints with probability greater than $1-\delta$.
By contrast, the NN scheme is supervised and presumes access to an \emph{oracle} that provides noise-realization labels.
Moreover, the learned mapping depends on the system dynamics and on the data-collection horizon $T$, typically necessitating re-training across regimes, and its performance is sensitive to the informativeness of the inputs used to excite the system.
Systematically quantifying this data-informativeness/identification-performance link for learned disturbance models remains an interesting topic for future work.
\color{black}

\subsection{Indirect Design Estimation}~\label{subsec:indirect_est}

We conclude this section by observing that an alternative way to extract disturbance information from  noisy data is by examining the difference between the observed state and the state prediction from the dynamics model.
Referring back to \eqref{eq:dynamics2}, and assuming knowledge of the system matrices $A$ and $B$, the disturbance would be given by
\begin{equation}~\label{eq:disturbance_from_dynamics}
	\xi_k^{(d)} = x_{k+1}^{(d)} - A x_{k}^{(d)} - B u_{k}^{(d)}, \quad k\in\dbracket{T-1}_{0}.
\end{equation}
Given the collected data, and under the rank condition \eqref{eq:PE}, an estimate $(\hat{B}, \hat{A})$ of the system matrices can be obtained as the unique solution to the (noiseless) least-squares problem
\begin{equation}~\label{eq:indirect_design}
	[\hat{B} \ \hat{A}] = \mathrm{argmin}_{B, A} \ \|X_{1,T} - [B \ A]\S\|_{\rm F} = X_{1,T} \S^\dagger.
\end{equation}
Concatenating \eqref{eq:disturbance_from_dynamics} over the entire sampling horizon yields the equality $\Xi_{0,T} = X_{1,T} - [B \ A]\S$ (see also \eqref{eq:dynamics_realizationN}).
We may then estimate the noise sequence from the estimated model parameters in \eqref{eq:indirect_design} as $\hat{\Xi}_{0,T} = X_{1,T} - [\hat{B} \ \hat{A}]\S$, or
\begin{equation}~\label{eq:CE_estimation}
	\hat{\Xi}_{0,T} = X_{1,T}(I_T - \S^\dagger \S),
\end{equation}
from which we may compute the disturbance covariance as in \eqref{eq:MLE_cov_soln}.
The procedure, then, is to first estimate the nominal model, then estimate the disturbance structure, and lastly solve the associated CS problem using these estimated model parameters.
%

Notice that the optimal noise realization under the assumption of a \textit{noiseless} system, \eqref{eq:CE_estimation}, is equivalent to that of the optimal ML noise realization, \eqref{eq:MLE_noise_soln}.
Thus, the CE estimation solution \eqref{eq:CE_estimation} is equivalent to the MLE solution \eqref{eq:MLE_noise_soln} under a known disturbance structure, implying that there may be deeper parallels between indirect and direct design methods in the context of noisy data.
For an overview of this observation, please see \cite{DeePC_indirect_direct_bridge}.

The estimated value of $\hat{\Xi}_{0,T}$ (and, perhaps, $\hat{\Sigma}_{\vct\xi}$) will differ, in general, from its true value ${\Xi}_{0,T}$.
For control design, 
we need to quantify the estimation error $\Delta\Xi_{0,T} \triangleq {\Xi}_{0,T} - \hat{\Xi}_{0,T}$.
%
%
In the next section, we therefore derive high-probability bounds on the estimation error $\Delta\Xi_{0,T}$ which is used in Section~\ref{sec:RDDDS} in a robust data-driven design by enforcing the running and terminal state constraints to hold within the given uncertainty sets, defined by these bounds, with high probability.


\section{Uncertainty Set Synthesis}~\label{sec:error_bounds}

Given the noise estimation techniques outlined in Section~\ref{sec:noise_estimation}, we now present two methods to derive bounds for the noise realization estimation errors
to be used later in a robust control design.
As mentioned earlier, this is a necessary step to account for the model mismatch due to noisy data.
The first method is based on the so-called \textit{Robust} Fundamental Lemma (RFL)~\cite{RFL}, which provides a stricter persistency of excitation condition, and guarantees bounded least-squares estimation errors for the indirect design.
The second method, based on the MLE noise estimation scheme, uses the statistical properties of the estimator to construct an upper bound on the noise estimation error with high confidence.

\subsection{Robust Fundamental Lemma}~\label{subsec:RFL}

As mentioned in Section~\ref{subsec:indirect_est}, an estimate  of the system matrices 
$\{\hat{A},\hat{B}\}$ using an input/output dataset  $\mathbb{D}$
can be obtained as the unique solution to the least-squares problem \eqref{eq:indirect_design}.
Now, recall from rom \eqref{eq:indirect_design} that $[\hat{B} \ \hat{A}] = X_{1,T} \S^\dagger$.
The PE condition $\mathrm{rank} (\S) = n+m$ implies that $\S$ is right invertible, and therefore it follows from \eqref{eq:dynamics_realizationN} that $ [B \ A] = (X_{1,T} - \Xi_{0,T}) \S^\dagger$.
It then follows that 
the model error can be upper-bounded as
\begin{equation}~\label{eq:model_uncertainty_bound}
	\|[\hat{B} \ \hat{A}] - [B \ A]\| = \left\|\Xi_{0,T} 
	\begin{bmatrix}
		U_{0,T} \\
		X_{0,T}
	\end{bmatrix}^\dagger \right\| \leq \frac{\sigma_{\max}(\Xi_{0,T})}{\sigma_{\min}(\S)},
\end{equation}
where $\sigma_{\min}(\S)>0$.
Since we do not have any control over the noise realization $\Xi_{0,T}$, we instead focus on the data matrix $\S$ defined in \eqref{eq:Smat:def}.

\begin{defn}[\cite{RFL}, Quantitative PE]~\label{def:alphaPE}
	Let $T > 0, z_k:\dbracket{T-1}_{0} \rightarrow \Re^{\sigma}, \alpha > 0$, and let $\ell > 0$, such that $T \geq \ell(\sigma + 1) - 1$.
	The input sequence $\{z_k\}_{k=0}^{T-1}$ is \emph{$\alpha$-persistency exciting} of order $\ell$ if $\sigma_{\min}(Z_{0,\ell,T-\ell+1}) \geq \alpha$.
\end{defn}

Note that this is a direct generalization of the familiar PE condition in Definition~\ref{def:2}.
Indeed, any $\alpha$-PE input sequence of order $\ell$ is also PE of order $\ell$.
Using this definition, the following theorem establishes sufficient conditions for the lower bound of the minimum singular value of $\S$.

\begin{thm}[\cite{RFL}, Robust Fundamental Lemma]~\label{thm:RFL}
    Let $T, \delta > 0$ and assume that the pair $(A, B)$ is controllable.
    Define the square matrix
    \begin{equation*}
        M \triangleq 
        \begin{bmatrix}
            A & B & 0_{n\times mn} \\
            0_{mn\times n} & 0_{mn\times m} & I_{mn} \\
            0_{m\times n} & 0_{m} & 0_{m\times mn}
        \end{bmatrix},
    \end{equation*}
    and let $\mathcal{Z}=\{z = [\xi^\intercal \ \eta^\intercal \  0_{nm}^\intercal]^\intercal \ | \ \xi\in\Re^{n}, \eta\in\Re^{m}, \|z\| = 1\}$.
    Define, for $z\in\mathcal{Z}$, the matrix $\Theta_z \triangleq [z \ M^\intercal z \ \cdots \ (M^\intercal)^{n} z]^\intercal$, and let $\kappa > 0$ such that%
    \footnote{The existence of such a $\kappa$ follows from the controllability of the pair $(A,B)$. See \cite[Lemma~1]{RFL}.}
    for all $z \in \mathcal{Z}$,
    $\sigma_{\min}(\Theta_{z}) \geq \kappa$.
    Let $\{x_k^{(d)}, u_k^{(d)}, x_{T}^{(d)}\}_{k=0}^{T-1}$ be an input/state trajectory of \eqref{eq:dynamics} and let $\{\xi_k^{(d)}\}_{k=0}^{T-1}$ be the process noise realization.
    If $\{u_k^{(d)}\}_{k=0}^{T-1}$ is $\delta {\sqrt{n+1}}/{\kappa}$-persistently exciting or order $n + 1$, then,
    \begin{equation}~\label{eq:min_sing_val_bound}
     \sigma_{\min}(\S) = 
        \sigma_{\min}\left(
        \begin{bmatrix}
            U_{0,T} \\
            X_{0,T}
        \end{bmatrix}\right) \geq \delta - \frac{\gamma\, \|\Xi_{0,n,T-n}\|}{\sqrt{n + 1}},
    \end{equation}
    where $\gamma > 0$ is an upper bound on the norm of the matrix 
    \begin{equation*}
        \Phi_{\xi} \triangleq 
        \begin{bmatrix}
            0 & 0 & \cdots & 0 \\
            \xi^\intercal & 0 & \cdots & 0 \\
            \xi^\intercal A & \xi^\intercal & \cdots & 0 \\
            \vdots & \vdots & \ddots & \vdots \\
            \xi^\intercal A^{n-1} & \xi^\intercal A^{n-1} & \cdots & \xi^\intercal
        \end{bmatrix},
    \end{equation*}
    that is, $\|\Phi_{\xi}\| \leq \gamma$ for all $\xi\in\Re^{n}$ such that $\|\xi\|\leq 1$,
\end{thm}

In a nutshell, Theorem~\ref{thm:RFL} says that if the input to the system satisfies the stricter PE condition of Definition~\ref{def:alphaPE}, then we are guaranteed a lower bound on the minimum singular value of the input/state data Hankel matrix.
\color{black}
This, in turn, provides a tighter upper bound through \eqref{eq:model_uncertainty_bound} on the system model estimation error $[\Delta B \ \Delta A] \triangleq [B \ A] - [\hat{B} \ \hat{A}]$ in the indirect design method of Section~\ref{sec:RDDMS} for the robust solution of the DD-MS problem.
Alternatively, we can use this model error bound in the direct robust DD-CS design of Section~\ref{sec:directCS} as follows.

First, re-write the realization dynamics \eqref{eq:dynamics_realization} as
\begin{equation}~\label{eq:realization_dynamics_expanded}
	X_{1,T} - (\Delta\Xi_{0,T} + \hat\Xi_{0,T}) = ([\hat{B} \ \hat{A}] + [\Delta B \ \Delta A]) \S.
\end{equation}
Further, by taking $\hat\Xi_{0,T}$ as the MLE solution in \eqref{eq:MLE_noise_soln}, and $[\hat B \ \hat A] = X_{1,T}\S^\dagger$ as the CE estimated model, \eqref{eq:realization_dynamics_expanded} yields
\begin{equation}~\label{eq:model_to_noise_errors}
	\Delta\Xi_{0,T} = -[\Delta B \ \Delta A] \S.
\end{equation}
Assuming now that the input sequence $\{u_k^{(d)}\}_{k=0}^{T-1}$ satisfies the conditions in Theorem~\ref{thm:RFL} for some chosen $\delta > 0$, we have that
\begin{align}
	\|\Delta\Xi_{0,T}\| &\leq \|[\Delta B \ \Delta A] \|  \|\S\| \nonumber \\
	&\leq \frac{\sigma_{\max}(\Xi_{0,T})}{\delta - \gamma \|\Xi_{0,n,T-n}\|/\sqrt{n+1}} \|\S\|. \label{eq:RFL_error_bound}
\end{align}
Thus, we obtain a bound of the form $\|\Delta\Xi_{0,T}\| \leq \rho(\delta)$, given the desired robustness level $\delta > 0$.

Unfortunately, the estimation error bound \eqref{eq:RFL_error_bound} cannot be computed easily, due to the unknown noise realization $\Xi_{0,T}$ and the constant $\gamma$, which is a function of the system model $A$.
	However, it may be possible to upper bound these quantities.
	For example, using techniques from random matrix theory (RMT), it can be shown from the Sudakov-Fernique inequality \cite{high_dim_probability} that
	\begin{equation}
		\label{eq:RMT_upper_bound}
		\E[\|\Xi_{0,T}\|] \leq \|\Sigma_{\vct\xi}^{1/2}\| (\sqrt{n} + \sqrt{T}).
	\end{equation}
We will not pursue this avenue further in this work.
Instead, in the next sections we establish computable bounds for $\|\Delta \Xi_{0,T}\|$ that can be derived from the problem data.

\subsection{Moment-Based Ambiguity Sets}~\label{subsec:moment-based}

In light of the discussion following~\eqref{eq:RFL_error_bound}, we are interested in practical bounds we can implement to ensure robust satisfaction of the constraints for the DD-MS (e.g., \eqref{eq:meanBCs})
and DD-CS (e.g., \eqref{eq:convexProblem_ineqConstraint2})
problems.
To do so, and equipped with the MLE noise realization estimate \eqref{eq:MLE_noise_soln}, we will use the statistical properties of $\hat\Xi_{0,T}$ to generate an ellipsoidal uncertainty set based on some degree of confidence $1 - \delta\in[0.5, 1)$.
In the context of the control design problem, this will imply that the resulting controller will steer the system to the desired final distribution \textit{for all} uncertainty estimates $\Delta\Xi_{0,T}\in \DeltaSet$, in some compact set $\DeltaSet$, with a prescribed degree of confidence level $1 - \delta$.

For simplicity, assume $\Sigma_{\vct\xi} \succ 0$ is known.
First, we re-write the MLE problem \eqref{eq:ML_problem} in terms of the vectorized parameters to be estimated $\xi \triangleq \mathrm{vec}(\Xi_{0,T}) = [\xi_0^\intercal, \cdots, \xi_{T-1}^\intercal]^\intercal\in\Re^{nT}$ as the optimization problem below, 
where for notational simplicity, we have dropped the superscript $(d)$ from $\xi^{(d)}$:
\begin{subequations}~\label{eq:MLE_vectorized}
	\begin{align}
		\min_{\xi} ~ &\mathcal{J}_{\mathrm{ML}}(\xi \ | \ \mathbb{D}) = \frac{1}{2}\xi^\intercal(I_{T}\otimes \Sigma_{\vct\xi}^{-1})\xi \\
		&   C(\xi) \triangleq (\Gamma \otimes I_{n})\xi - \lambda = 0,
  \label{eq:MLE_vectorized:B}
	\end{align}
\end{subequations}
where $\Gamma \triangleq I_{T} - \S^\dagger \S\in\Re^{T\times T}$, and $\lambda \triangleq \mathrm{vec}(X_{1,T}\Gamma)$.
It can be shown~\cite{MLE_convergence} that, as the number of samples grows, the unconstrained ML noise estimate $\hat\xi$ is asymptotically efficient and converges to a normal distribution as $\sqrt{T} \, \Delta \xi \overset{d}{\rightarrow} \mathcal{N}(0, \mathcal{I}^{-1})$, where $\Delta \xi \triangleq \xi - \hat\xi$ and where
$\mathcal{I} = \E_{\vct\xi}\left[\frac{\partial^2}{\partial \xi^2} \mathcal{J}_{\mathrm{ML}}(\xi \ | \ \mathbb{D})\right]$ is the \textit{Fisher Information Matrix} (FIM), which is given by $\mathcal{I} = I_{T}\otimes \Sigma_{\vct\xi}^{-1}$ in the unconstrained case.

\color{black}
In our setting, the constraint Jacobian $J(\xi)\triangleq \partial C(\xi)/\partial\xi=\Gamma\otimes I_n$ is generally \emph{rank-deficient} because $\Gamma$ is an orthogonal projector. 
The constraint manifold is nevertheless smooth, and asymptotic efficiency holds on the tangent space if we enforce only an \emph{independent} set of equality constraints. 
Concretely, 
let $V_\perp\in\Re^{T\times p}$ have orthonormal columns spanning $\operatorname{range}(\Gamma)$ (so that $p=\operatorname{rank}(\Gamma) = T - (n+m)$), and define the full row-rank Jacobian
\begin{equation*}
    J_r \triangleq (V_\perp^\intercal\otimes I_n)\in\Re^{(p n)\times (nT)},
\end{equation*}
which satisfies $\operatorname{range}(J_r)=\operatorname{range}(J)$ 
and encodes the same constraint manifold.
Under standard regularity conditions, and with $J_r$ used as the effective constraint Jacobian,
the constrained MLE is $\sqrt{T}$-consistent, asymptotically normal, and asymptotically efficient on the tangent space of the constraints~\cite{constrained_MLE_og,constrained_MLE_newer}.

It follows from \cite{CR_bound_MLE} that the asymptotic covariance of the constrained MLE error is the projection of $\mathcal{I}^{-1}$ onto the tangent space. 
Equivalently, one may write
\begin{subequations}~\label{eq:constrained_MLE_error_cov}
    \begin{align}
        \Sigma_{\Delta}
        &= \mathcal{I}^{-1} - \mathcal{I}^{-1}J_r^\intercal\big(J_r\mathcal{I}^{-1}J_r^\intercal\big)^{-1}J_r\mathcal{I}^{-1} \\
        &= \mathcal{I}^{-1} - \mathcal{I}^{-1}J^\intercal\big(J\mathcal{I}^{-1}J^\intercal\big)^{\dagger}J\mathcal{I}^{-1}.
    \end{align}
\end{subequations}

\color{black}
Using the MLE error covariance matrix \textcolor{black}{\eqref{eq:constrained_MLE_error_cov}}, we can construct high-confidence uncertainty sets for use later in a robust control design
(Section~\ref{sec:RDDDS}).
To this end, we first compute the analytical form of the error covariance for the constrained MLE problem \eqref{eq:ML_problem}.
\begin{lem}~\label{lem:error_estimate_distribution}
    The distribution of the error of the constrained ML estimator \eqref{eq:ML_problem} for the unknown noise realization $\xi = \mathrm{vec}(\Xi_{0,T})$ 
    converges, as $T \rightarrow \infty$, to the normal distribution $\mathcal{N}(0,\Sigma_{\Delta})$, that is, $\Delta\vct\xi \sim\mathcal{N}(0,\Sigma_{\Delta})$, where $\Sigma_{\Delta} = \S^\dagger \S \otimes \Sigma_{\vct\xi}$.
\end{lem}

\begin{pf}
    See Appendix~\ref{AppC}.
\end{pf}

Given the {asymptotic}
noise estimation error covariance $\Sigma_{\Delta}$, we can construct an associated high confidence uncertainty set for the random matrix $\Delta\Xi_{0,T}$ by considering the quantile of the error distribution.
To this end, we first present the original construction in \cite{DD_CS_noise_conservative} based on the full error covariance of the joint random vector $\Delta\xi = [\Delta \xi_0^\intercal, \ldots, \Delta\xi_{T-1}^\intercal]^\intercal$.
\color{black}
We then show that this uncertainty set is too loose, and its overapproximation does not scale well with the sampling horizon $T$.
To overcome this issue, we present a novel uncertainty set synthesis scheme that generates a much less conservative, 
yet still feasible, upper bound on the estimation error that is used in the robust control design of Section~\ref{sec:RDDDS}.

\begin{prop}~\label{prop:uncertainty_set_general}
	Assume that the {noise} error estimate is normally distributed as $\Delta\vct\xi\sim\mathcal{N}(0,\Sigma_{\Delta})$.
    Then, given some level of risk $\delta\in (0, 0.5]$, the set $\DeltaSet = \{\|\Delta\Xi_{0,T}\| \leq \rho\}$, where 
    $\rho^2 = Q_{\vct\rchi_{nT}^2}(1-\delta)/\lambda_{\min}^{+}(\Sigma_{\Delta}^{\dagger})$, contains the $(1-\delta)$-quantile of $\Delta\Xi_{0,T}$. 
\end{prop}
\begin{pf}
    See Appendix~\ref{AppD}.
\end{pf}

\color{black}
For our problem, the error covariance of the noise realization estimate $\Sigma_\Delta$ is given in Lemma~\ref{lem:error_estimate_distribution}.
\color{black}
Hence, we have the following corollary.

\begin{cor}~\label{cor:uncertainty_set_MLE}
    For the MLE problem \eqref{eq:ML_problem}, the associated $(1-\delta)$-quantile uncertainty set $\DeltaSet = \{\|\Delta\Xi_{0,T}\| \leq \rho(\delta)\}$ has the bound $\rho (\delta) \triangleq \|\Sigma_{\vct\xi}^{1/2}\| {Q_{\vct\rchi_{nT}^2}^{1/2}(1-\delta)}$, 
    that is, $\mathbb{P}\{ \Delta \Xi_{0,T} \in \DeltaSet\} \ge 1 - \delta$.
\end{cor}

\begin{pf}
    See Appendix~\ref{AppE}.
\end{pf}

In summary, using the MLE scheme~\eqref{eq:MLE_soln} to estimate the unknown noise realizations of the LTI system \eqref{eq:dynamics} from the collected data $\mathbb{D}$, we are able to tractably compute a confidence ellipsoidal set $\DeltaSet$ from Corollary~\ref{cor:uncertainty_set_MLE}, 
to be used in the next section for (high-probability) robust satisfaction of the mean constraints \eqref{eq:meanProblem_eqConstraint1} and the covariance constraints \eqref{eq:convexProblem_ineqConstraint2}.

Notice, however, that due to the singularity of the error covariance matrix $\Sigma_{\Delta}$ (Lemma~\ref{lem:error_estimate_distribution}), the effective number of degrees of freedom is actually reduced.
As a result, this naïve method overestimates the size of the uncertainty set by not accounting for the reduced variability dictated by the singular covariance matrix.
Corollary~\ref{cor:uncertainty_set_MLE} therefore results in an unnecessarily conservative uncertainty bound compared to that of \eqref{eq:unc_set_subspace} below.
We first provide a formal definition of a normal distribution that takes into account singular covariance matrices.
\begin{prop}[\cite{Kuhn_inverse_covariance}]~\label{def:normal_dist}
    Let $\P$ be a normal distribution on $\Re^{p}$ with mean $\mu\in\Re^{p}$ and covariance matrix $\Sigma\succeq 0$, 
    that is, $\P = \mathcal{N}(\mu, \Sigma)$, and let $r = \mathrm{rank}(\Sigma)$.
    Then, $\P$ is supported on $\mathrm{supp}(\P) \triangleq \{\mu + Ev : v \in \Re^{r}\}$, and its density with respect to the Lebesgue measure on $\mathrm{supp}(\P)$ is given by
    \begin{equation}~\label{eq:normal_dist}
        \rho_{\P}(\xi) \triangleq \frac{1}{\sqrt{(2\pi)^{r}\det(D)}} e^{-\frac{1}{2}(\xi - \mu)^\intercal ED^{-1}E^\intercal (\xi - \mu)},
    \end{equation}
  where  $D \succ 0$ is the diagonal matrix of the positive eigenvalues of $\Sigma$, and $E\in\Re^{p\times r}$ is the matrix whose columns correspond to the orthonormal eigenvectors of the positive eigenvalues of $\Sigma$.
\end{prop}

Next, we use this result to construct a confidence ellipsoid for a normal distribution with a singular covariance matrix by recognizing that $(\xi-\mu)^\intercal ED^{-1}E^\intercal (\xi - \mu)$ is a $\vct\rchi_{r}^2$ random variable with $r$ degrees of freedom~\cite{DavisKapadiaWebster1969}.

\begin{prop}~\label{def:conf_ellipsoid}
    Given the normal distribution $\P = \mathcal{N}(\mu, \Sigma)$ with mean $\mu\in\Re^{p}$ and covariance matrix $\Sigma\succeq 0$, the associated uncertainty set
    \begin{equation}~\label{eq:unc_set_proper}
        \mathcal{U}_{\mathcal{N}} \triangleq \{\xi\in\mathrm{supp}(\P) : (\xi - \mu)^\intercal ED^{-1} E^\intercal(\xi - \mu) \leq Q_{\vct\rchi_r^2}(1-\delta)\}
    \end{equation}
    contains the $(1-\delta)$-quantile of the distribution $\P$, that is, $\mathbb{P}(\xi\in\mathcal{U}_{\mathcal{N}}) \geq 1 - \delta$.
\end{prop}
The uncertainty set for the joint distribution $\Delta\vct\Xi_{0,T}$ from MLE can now be constructed by recognizing that $\mathrm{rank}(\Sigma_{\Delta}) = n(n + m)$ since\footnote{In general, $\Sigma_{\Delta}$ becomes a very low rank matrix, since when $T$ is large, $Tn \gg n(n + m)$.} $\Sigma_{\Delta} = \mathcal{S}^\dagger \mathcal{S} \otimes \Sigma_{\vct\xi}$ from Lemma~\ref{lem:error_estimate_distribution}, and $\mathrm{rank}(\mathcal{S}^\dagger\mathcal{S}) = n + m$.
Following a {similar argument as in the proof of} Corollary~\ref{cor:uncertainty_set_MLE}, we arrive at the set
\begin{equation}~\label{eq:unc_set_subspace}
    \DeltaSet^\star = \Big\{\|\Delta\Xi_{0,T}\| \leq \|\Sigma_{\vct\xi}^{1/2}\| Q_{\vct\rchi_{n(n+m)}^2}^{1/2}\big(1-\delta\big)\Big\},
\end{equation}
which guarantees that $\P(\Delta\Xi_{0,T} \in \DeltaSet^\star) \geq 1 - \delta$.
Since, in general, $r = n(n+m) \ll nT$, it follows~\cite{chi_square_monotonicity} that $Q_{\vct\rchi_r^2}(1-\delta) < Q_{\vct\rchi_{Tn}^2}(1-\delta)$, and hence $\DeltaSet^\star \subseteq \DeltaSet$.

\begin{rem}
    The above analysis relied on 
    the \emph{asymptotic} properties of the MLE estimator to obtain the error covariance of the noise realization estimates through Lemma~\ref{lem:error_estimate_distribution}.
    One natural question is whether the derived bound still holds for the case of a finite $T$-long input/state dataset. $\mathbb{D}$.
    In Appendix~\ref{AppF} we answer this question affirmatively and we show that, in fact, the finite-sample norm estimate coincides with the uncertainty set radius $\rho = \|\Sigma_{\vct\xi}^{1/2}\| Q_{\vct\rchi_{n(n+m)}^2}^{1/2}\big(1-\delta\big)$ in \eqref{eq:unc_set_subspace} as long as the data is persistently exciting.
\end{rem}



\section{Robust DD-DS}~\label{sec:RDDDS}

Equipped with the machinery to efficiently estimate and bound the uncertainty due to the noise, we are now in a position to tackle the uncertain convex programs in \eqref{eq:meanProblem} and \eqref{eq:convexProblem}.
{The analysis in this section will lead to robust versions of the DD-MS and DD-CS problems.}
To this end, notice that the DD-MS problem essentially becomes a robust control problem with unstructured model uncertainty, albeit only with high probability guarantees.
Hence, in Section~\ref{sec:RDDMS}, we use techniques from system-level synthesis (SLS)~\cite{SLS_OG}
to tractably enforce terminal constraint satisfaction (with high probability) for all bounded model uncertainties arising from the methods in Section~\ref{sec:error_bounds}.
We note that this procedure is similar to the \textit{Coarse-ID} proposed in \cite{LQR_coarseID}, which used SLS to solve a robust control problem using uncertainty bounds constructed during the model identification step.
Therein the authors employed techniques from RMT to construct tight uncertainty sets, which, as mentioned in Section~\ref{subsec:RFL}, provide a promising framework for analyzing the estimation errors arising from both noise and model estimation.
The main difference between our work and \cite{LQR_coarseID} is that we employ an indirect design technique \textit{solely} for the mean steering problem, while other robust optimization methods {(i.e., the robust counterpart of uncertain LMIs)}, 
based on a direct design, 
are utilized to address the covariance steering problem.
In this regard, the DD-CS problem requires robust satisfaction of LMI constraints over the planning horizon that encode the covariance propagation constraints under the chosen feedback control strategy.
In Section~\ref{subsec:RDDCS}, we will form the robust counterpart of these semi-infinite constraints, and use techniques from robust optimization to tractably enforce these as equivalent, deterministic LMIs.
First, we address the robust DD-MS control design problem.
%
%
\subsection{Robust Problem Formulation for DD-MS}~\label{sec:RDDMS}

Solving the DD-MS program~\eqref{eq:meanProblem} by
replacing the true $\Xi_{0,T}$
with its estimate $\hat\Xi_{0,T}$ 
will result in optimal controllers that do \textit{not} satisfy the terminal constraint $\mu_{N} = \mu_{f}$ due to the inaccuracy in the estimated model from the indirect design step \eqref{eq:indirect_design}.
The true mean dynamics are
\begin{equation}
	\mu_{k+1} = (\hat{A} + \Delta A)\mu_k + (\hat{B} + \Delta B)v_k,
\end{equation}
where the nominal matrices $[\hat{B} \ \hat{A}] = X_{1,T} \mathcal{S}^\dagger$ are computed from CE estimation, and where the model deviations $\Delta A, \Delta B$ are bounded as 
\begin{equation}
\|[\Delta B \ \Delta A]\| \leq    \rho(\delta) / \sigma_{\min}(\mathcal{S}) \triangleq    \alpha(\delta),
\end{equation}
where we have used the fact
\begin{equation}
    \|\Delta\Xi_{0,T}\| = \|[\Delta B \ \Delta A] \mathcal{S}\| \geq \sigma_{\min}(\mathcal{S}) \|[\Delta B \ \Delta A] \|,
\end{equation}
where $\| \Delta \Xi_{0,T} \| \le \rho (\delta)$ from Corollary~\ref{cor:uncertainty_set_MLE}.

Note, however, that enforcing the terminal constraint $\mu_N = \mu_f$  \textit{for all} uncertainties $\|[\Delta B \ \Delta A]\| \leq \alpha(\delta)$ is intractable, in general.
Instead, we relax the pointwise terminal constraint to a terminal set given by a polytope such that $\mu_{N} \in \mathcal{X}_{f} \triangleq \{x : F_{x_N}x \leq b_{x_N}\}$, and require robust satisfaction of the constraint $\mu_{N} \in \mathcal{X}_{f}$, for all $\|\Delta A\| \leq \varepsilon_{A}$ and for all $\|\Delta B\| \leq \varepsilon_{B}$, for some $\varepsilon_{A}, \varepsilon_{B} > 0$.
Along these lines, and in order to enhance tractability, we also impose polyhedral constraints on the transient motion of the mean state and the feed-forward input as $\mu_k \in \mathcal{X}_{k} \triangleq \{x : F_x x \leq b_x\}$ and $v_k \in \mathcal{U}_k \triangleq \{u : F_u u \leq b_u\}$.
Lastly, instead of the open-loop control $v_k$, we introduce a feedback mean control
%
%
in terms of the mean state history as follows $v_k = \sum_{i=0}^{k}L_{k,i}\mu_i$.

For notational convenience, let the \textit{nominal} mean state be denoted by $\bar{\mu}_k$ which satisfies the error-free dynamics $\bar{\mu}_{k+1} = \hat{A} \bar\mu_k + \hat{B} \bar v_k$ and let $\bar{v}_k = 
\sum_{i=0}^{k}L_{k,i}\bar{\mu}_i$.
In summary, the robust DD-MS (R-DD-MS) problem is posed as follows.
\begin{empheqboxed}
    \begin{subequations}~\label{eq:R-DD-MS}
        \begin{equation}
            \min_{L_{k,i}} \bar{\mathcal{J}}_{\mu} = \sum_{k=0}^{N} \Big((\bar\mu_k - x_k^{r})^\intercal Q_k (\bar\mu_k-x_k^{r}) + \bar{v}_k^\intercal R_k \bar{v}_k\Big) \label{eq:mean_cost} 
        \end{equation}
        such that, for all $k\in\dbracket{N-1}$ and for all $\|\Delta A\| \leq \varepsilon_A$ and $\|\Delta B\| \leq \varepsilon_B$,
        \begin{align}
            &\bar\mu_{k+1} = \hat{A} \bar\mu_k + \hat{B} \bar{v}_k, \label{eq:mean_nominal} \\
            &\mu_{k+1} = (\hat{A} + \Delta A)\mu_k + (\hat{B} + \Delta B)v_k, \label{eq:mean_full} \\
            &v_k = \sum_{i=0}^{k}L_{k,i}\mu_i, \\
            &\mu_k\in\mathcal{X}_k, \ v_k \in \mathcal{U}_k, \ \mu_{N} \in \mathcal{X}_{f}. \label{eq:constraints_RDDMS}
        \end{align}
    \end{subequations}
\end{empheqboxed}

There are numerous methods in the robust control literature to tackle the R-DD-MS problem in \eqref{eq:R-DD-MS}, ranging from the early works of tube-MPC \cite{Tube_MPC_OG} to disturbance-feedback with lumped uncertainty~\cite{RMPC_lumped}, to the more modern methods using system level synthesis (SLS)~\cite{SLS_OG}.
Below, we use SLS to solve the R-DD-MS problem by reformulating the semi-infinite program \eqref{eq:R-DD-MS} as a tractable SDP.

\subsection{Solution to the Robust DD-MS Problem via SLS}

The SLS approach to robust control aims at transforming the optimization problem over feedback control laws to one over closed-loop system \textit{responses}, i.e., linear maps from the uncertainty process to the states and inputs in the closed loop.
To this end, and for the nominal dynamics \eqref{eq:mean_nominal}, define the augmented state and control inputs as $\bar\mu = [\bar\mu_0^\intercal, \cdots, \bar\mu_{N}^\intercal]^\intercal, \bar v = [\bar v_0^\intercal, \cdots, \bar v_{N}^\intercal]^\intercal$, and the vector%
\footnote{This is a special case of the more general expression $w = [\mu_0^\intercal, w_0^\intercal, \ldots, w_{N-1}^\intercal]^\intercal$, where $w_{k} \in \mathcal{W}$ are additive bounded uncertainties to the dynamics~\eqref{eq:mean_nominal}.
}
$\tilde{w} \triangleq [\mu_0; w] = [\mu_0; 0_{Nn}]$.
Let the control input $\bar{v} = L \bar{\mu}$, where
\begin{equation} \label{Lmat;def}
	L = 
	\begin{bmatrix}
		L_{0,0} & & & \\
		L_{1,0} & L_{1,1} & & \\
		\vdots & \ddots & \ddots & \\
		L_{N, 1} & \cdots & L_{N, N-1}  & L_{N, N}
	\end{bmatrix},
\end{equation}
and concatenate the dynamics matrices as $\hat{\mathcal{A}} \triangleq \mathrm{blkdiag}(I_{N},0)\otimes \hat{A}$ and $\hat{\mathcal{B}} \triangleq \mathrm{blkdiag}(I_{N}, 0) \otimes \hat{B}$.
Let $Z\in\Re^{n(N+1)\times n(N+1)}$ be the block-downshift operator, that is, a matrix with the identity matrix on the first block sub-diagonal and zeros elsewhere.
Under the feedback controller $L$, the closed-loop behavior of the nominal system \eqref{eq:mean_nominal} can be represented compactly as
\begin{equation}~\label{eq:CL_mean_nominal}
	\bar\mu = Z(\hat{\mathcal{A}} + \hat{\mathcal{B}} L)\bar\mu + \tilde{w}.
\end{equation}
The closed-loop map from $\tilde{w} \mapsto (\bar\mu, \bar v)$ is given by
\begin{equation}~\label{eq:CL_nominal_system_map}
	\begin{bmatrix}
		\bar \mu \\ 
		\bar v
	\end{bmatrix} = 
	\begin{bmatrix}
		(I - Z(\hat{\mathcal{A}} + \hat{\mathcal{B}} L))^{-1} \\
		L(I - Z(\hat{\mathcal{A}} + \hat{\mathcal{B}} L))^{-1}
	\end{bmatrix} \tilde{w} = 
	\begin{bmatrix}
		\bar{\Phi}_{x} \\
		\bar{\Phi}_{u}
	\end{bmatrix} \tilde{w},
\end{equation}
where the matrices $\bar{\Phi}_{x}$ and $\bar{\Phi}_{u}$ are the \textit{nominal} system responses under the action of the feedback controller $L$ in \eqref{Lmat;def} on the LTI system~\eqref{eq:mean_nominal}.
The essence of the SLS approach is to treat these closed-loop system maps as the decision variables in the resulting optimization problem.
In order to satisfy the closed-loop dynamics in \eqref{eq:CL_nominal_system_map}, the matrices $\bar{\Phi}_{x}$ and $\bar{\Phi}_{u}$ must be constrained to an affine subspace parameterizing all system responses, similar to the subspace relations in \eqref{eq:consistency_equation} that encode the LTI dynamics of the realization data.
The next theorem formalizes this intuition and provides the corresponding controller.
\begin{thm}[\cite{SLS_OG}]~\label{thm:system_response}
    Consider the nominal system dynamics \eqref{eq:mean_nominal} with state feedback law $\bar v = L \bar\mu$, where $L$ is a block-lower triangular matrix.
    Then, the following are true:
    \begin{itemize}
        \item[i)] 
        The affine subspace defined by
        \begin{equation}~\label{eq:affine_subspace_maps}
            \begin{bmatrix}
                      I - Z\hat{\mathcal{A}} & -Z\hat{\mathcal{B}}
            \end{bmatrix}
            \begin{bmatrix}
                \bar{\Phi}_{x} \\
                \bar{\Phi}_{u}
            \end{bmatrix} = I, 
        \end{equation}
        parameterizes all possible system responses \eqref{eq:CL_nominal_system_map}.
        
        \item[ii)] 
        For any block-lower triangular matrices $\bar{\Phi}_{x}$ and 
        $\bar{\Phi}_{u}$ satisfying \eqref{eq:affine_subspace_maps}, the controller $L = \bar{\Phi}_{u} \bar{\Phi}_{x}^{-1}$ achieves the desired response.
    \end{itemize}
\end{thm}
%

Thus, instead of requiring satisfaction of the dynamic constraints \eqref{eq:mean_nominal}, we can, equivalently, require satisfaction of the system map constraints \eqref{eq:affine_subspace_maps} that achieve the desired system response.
Moreover, this framework can be extended to handle model uncertainty as well, through the following theorem.
\begin{thm}[\cite{SLS_OG}]~\label{thm:system_response_uncertainty}
	Let $\bar\Lambda$ be an arbitrary block-lower triangular matrix, and suppose that $\Phi_{x}$ and $\Phi_{u}$
    {are  block-lower triangular matrices that}
    satisfy
	\begin{equation}~\label{eq:affine_subspace_maps_uncertainty}
		\begin{bmatrix}
			I_{(N+1)n} - Z\mathcal{A} & -Z\mathcal{B}
		\end{bmatrix}
		\begin{bmatrix}
			\Phi_{x} \\
			\Phi_{u}
		\end{bmatrix} = I_{(N+1)n} - \bar\Lambda.
	\end{equation}
	If $(I - \bar\Lambda_{i,i})^{-1}$ exists for all $i = 0,\ldots, N$, then the controller $L = \Phi_{u}\Phi_{x}^{-1}$ achieves the system response
	\begin{equation}~\label{eq:CL_system_response_uncertainty}
		\begin{bmatrix}
			\mu \\
			v
		\end{bmatrix} = 
		\begin{bmatrix}
			\Phi_{x} \\
			\Phi_{u}
		\end{bmatrix}(I - \bar\Lambda)^{-1} \tilde{w}.
	\end{equation}
\end{thm}

%
To see how Theorem~\ref{thm:system_response_uncertainty} allows us to encode model uncertainty into the SLS framework, note that the nominal responses also approximately satisfy \eqref{eq:affine_subspace_maps} with respect to the \textit{true} model \eqref{eq:mean_full} with an extra perturbation term given by
\begin{align}
	\begin{bmatrix}
		I - Z\mathcal{A} & -Z\mathcal{B}
	\end{bmatrix}
	\begin{bmatrix}
		\bar{\Phi}_{x} \\
		\bar{\Phi}_{u}
	\end{bmatrix} &= I - Z
	\begin{bmatrix}
		\Delta A & \Delta B
	\end{bmatrix}
	\begin{bmatrix}
		\bar{\Phi}_{x} \\
		\bar{\Phi}_{u}
	\end{bmatrix} \nonumber \\
	& \triangleq I - \Lambda\bar{\Phi},
\end{align}
where in the first equality we use the fact that $\bar{\Phi}_{x}$ and
$\bar{\Phi}_{u}$ satisfy \eqref{eq:affine_subspace_maps}, and in the second equality, we define $\bar{\Phi} \triangleq [\bar{\Phi}_{x}; \bar{\Phi}_{u}]$, and $\Lambda \triangleq Z[\Delta A \ \Delta B]$.
As a result, invoking Theorem~\ref{thm:system_response_uncertainty} we conclude that the controller $L = \bar{\Phi}_{u}\bar{\Phi}_{x}^{-1}$, computed using the system estimates $\{\hat{\mathcal{A}}, \hat{\mathcal{B}}\}$, achieves the response \eqref{eq:CL_system_response_uncertainty} on the actual system $\{\mathcal{A}, \mathcal{B}\}$, with $\bar\Lambda = \Lambda \bar{\Phi}$.

Next, we reformulate the objective function \eqref{eq:mean_cost}
by, equivalently, re-writing it in terms of the augmented state and input as
\begin{align}
    \bar{\mathcal{J}}_{\mu} &= \left\|
    \begin{bmatrix}
        \mathcal{Q}^{1/2} & 0 \\
        0 & \mathcal{R}^{1/2}
    \end{bmatrix}
    \begin{bmatrix}
        \bar{\mu} \\
        \bar{v}
    \end{bmatrix}\right\|^{2} - 2
    \begin{bmatrix}
        \mathcal{Q} x^{r} \\
        0_{(N+1)m}
    \end{bmatrix}^\intercal 
    \begin{bmatrix}
        \bar{\mu} \\ 
        \bar{v}
    \end{bmatrix} + \left\|\mathcal{Q}^{1/2}x^{r}\right\|^{2} \nonumber \\
    &= \left\|\mathcal{M}^{1/2}\bar{\Phi} \tilde{w} \right\|^{2} - 2
    \begin{bmatrix}
        \mathcal{Q} x^{r} \\
        0_{(N+1)m}
    \end{bmatrix}^\intercal \bar{\Phi} \tilde{w} +\left\|\mathcal{Q}^{1/2}x^{r}\right\|^{2}, \label{eq:cost_SLS_intermediate}
\end{align}
where $\mathcal{Q} \triangleq \mathrm{blkdiag}(Q_0,\ldots,Q_{N}),$ 
$\mathcal{R} \triangleq \mathrm{blkdiag}(R_0, \ldots, R_{N})$, and $\mathcal{M} \triangleq \mathrm{blkdiag}(\mathcal{Q},\mathcal{R})$.
Note that from the definition of $\tilde{w}$ the only non-zero entry in $\tilde{w}$ is its first block.
\color{black}
Hence, by partitioning $\bar{\Phi} = [\bar{\Phi}^{0} \ \bar{\Phi}^{\tilde{w}}]$, where 
$\bar{\Phi}^{0} \in \Re^{(N+1)(n+m)\times n}$ and $\bar{\Phi}^{\tilde{w}}\in\Re^{(N+1)(n+m)\times Nn}$, the cost \eqref{eq:cost_SLS_intermediate} simplifies to
\color{black}
\begin{equation} \label{eq:cost_SLS}
    \bar{\mathcal{J}}_{\mu} = \left\|\mathcal{M}^{1/2} \bar{\Phi}^{0}\mu_0\right\|^{2} - 2
    \begin{bmatrix}
        \mathcal{Q} x^{r} \\
        0_{(N+1)m}
    \end{bmatrix}^\intercal \bar{\Phi}^{0}\mu_0 +\left\|\mathcal{Q}^{1/2}x^{r}\right\|^{2}.
\end{equation}
Similarly, we can simplify the uncertain system response in \eqref{eq:CL_system_response_uncertainty} as
\begin{align}
    \begin{bmatrix}
        \mu \\
        v
    \end{bmatrix} &= (\bar\Phi + \bar\Phi\Lambda (I - \bar\Phi \Lambda)^{-1} \bar\Phi) \tilde{w} \nonumber \\
    &=\bar\Phi^{0}\mu_0 + \bar\Phi \Lambda (I - \bar\Phi \Lambda)^{-1} \bar\Phi^{0} \mu_0, \label{eq:SLS_expansion}
\end{align}
where the first equality comes from the Woodbury matrix identity \cite{woodbury1950inverting}.
Lastly, we concatenate all the constraints $x_k\in\mathcal{X}_k,u_k\in\mathcal{U}_k,x_{N}\in\mathcal{X}_{f}$ together in the compact form $F[\mu; v] \leq b$, where $F \triangleq \mathrm{blkdiag}(F_{x},\cdots, F_{x}, F_{x_{N}}, F_{u}, \cdots, F_{u})$ and $b \triangleq [b_x^\intercal, \cdots, b_x^\intercal, b_{x_N}^\intercal, b_{u}^\intercal, \cdots, b_{u}^\intercal]^\intercal$.
In summary, the R-DD-MS problem \eqref{eq:R-DD-MS} can be written as the following program
\begin{empheqboxed}
    \begin{subequations}~\label{eq:R-DD-MS_SLS}
	\begin{align}
		&\min_{\bar\Phi_{x}, \bar\Phi_{u}} &&\|\mathcal{M}^{1/2}\bar\Phi^{0}\mu_0\|_{2}^{2} - 2[\mathcal{Q}x^{r}; 0_{(N+1)m}]^\intercal \bar\Phi^{0} \mu_0 \nonumber \\
            &&&\hspace{3cm} +\left\|\mathcal{Q}^{1/2}x^{r}\right\|^{2}, \\
		&\ \ &&
		\begin{bmatrix}
			I - Z\hat{\mathcal{A}} & -Z\hat{\mathcal{B}}
		\end{bmatrix}
		\bar\Phi = I, \\
		&&&F(I + \bar\Phi \Lambda (I - \bar\Phi \Lambda)^{-1})\bar\Phi^{0} \mu_0 \leq b. \label{eq:R-DD-MS_constraints}
	\end{align}
    \end{subequations}
\end{empheqboxed}

The main difficulty in \eqref{eq:R-DD-MS_SLS} is in the robust constraints \eqref{eq:R-DD-MS_constraints}, which are nonlinear in $\bar\Phi$.
Following the work of \cite{RMPC_SLS_hyperparemeters}, we can upper bound the LHS of the constraints and formulate sufficient conditions such that \eqref{eq:R-DD-MS_constraints} holds, for all $\|\Delta A\| \leq \varepsilon_A$ and
$\|\Delta B\| \leq \varepsilon_B$.
    To this end, define the partitions $\bar{\Phi}^{0} = [\bar{\Phi}_{x}^{0}; \bar{\Phi}_{u}^{0}]$ and $\bar{\Phi}^{\tilde{w}} = [\bar{\Phi}_{x}^{\tilde{w}}; \bar{\Phi}_{u}^{\tilde{w}}]$, where $\bar{\Phi}_{x}^{0} \in \Re^{(N+1)n\times n}, \bar{\Phi}_{u}^{0} \in \Re^{(N+1)m\times n}, \bar{\Phi}_{x}^{\tilde{w}} \in \Re^{(N+1)n\times Nn}, \bar{\Phi}_{u}^{\tilde{w}} \in \Re^{(N+1)m\times Nn}$.
The convex approximation is stated in the following theorem.

\begin{thm}~\label{thm:robust_control_SLS}
    Let $\tau,\gamma > 0$ and $\theta\in (0,1)$.
    Consider the following convex optimization problem
    \begin{subequations}~\label{eq:R-DD-MS_SLS_convex}
        \begin{align}
            &\min_{\bar\Phi_{x}, \bar\Phi_{u}} &&\|\mathcal{M}^{1/2}\bar\Phi^{0}\mu_0\|_{2}^{2} - 2[\mathcal{Q}x^{r}; 0_{(N+1)m}]^\intercal \bar\Phi^{0} \mu_0 \\
            &\ ~ \ &&
            \begin{bmatrix}
                I - Z\hat{\mathcal{A}} & -Z\hat{\mathcal{B}}
            \end{bmatrix}
            \bar\Phi = I, \\
            &&& \hspace*{-7mm} F_j^\intercal \bar\Phi^{0}\mu_0 + \|F_j^\intercal \bar\Phi^{\tilde{w}}\| \frac{1 - \tau^N}{1 - \tau}\gamma \leq b_j, \ \forall j\in\dbracket{J}, \\
            &&&\left\|
            \begin{bmatrix}
                \frac{\varepsilon_{A}}{\theta}\bar\Phi_{x}^{\tilde{w}} \\
                \frac{\varepsilon_{B}}{1-\theta}\bar\Phi_{u}^{\tilde{w}}
            \end{bmatrix}\right\| \leq \tau, \quad \left\|
            \begin{bmatrix}
                \frac{\varepsilon_{A}}{\theta}\bar\Phi_{x}^{0} \\
                \frac{\varepsilon_{B}}{1-\theta}\bar\Phi_{u}^{0}
            \end{bmatrix} \mu_0\right\| \leq \gamma,
        \end{align}
    \end{subequations}
    where $F_j \in \Re^{(N+1)(n+m)}, b_j\in\Re$ denote the $j$th row and element of $F$ and $b$, respectively.
    Let the solution to this optimization problem be $\bar\Phi = [\bar\Phi_{x}; \bar\Phi_{u}] = [\bar\Phi^{0} \ \bar\Phi^{\tilde{w}}]$
    where $\bar\Phi_{x}$ and $\bar\Phi_{u}$ are lower-block triangular matrices.
    Then, 
    the controller $L = \bar{\Phi}_{u} \bar{\Phi}_{x}^{-1}$ guarantees the constraint satisfaction in~\eqref{eq:R-DD-MS_constraints} for all possible model uncertainties $\|\Delta A\| \leq \varepsilon_A$ and $\|\Delta B\| \leq \varepsilon_B$ in~\eqref{eq:R-DD-MS}, that is, 
    the solution to \eqref{eq:R-DD-MS_SLS_convex} is a conservative convex approximation of \eqref{eq:R-DD-MS_SLS}.
\end{thm}

\begin{pf}
    See Appendix~\ref{AppG}.
\end{pf}

\subsection{Solution of the Robust DD-CS Problem}~\label{subsec:RDDCS}

In this section, we reformulate the uncertain CS program \eqref{eq:convexProblem} so that it is amenable to a tractable convex matrix feasibility problem.
The original constraints in \eqref{eq:convexProblem_ineqConstraint2}, when reformulated as the LMI constraints $G_k^{\Sigma} \succeq 0$ in \eqref{eq:convex_program_LMI_2}, may be robustly satisfied with high-probability, using the decomposition $\Xi_{0,T} = \hat{\Xi}_{0,T} + \Delta\Xi_{0,T}$ along with the established estimation error bounds, as the semi-infinite uncertain LMIs
\begin{equation}~\label{eq:uncertain_LMI_cov}
    \hat{G}_k^{\Sigma} + \Delta G_k^{\Sigma}(\Delta \Xi_{0,T}) \succeq 0, \quad \forall \|\Delta\Xi_{0,T}\| \leq \rho(\delta), 
\end{equation}
where,
\begin{equation}
    \hat{G}_{k}^\Sigma = 
    \begin{bmatrix}
        \Sigma_{k+1} - \Sigma_{\vct\xi} & (X_{1,T} - \hat{\Xi}_{0,T})S_k \\
        S_k^\intercal(X_{1,T} - \hat{\Xi}_{0,T})^\intercal & \Sigma_{k} \label{eq:nominal_matrix}
    \end{bmatrix} \succeq  0,
\end{equation}
is the nominal covariance LMI, and
\begin{equation}
	\Delta G_k^{\Sigma} = 
	\begin{bmatrix}
		0_{n} & -\Delta \Xi_{0,T} S_k \\
		-S_k^\intercal \Delta \Xi_{0,T}^\intercal & 0_{n} 
	\end{bmatrix} \succeq 0 \label{eq:pertburation_matrix}
\end{equation}
is the perturbation to the covariance LMI.
Next, we rewrite the perturbation matrix \eqref{eq:pertburation_matrix} as
\begin{equation}
	\Delta G_k^{\Sigma} = \Theta^\intercal (S_k) \Delta \Xi_{0,T}^\intercal \Pi + \Pi^\intercal \Delta \Xi_{0,T} \Theta(S_k), \label{eq:uncertain_LMI_decomposition}
\end{equation}
where $\Theta^\intercal(S_k) \triangleq [0_{n,T}; -S_k^\intercal]$ and $\Pi^\intercal = [I_{n};0_{n}]$.
Finally, using \cite{robust_optimization}, we may equivalently 
represent the uncertain LMI \eqref{eq:uncertain_LMI_cov} as the following standard LMI
\begin{equation}~\label{eq:robust_covariance_prop_constraints}
    \begin{bmatrix}
        \lambda I_{T} & \rho(\delta)\, \Theta(S_k) \\
        \rho(\delta)\, \Theta^\intercal (S_k) & \hat{G}_k^{\Sigma}(\Sigma_k, \Sigma_{k+1}, S_k) - \lambda \Pi^\intercal \Pi 
    \end{bmatrix} \succeq 0,
\end{equation}
in terms of the decision variables $S_k, \lambda, \Sigma_k$ and $\Sigma_{k+1}$.

\section{Parametric Uncertainty DD-CS}~\label{sec:PUDDCS}

\color{black}
Rather than deriving bounds on $\Delta\Xi_{0,T}$ and designing for the worst-case disturbance (Sections~\ref{sec:error_bounds} and \ref{sec:RDDDS}), in this section we treat the parametric disturbances as stochastic ones obeying a known probability law $\vct\xi_k^{(d)} \sim \mathcal{N}(0, \Sigma_{\vct \xi})$.
%
%
For simplicity, in this section, we assume that $D$ is known.
We pose a data-driven density steering problem with probabilistic parametric uncertainty (PU-DD-DS).
In other words, we treat the noise vectors $\vct\xi_k^{(d)}$ as random variables with known probability distributions rather than attempting to estimate their specific realization.
This approach has the advantage of allowing us 
to design a controller that is inherently robust to the entire distribution of possible noise realizations, rather than being robustly optimized for a single estimated noise instance.

\color{black}

\subsection{Solution of the PU-DD-DS Problem}

The dynamics of the actual system may be written as
\begin{equation}
    \vct x_{k+1} = (\hat{A} + \Delta A(\vct\Xi_{0,T}))\vct x_k + (\hat{B} + \Delta B(\vct\Xi_{0,T}))\vct u_k + \vct \xi_k, \label{eq:PU_dynamics}
\end{equation}
where \textcolor{black}{%
$[\hat{B} \ \hat{A}] = X_{1,T} \mathcal{S}^\dagger$ is the CE estimate from \eqref{eq:indirect_design}, and $[\Delta B \ \Delta A] = -\vct\Xi_{0,T}\mathcal{S}^\dagger$ is the error from the CE design.%
}

Partition the matrix $\mathcal{S}^\dagger = [\mathcal{S}_{1}, \mathcal{S}_{2}] \in \mathbb{R}^{T\times (m+n)}$ in \eqref{eq:Smat:def} into $\mathcal{S}_{1} \in \mathbb{R}^{T\times m}$ and $\mathcal{S}_{2} \in \mathbb{R}^{T\times n}$, and let $\tau_i\in\Re^{m}$ denote the $i$th row of $\mathcal{S}_1$ 
and $\sigma_i\in\Re^{n}$ denote the $i$th row of $\mathcal{S}_2$.
Since $\vct\Xi_{0,T} = [\vct\xi_0^{(d)}, \ldots, \vct\xi_{T-1}^{(d)}]$,
the dynamics \eqref{eq:PU_dynamics} can be written as
\begin{equation}
	\vct x_{k+1} = \left( \hat{A} - \sum_{i=0}^{T-1}\vct\xi_i^{(d)} \tau_i^\intercal\right) \vct x_k + \left( \hat{B} - \sum_{i=0}^{T-1}\vct\xi_i^{(d)}\sigma_i^\intercal\right)  \vct u_k + \vct \xi_k. \label{eq:PU_dynamics_2}
\end{equation}
Forming the outer product $\tilde{\vct x}_{k+1}\tilde{\vct x}_{k+1}^\intercal$, where $\tilde{\vct x}_k \triangleq \vct x_{k} - \mu_k$, and taking expectations yields the 
following expression for the covariance dynamics
\begin{equation}~\label{eq:PU_cov_dynamics}
    \Sigma_{{k+1}} = \hat{\Sigma}_{x_{k+1}} + \Delta \Sigma_{x_{k+1}}^{(1)} + \Delta \Sigma_{x_{k+1}}^{(2)},
\end{equation}
where,
\begin{subequations}
    \begin{align}
        \hspace*{-5mm}
        \hat{\Sigma}_{x_{k+1}} &= \hat{A}  \Sigma_{k} \hat{A}^\intercal + \hat{A}\Sigma_{x_k,u_k}\hat{B}^\intercal \nonumber \\
        &\hspace{1.2cm}+ \hat{B}\Sigma_{x_k,u_k}^\intercal \hat{A}^\intercal + \hat{B}\Sigma_{u_k}\hat{B}^\intercal + \Sigma_{\vct\xi}, \\
        \Delta \Sigma_{x_{k+1}}^{(1)} &= \sum_{i=0}^{T-1}\Big(\sigma_i^\intercal \Sigma_{k} \sigma_i + \sigma_i^\intercal\Sigma_{x_k, u_k} \tau_i \nonumber \\
        &\hspace{1.5cm}+ \tau_i^\intercal\Sigma_{x_k, u_k}^\intercal \sigma_i + \tau_i^\intercal\Sigma_{u_k}\tau_i\Big)\Sigma_{\vct\xi}, \\
        \Delta \Sigma_{x_{k+1}}^{(2)} &= \sum_{i=0}^{T-1} 
        \textcolor{black}{%
        (\sigma_i^\intercal \mu_k + \tau_i^\intercal v_k)^2 %
        } \Sigma_{\vct \xi},
        \label{eq:nonlinear_covariance_delta}
    \end{align}
\end{subequations}
where $\Sigma_{u_k} \triangleq \E[\big(\vct u_k - \E[\vct u_k]\big)\big(\vct u_k - \E[\vct u_k]\big)^\intercal]$ and $\Sigma_{x_k,u_k} \triangleq \E[\big(\vct x_k - \E[\vct x_k]\big)\big(\vct u_k - \E[\vct u_k]\big)^\intercal]$,
and
where we have used the fact that the noise follows an i.i.d normal distribution, and $\E[\vct\xi_{k}\vct\xi_{j}^\intercal] = \delta_{kj}I_{n}$ 
and 
$\E[\vct\xi_{i}^{(d)}\vct\xi_k^\intercal] = 0$.
With the affine feedback controller \eqref{eq:controlLaw} the covariance matrices are given by
\begin{equation}~\label{eq:controlCovariances}
    \Sigma_{u_k} = K_k \Sigma_{k} K_k^\intercal, \quad \Sigma_{x_k, u_k} = \Sigma_{k} K_k^\intercal. 
\end{equation}
Note that with multiplicative uncertainties, the mean and covariance designs become coupled
through the extra term \eqref{eq:nonlinear_covariance_delta} entering equation~\eqref{eq:PU_cov_dynamics};
this is in contrast to the case of only additive disturbances,
where the mean and covariance subproblems are decoupled.

For the uncertain mean dynamics, we have the following expression
\begin{equation*}
    \mu_{k+1} = (\hat{A} + \Delta A(\Xi_{0,T}))\mu_{k} + (\hat{B} + \Delta B(\Xi_{0,T}))v_{k}.
\end{equation*}
To proceed, we follow an CE approach, by neglecting the model error matrices, which yields the approximate mean dynamics
\begin{equation}~\label{eq:approx_mean_dynamics}
    \mu_{k+1} \approx \hat{A} \mu_k + \hat{B} v_k,
\end{equation}
with the additional caveat that the true terminal mean constraints may not be satisfied in practice, due to the model mismatch.
Future work will investigate ways to incorporate robust design techniques, such as those in Section~\ref{sec:RDDMS}, in the context of PU-DD-DS.
In summary, the PU-DD-DS problem is given as follows.

\begin{prob}[PU-DD-DS]~\label{prob:PU-DD-DS}
    Given the dataset $\mathbb{D}$ corresponding to the unknown linear system (\ref{eq:dynamics}) with the nominal model $[\hat{B} \ \hat{A}] = X_{1,T} \mathcal{S}^\dagger$, find the optimal control sequence $\{\vct u_k\}_{k=0}^{N-1}$
    that minimizes the cost (\ref{eq:cost_function}), subject to the (approximate) mean and (exact) covariance dynamics \eqref{eq:approx_mean_dynamics}, \eqref{eq:PU_cov_dynamics}, respectively, initial state (\ref{eq:init_condition}) and terminal boundary conditions $\mu_N = \mu_f$, 
    $\Sigma_{N} \preceq \Sigma_{f}$.
\end{prob}

Next, we provide a convex reformulation of Problem~\ref{prob:PU-DD-DS}, which is similar to the derivation in \cite{CS_mulitplicative_knaup}.
To this end, first notice that the third term \eqref{eq:nonlinear_covariance_delta} is quadratic in the decision variables $\{\mu_k, v_k\}$.
Additionally, the control parameterization \eqref{eq:controlCovariances} results in a nonlinear program in the decision variables $\{\Sigma_{u_k}, K_k, \Sigma_{x_k}\}$.
To remedy the former issue, we relax the equality constraint \eqref{eq:nonlinear_covariance_delta} by 
introducing the new decision variables $\Sigma_{ik}^{\Delta}$ for each $i \in \dbracket{T-1}_{0}$ and $k \in \dbracket{N-1}_{0}$ such that%
\begin{equation}
	\Sigma_{ik}^{\Delta} \succeq (\sigma_i^\intercal\mu_k + \tau_i^\intercal v_k)\Sigma_{\vct\xi}(\sigma_i^\intercal\mu_k + \tau_i^\intercal v_k),
\end{equation}
which, using the Schur complement, can be recast as the LMI
\begin{equation}
	\begin{bmatrix}
		\Sigma_{ik}^{\Delta} & (\sigma_i^\intercal \mu_k + \tau_i^\intercal v_k) I_{n} \\
		(\sigma_i^\intercal \mu_k + \tau_i^\intercal v_k) I_{n} & \Sigma_{\vct\xi}^{-1}
	\end{bmatrix} \succeq 0. \label{eq:PU_relaxation1}
\end{equation}

To address the nonlinear dependence in the control parameterization,
and in a similar manner to the theory developed in Section~\ref{subsec:DDCS_problem}, define the new decision variables $U_k \triangleq K_k \Sigma_{x_k} = \Sigma_{x_k,u_k}^\intercal$.
The control covariance thus becomes $\Sigma_{u_k} = U_k \Sigma_{x_k}^{-1} U_k^\intercal$, which is still nonlinear in the decision variables.
To this end, we relax this equality constraint by introducing yet another new decision variable $Y_k$ such that
\begin{equation}
	Y_k \succeq U_k \Sigma_{x_k}^{-1} U_k^\intercal \implies 
	\begin{bmatrix}
		Y_k & U_k \\
		U_k^\intercal & \Sigma_{x_k}
	\end{bmatrix} \succeq 0. \label{eq:PU_relaxation2}
\end{equation}

In summary, the relaxed PU-DD-DS problem is given by the SDP in \eqref{eq:PUconvexProblem}
\begin{empheqboxed}
    \begin{subequations}~\label{eq:PUconvexProblem}
        \begin{align}
            &\hspace*{-4mm}\ \min_{\Sigma_k, U_k, Y_k, \Sigma_{ik}^{\Delta}} \sum_{k=0}^{N-1}\big(\mu_k^\intercal Q_k \mu_k + v_k^\intercal R_k v_k \label{eq:PUconvexProblem_cost} \\
            &\hspace{3.2cm} +\mathrm{tr}(Q_k\Sigma_k) + \mathrm{tr}(R_k Y_k)\big), \nonumber
        \end{align}
        such that, for all $k\in\dbracket{N-1}_{0}, \ i \in \dbracket{T-1}_{0}$,
        \begin{align}
            &\mu_{k+1} = \hat{A} \mu_k + \hat{B} v_k \\
            &\Sigma_{x_{k+1}} = \hat{\Sigma}_{x_{k+1}} + \sum_{i=0}^{T-1}\Sigma_{ik}^{\Delta} + \sum_{i=0}^{T-1}
            \Big(
            \sigma_i^\intercal \Sigma_{x_k}\sigma_i + \sigma_i^\intercal U_k^\intercal \tau_i \nonumber \\
            &\hspace{2.7cm} + \tau_i^\intercal U_k \sigma_i + \tau_i^\intercal Y_k \tau_i
            \Big)
            \Sigma_{\vct\xi}, \\
            &\hat{\Sigma}_{x_{k+1}} = \hat{A}\Sigma_{x_k}\hat{A}^\intercal + \hat{A}U_k^\intercal \hat{B}^\intercal + \hat{B} U_k \hat{A}^\intercal \nonumber \\
            &\hspace{3.5cm} + \hat{B} Y_k \hat{B}^\intercal + \Sigma_{\vct\xi}, \\
            &Y_k \succeq U_k \Sigma_{x_k}^{-1} U_k^\intercal, \\
            &\Sigma_{ik}^{\Delta} \succeq (\sigma_i^\intercal \mu_k + \tau_i^\intercal v_k)\Sigma_{\vct\xi}(\sigma_i^\intercal \mu_k + \tau_i^\intercal v_k)^\intercal, \label{eq:mean_cov_constraints}\\
            &\Sigma_f - \Sigma_{x_{N}} \succeq 0, \\
            &\mu_f - \mu_{N} = 0.
        \end{align}
    \end{subequations}
\end{empheqboxed}

\section{Numerical Example}~\label{sec:sims}

In this section, we compare all previous methods developed in the previous sections, using the following linear system from~\cite{capture_propagate_control}
\begin{equation*}
	A = \frac{1}{2}
	\begin{bmatrix}
		1 & -1 \\
		2 & 1
	\end{bmatrix}, \quad
	B = I_{2}, \quad D = 0.1 I_{2}.
\end{equation*}
The initial state is normally distributed with mean $\mu_0 = [2, 10]^\intercal$ and covariance $\Sigma_0 = (1/3)^2 I_{2}$.
The target terminal distribution has mean $\mu_f = [-1; -1]^\intercal$ and covariance $\Sigma_{f} = 0.25 \Sigma_0$.
For the objective function, we set $Q = I_{2}$ and $R = 10I_{2}$.
For the robust mean design (Section~\ref{sec:RDDMS}) we set the terminal constraint set as $\mathcal{X}_{f} = \{x : |\mu_{f,i} - x_{i}| \leq 0.5,~i=1,2\}$, where $x_{i}\in\Re$ denotes the $i$th element of the state.

\subsection{Noise Realization Estimation Analysis}~\label{subsec:noise_estimation_analysis}

\begin{figure}[!htb]
	\centering
	\includegraphics[scale=0.35]{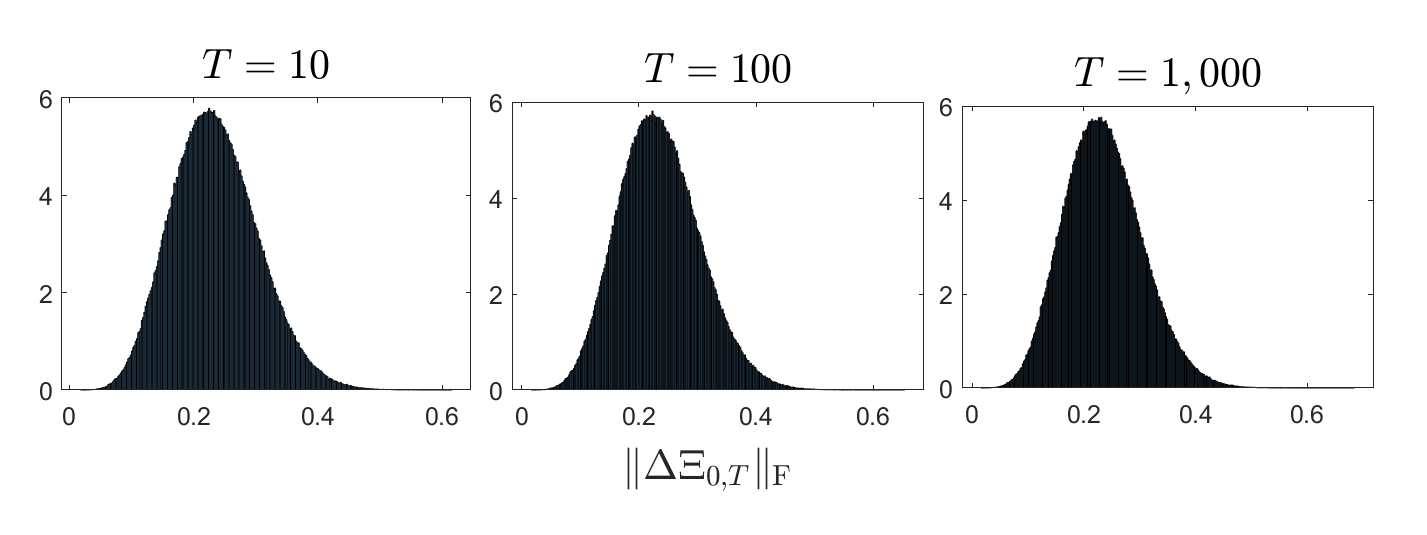}
	\caption{Empirical distribution of norm of joint estimation errors $\Delta\Xi_{0,T}$, for varying sampling horizons $T$.}
	\label{fig:noise_estimation_errors_norm}
\end{figure}

\begin{figure*}[!htb]
    \centering
    \includegraphics[scale=0.5, trim={0cm 0cm 0cm 0cm}, clip]{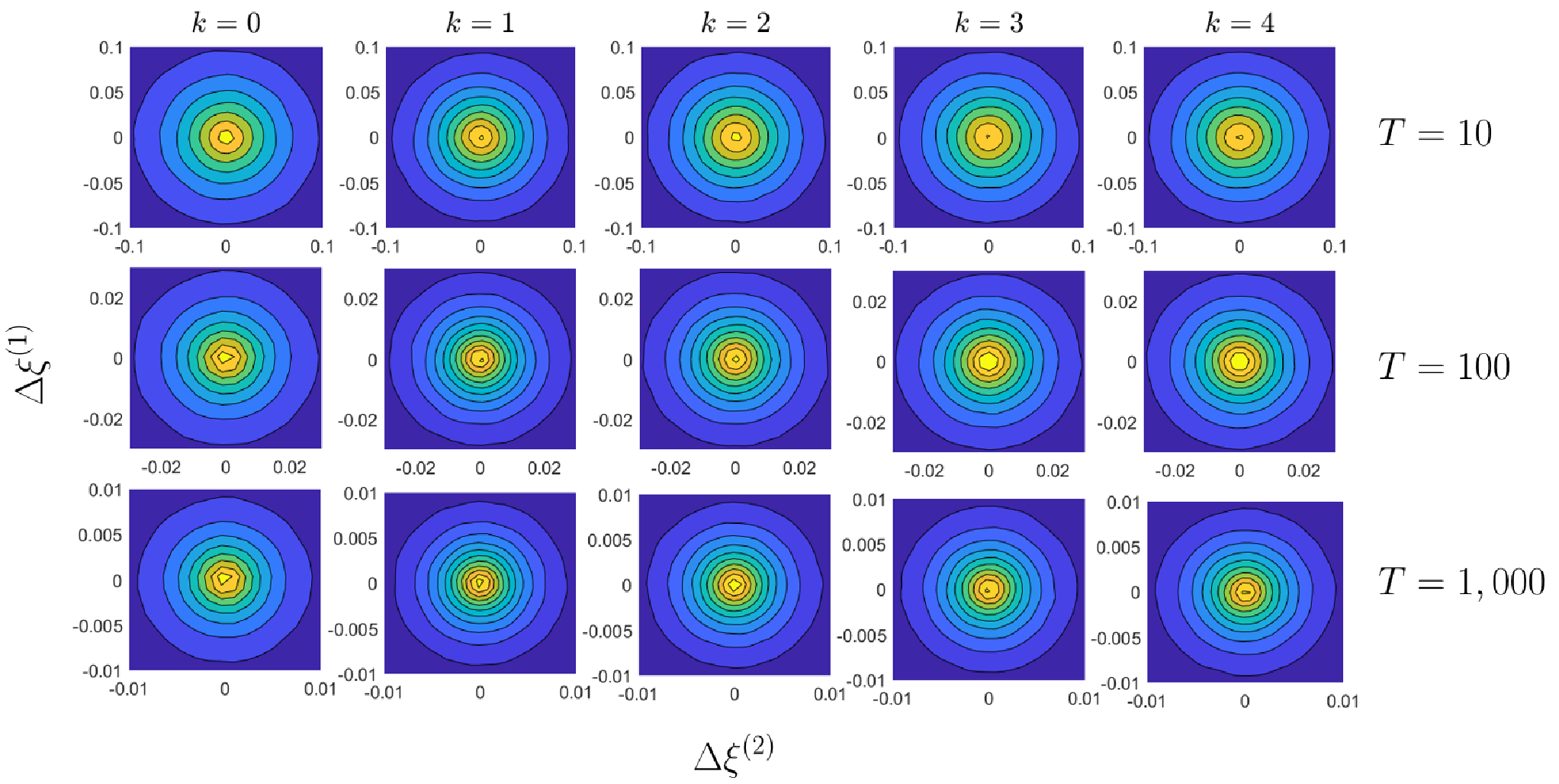}
    \caption{Empirical distribution of noise realization estimation errors $\Delta\xi_{k}$, for varying sampling horizons $T$. Each row corresponds to a different sampling horizon: $T = 10$ (top row), $T = 100$ (middle row), and $T = 1,000$ (bottom row).}
    \label{fig:noise_estimation_errors_individual}
\end{figure*}

We first start with an analysis of the estimation errors $\Delta\Xi_{0,T}$ resulting from the MLE problem \eqref{eq:ML_problem}.
To this end, we ran the noise estimation procedure for $10^{6}$ random trials, where the data was randomly generated for each trial from inputs $\vct u_k^{(d)} \sim U[-1, 1]$, initial state $\vct x_0^{(d)} \sim \mathcal{N}(0, I_2)$, and disturbances $\vct w_k^{(d)} \sim \mathcal{N}(0, DD^\intercal)$.
Figure~\ref{fig:noise_estimation_errors_individual} shows the distribution of the first five noise estimation errors $\Delta\xi_k$ for varying sampling horizon lengths $T\in\{10, 100, 1000\}$.
Notably, we see that, indeed, as we gather more noisy data, the covariance of the estimation errors of each individual noise term converges to its exact value.
%
The total \textit{magnitude} of the joint estimation error, $\|\Delta\Xi_{0,T}\|$, as shown in Figure~\ref{fig:noise_estimation_errors_norm}, however, does not converge to zero.
This is explained by noting that while the individual estimation errors converge with more samples, the compounded error remains fixed due to the increasing number of elements to be estimated (see Appendix~\ref{AppF} for a more formal derivation of the normed error distribution).
Additionally, Table~\ref{table:unc_set_upper_bounds} shows the efficacy of the various upper bounds constructed in Section~\ref{sec:error_bounds}.
\renewcommand{\arraystretch}{1.3}
\begin{table}[!htb]
	\caption{Upper bounds for noise uncertainty set constructions compared to true error quantile $\rho^\star$ for various confidence levels~$\delta$.}
	\label{table:unc_set_upper_bounds}
	\centering
	\begin{tabular}{c|ccccc}
		\toprule
		\midrule
            $\delta$ & 0.1 & 0.2 & 0.3 & 0.4 & 0.5 \\
            \midrule 
            $\rho^\star$ & 0.326 & 0.292 & 0.269 & 0.249 & 0.231 \\
            Tight & 0.326 & 0.293 & 0.269 & 0.249 & 0.231 \\
            Loose $(T = 10)$ & 0.533 & 0.500 & 0.477 & 0.458 & 0.440 \\
            Loose $(T = 100)$ & 1.503 & 1.472 & 1.449 & 1.430 & 1.412 \\
            Loose $(T = 1,000)$ & 4.562 & 4.531 & 4.509 & 4.489 & 4.471 \\
		\midrule
		\bottomrule
	\end{tabular}
\end{table}

As mentioned in Section~\ref{subsec:moment-based},
the uncertainty set \eqref{eq:unc_set_subspace} based on the subspace decomposition of the singular joint Gaussian density (\textit{Tight} row) has the \textit{smallest} conservativeness of all the three alternatives, as it is almost exactly equal to the true quantile.
Additionally, the original uncertainty set construction (\textit{Loose} row) from our previous work in \cite{DD_CS_noise_conservative} (see Corollary~\ref{cor:uncertainty_set_MLE}) provides unnecessarily too loose bounds, and degrades rather quickly for large sampling horizons.
We thus confirm that the uncertainty set constructed from the subspace density \eqref{eq:normal_dist} and associated confidence set in Proposition~\ref{def:conf_ellipsoid}
has the tightest overapproximation to the true quantile, and the original construction in \cite{DD_CS_noise_conservative} is the most conservative due to the spurious extra degrees of freedom.
For the rest of the analysis, in this section, we choose to use the least conservative upper bound $\rho(\delta) = \|\Sigma_{\vct\xi}^{1/2}\| Q^{1/2}_{\vct\rchi^{2}_{n(n + m)}}(1-\delta)$ from \eqref{eq:unc_set_subspace}.

To provide a comparison between the degree of accuracy in the noise estimation between the NN estimator and the MLE estimator, we implemented a feed-forward ReLU NN on simulated input/state/true noise data from the defined linear stochastic system.
The data collection horizon was fixed to $T = 15$ time steps.
The network has an input size of $47 = n(T+1) + mT$ and an output size of $30 = nT$.
Two ReLU activation layers with 500 neurons were chosen, with an added final linear layer for the outputs.
Finally, a standard mean-squared error (MSE) loss was used for training and validation.
The ReLU network was implemented in \texttt{PyTorch Lightning} using the \texttt{ADAM} optimizer~\cite{Kingma2014AdamAM} with a variable learning rate.
\begin{figure}[!htb]
    \centering
    \includegraphics[width=\linewidth]{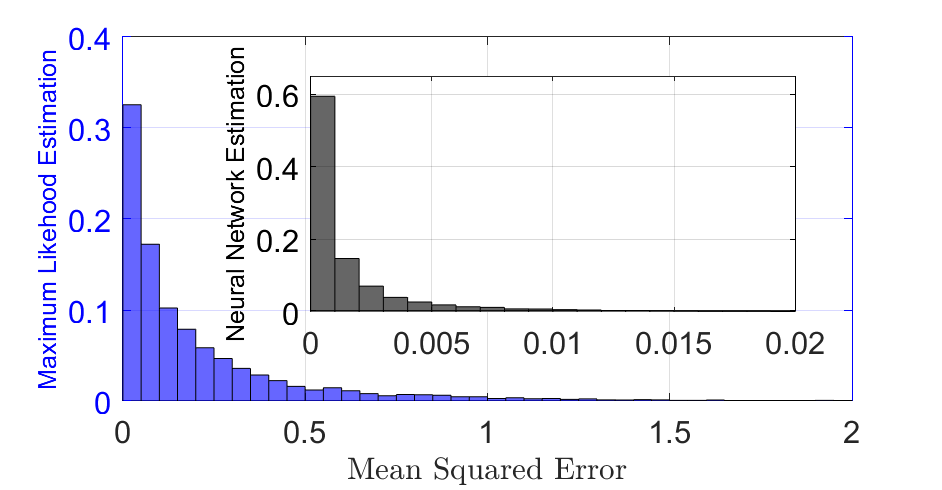}
    \caption{MSE distribution $\mathbb{E}[\|\Xi_{0,T} - \hat{\Xi}_{0,T}\|^{2}]$ of estimation errors between neural network estimation (black) and maximum likelihood estimation (blue).}
    \label{fig:MSE_estimation_comparison}
\end{figure}
Figure~\ref{fig:MSE_estimation_comparison} displays the distribution of the MSE between the NN and MLE estimators, with the mean taken per trial, and with 10,000 total trials forming the empirical distribution.

Notably, it is apparent that the trained NN model is more accurate in estimating the noise realization, as it was trained with respect to data from this exact dynamics model and with exact noise estimates.
However, as mentioned in Section~\ref{subsec:NN}, the NN estimator does not generalize well as the NN would need to be re-trained for a new dynamics model, or even for different data-collection horizons.
The MLE estimator, on the other hand, although has worse average performance, is applicable to any dynamics model, time horizon, and even noise distribution.

\subsection{DD-MS Robustness Analysis}

Next, we turn our attention to analyzing the effect of an unknown model on the resulting optimal mean trajectories.
To this end, we compare the CE (i.e., using the subspace predictor $[\hat{B} \ \hat{A}] = X_{1,T} \mathcal{S}^\dagger$) design with the robust design outlined in Section~\ref{sec:RDDMS}.
We remark that the MLE noise estimation scheme in Section~\ref{subsec:MLE} does not influence the nominal model estimate since $\hat{\Xi}_{0,T}\mathcal{S}^\dagger = X_{1,T}(I_{T} - \mathcal{S}^\dagger\mathcal{S})\mathcal{S}^\dagger = 0$.
To compare the two designs, we ran a set of 500 trials with data generated randomly for each trial in a similar vein to the noise estimation study.
We use the upper bound $\alpha = \rho(\delta) / \sigma_{\min}(\mathcal{S})$, where $\rho(\delta)$
%
as in \eqref{eq:unc_set_subspace}, for the model estimation error, with $\delta = 0.1$ 
for the robust DD-MS design, and a data collection horizon $T = 30$.

\begin{figure}
    \centering
    \includegraphics[width=\linewidth]{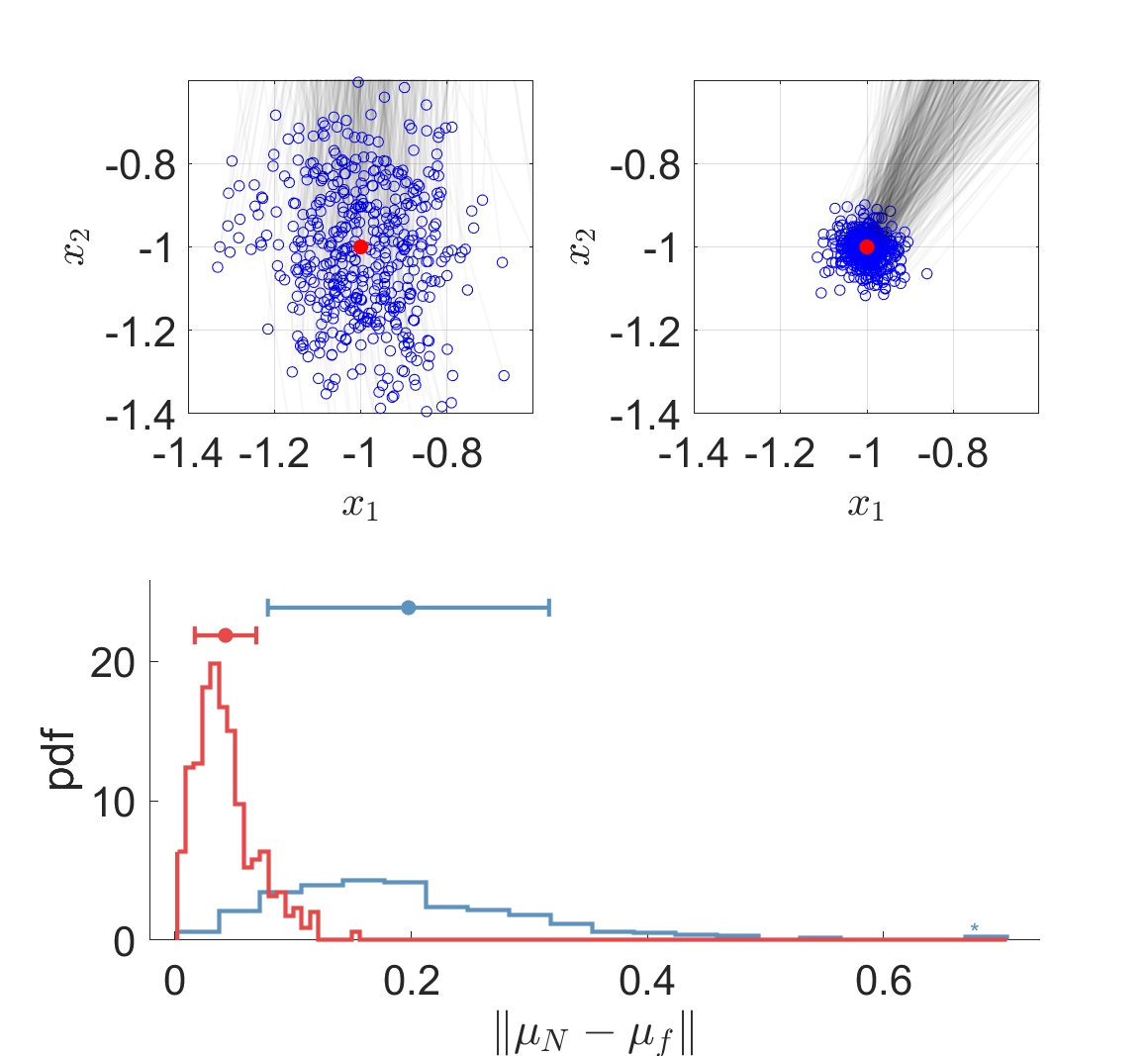}
    \caption{Comparison of (left) DD-MS and (right) R-DD-MS terminal splashpoints for 500 randomized trials of data collected from the true model over a horizon $T=30$, with empirical error distribution (bottom).}
    \label{fig:MS_trajs_true}
\end{figure}

Figure~\ref{fig:MS_trajs_true} shows the difference between the optimal mean trajectories for the two designs with respect to the \textit{true} model $\{A, B\}$.
Clearly, the robust design has better control over the dispersion of the terminal mean trajectories than that of the CE design (as done in \cite{DD_CS_noise_conservative}), as intended.

\begin{figure}[!htb]
    \centering
    \hspace*{0cm}
    \begin{subfigure}{0.45\columnwidth}
        \centering
        \hspace{-0.5cm}
                \includegraphics[width=\linewidth, trim={0.1cm 0.1cm 0.6cm 0.8cm}, clip]{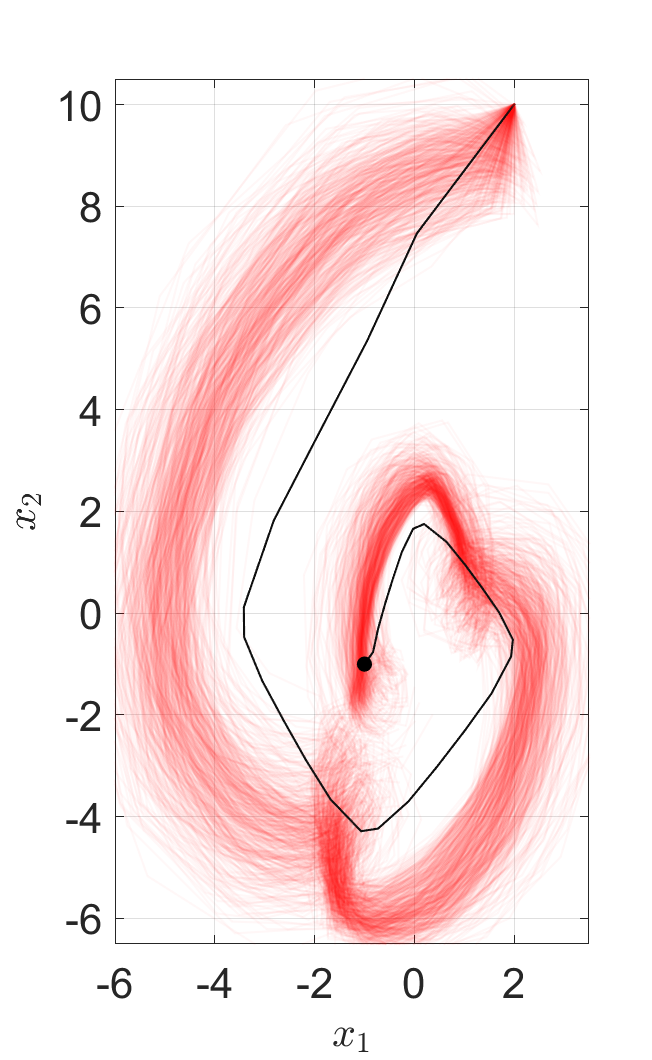}
        \caption{DD-MS trajectories.}
        \label{fig:DD-MS_trajs_random}
    \end{subfigure}
    \hspace{5pt} 
    \begin{subfigure}{0.45\columnwidth}
        \centering
        \hspace{-0.5cm}
        \includegraphics[width=\linewidth, trim={0.1cm 0.1cm 0.6cm 0.8cm}, clip]{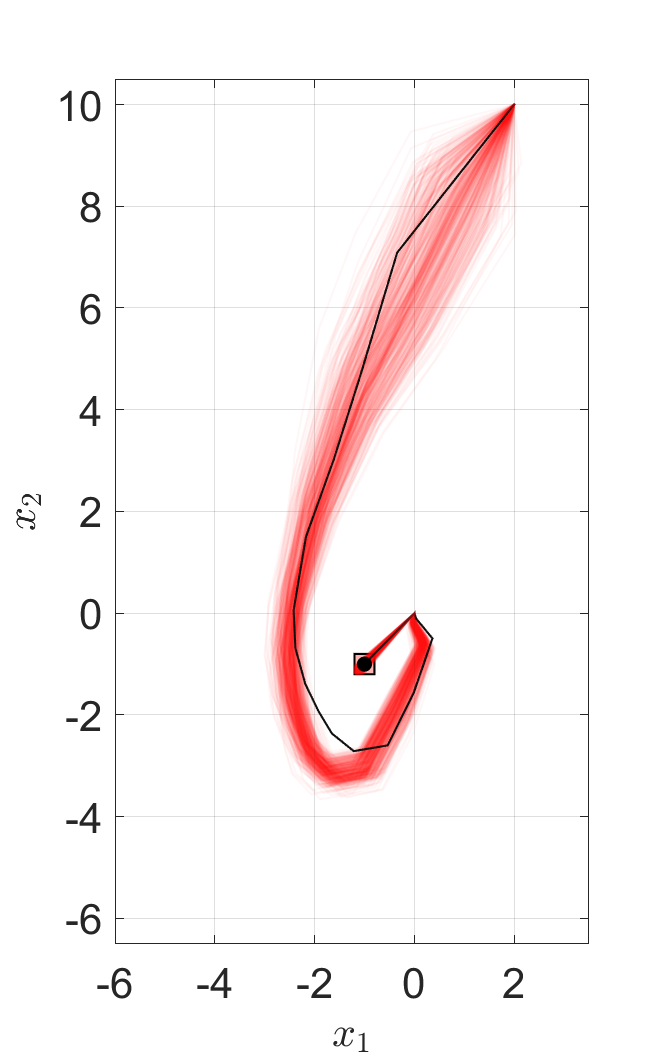}
        \caption{R-DD-MS trajectories.}
        \label{fig:R-DD-MS_trajs_random}
    \end{subfigure}
    \caption{Comparison of (R)DD-MS optimal trajectories for 500 randomized trials of data collected from random models with $\|[\Delta B \ \Delta A]\| \leq \alpha$ over a horizon $T = 30$.}
    \label{fig:MS_trajs_random}
\end{figure}

To see how robust the two data-driven methods are to \textit{random} models, not just simply comparing to a fixed known ground truth, we also ran a single iteration of the data-collection and control design scheme with a fixed robustness level $\alpha = 0.2$, and, subsequently, ran the optimal mean controllers on 500 independent random models that were generated from the requirement $\|[B \ A] - [\hat{B} \ \hat{A}]\| \leq \alpha$.
The resulting optimal mean trajectories are shown in Figure~\ref{fig:MS_trajs_random}.
Notably, the nominal mean controller now performs considerably worse when compared with that of the robust design on randomly perturbed models.
In essence, Figures~\ref{fig:MS_trajs_true}-\ref{fig:MS_trajs_random} show the robustness properties of DD-MS and R-DD-MS from two perspectives: the first shows robustness to random datasets on a known, underlying model, while the second shows robustness to a single dataset on randomly perturbed models.

\begin{wrapfigure}[22]{r}{0.45\columnwidth}
	\centering
        \hspace*{-0.3cm}
	\includegraphics[scale=0.4, trim={0.2cm 0cm 1.2cm 0.8cm}, clip]{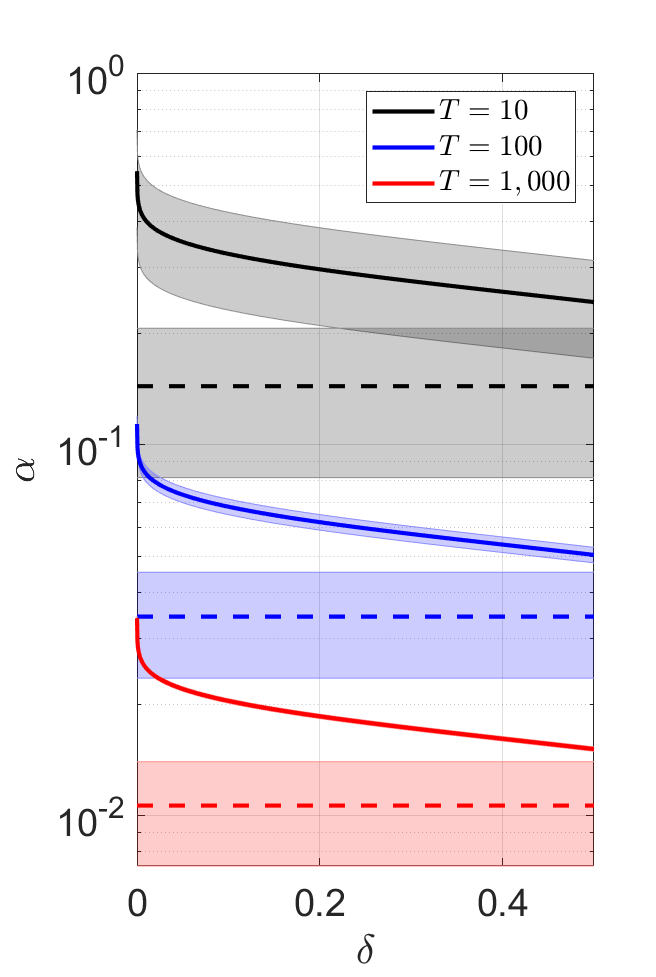}
	\caption{Model estimation error upper bounds (solid) and true estimation error (dashed).}
	\label{fig:model_uncertainty_bounds}
\end{wrapfigure}

\textcolor{black}{%
For completeness, we wish to quantitatively understand the extent of conservatism of the bound $\alpha$ in R-DD-MS as well as the effect of the sampling horizon $T$ on the resulting nominal model inaccuracies.%
}
To this end, we ran a series of one million random trials and computed the values of $\alpha = \rho(\delta) / \sigma_{\min}(\mathcal{S})$ as well as the CE estimated model $[\hat{B} \ \hat{A}] = X_{1,T} \mathcal{S}^\dagger$ for each trial.
\setlength{\tabcolsep}{1pt}  
\begin{table}[!htb]
    \caption{Percentage of feasible solutions to R-DD-MC for varying model uncertainty $\delta$ and terminal constraint space size $|\mathcal{X}_{f}|$.}
    \label{table:feasibility_DDMS}
    \centering
    \begin{tabular}{c@{\hspace{3pt}}|@{\hspace{-3pt}}c@{\hspace{-8pt}}c@{\hspace{-8pt}}c@{\hspace{-8pt}}c@{\hspace{-8pt}}c@{\hspace{-8pt}}c}
        \toprule
        \midrule
        & \multicolumn{6}{c}{Box Width} \\
        \midrule
        $\alpha$ & 0.5 & 0.4 & 0.3 & 0.2 & 0.1 & 0.05 \\
        \midrule
        0.05 & \ApplyGradient{100} & \ApplyGradient{100} & \ApplyGradient{100} & \ApplyGradient{100} & \ApplyGradient{100} & \ApplyGradient{99.4} \\ 
        0.1 & \ApplyGradient{100} & \ApplyGradient{100} & \ApplyGradient{100} & \ApplyGradient{100} & \ApplyGradient{96.2} & \ApplyGradient{50.6} \\
        0.15 & \ApplyGradient{100} & \ApplyGradient{100} & \ApplyGradient{100} & \ApplyGradient{100} & \ApplyGradient{43.8} & \ApplyGradient{17.4} \\
        0.2 & \ApplyGradient{100} & \ApplyGradient{100} & \ApplyGradient{100} & \ApplyGradient{71.0} & \ApplyGradient{13.4} & \ApplyGradient{8.00} \\
        0.25 & \ApplyGradient{100} & \ApplyGradient{99.4} & \ApplyGradient{42.4} & \ApplyGradient{8.40} & \ApplyGradient{6.20} & \ApplyGradient{4.60} \\
        \midrule
        \bottomrule
    \end{tabular}
\end{table}
Using this data, we plot the mean and variance of $\alpha$ as well as $\alpha^\star \triangleq \|[\Delta B \ \Delta A]\|$ for each risk level $\delta$ and for different sampling horizons $T$.
The results are shown in Figure~\ref{fig:model_uncertainty_bounds}.
For clarity, the dependency of $\alpha = \alpha(\rho)$ on $\delta$ and $T$ comes
from the fact that $\rho$ is a function of $\delta$ and $\sigma_{\min}(\mathcal{S})$ is a function of $T$.
We see that $\rho$ decreases with increasing $\delta$ due to a smaller confidence interval, and, similarly, $\sigma_{\min}(\mathcal{S})$ increases with $T$ due to a more expressive dataset.
Indeed, the former point is the entire motivation for the robust fundamental lemma, which aims to quantitatively provide an upper bound on this increase.
We see that, overall, the constructed uncertainty bounds provide a tight over-approximation
of the true normed estimation error, and at 1,000 samples (red curve), we achieve an error of around 2\% with respect to the true model with probability 95\%.
With smaller sample sizes (black curve), however, these errors get quite large with greater dispersions, and thus we see that robust mean designs are necessary for precise control with sparse data.
\setlength{\intextsep}{5pt}     
\setlength{\columnsep}{10pt}    

We conclude the discussion on the DD-MS design by focusing on the specific parameters involved, namely, the robustness level $\alpha$ and the terminal constraint box $\mathcal{X}_{f}$.
One common issue with robust MPC frameworks is the design of the terminal set in order to ensure recursive feasibility \cite{MPC_book}.
In this work, however, we simply want a small enough terminal set, centered around the desired terminal mean that we can robustly steer the system trajectories to.
As such, it is not guaranteed that the robust control problem will be feasible with a given terminal constraint set and model uncertainty $\alpha(\delta  |  \mathbb{D})$ under the dataset $\mathbb{D}$ and confidence level $\delta$. 
Table~\ref{table:feasibility_DDMS} shows a quantitative comparison of the percentage of feasible solutions for varying levels of robustness and terminal set sizes.
We see that, when the terminal box is large and when there is not much robustness, all problems are feasible.
However, as we increase the level of robustness and decrease the size of the terminal box, many more problems become infeasible.
%
Thus, the control designer has a trade-off between the desired accuracy of the nominal model with the level of precision in the terminal state.
\begin{figure*}[!htb]
    \centering
    \hspace*{0.6cm}
    \begin{subfigure}{0.4\columnwidth}
        \centering
        \hspace*{-1.3cm}
        \includegraphics[scale=0.65, trim={0.1cm 1cm 0.8cm 1.5cm}, clip]{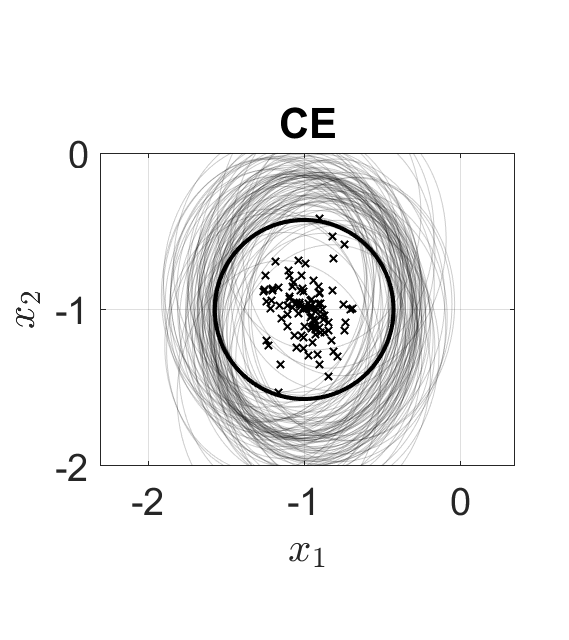}
        \caption{Non-robust design.}
        \label{fig:CE_splashpoints}
    \end{subfigure}
    \hspace{2cm} 
    \begin{subfigure}{0.5\columnwidth}
        \centering
        \hspace*{-1cm}
        \includegraphics[scale=0.65, trim={0.1cm 1cm 0.8cm 1.5cm}, clip]{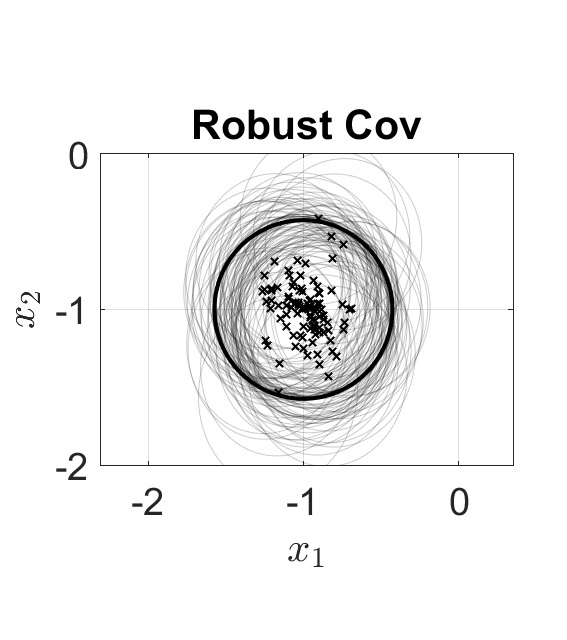}
        \caption{Robust covariance design.}
        \label{fig:robust_cov_splashpoints}
    \end{subfigure}
    \hspace{1cm} 
    \begin{subfigure}{0.6\columnwidth}
        \centering
        \hspace*{-0.5cm}
        \includegraphics[scale=0.65, trim={0.1cm 1cm 0.8cm 1.5cm}, clip]{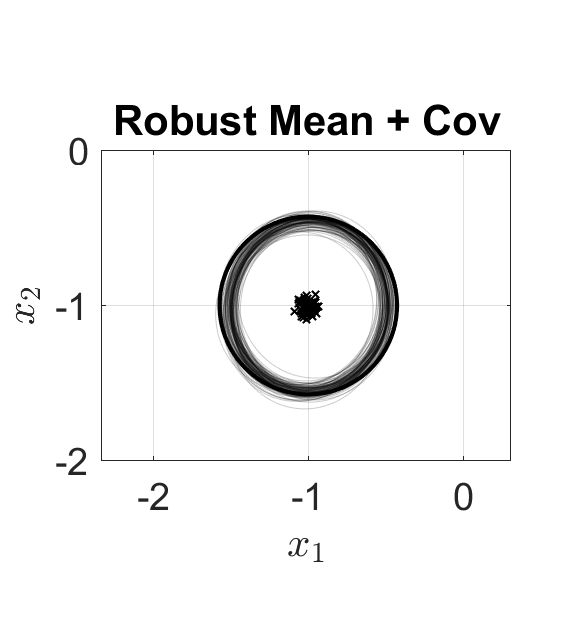}
        \caption{Robust mean/covariance design.}
        \label{fig:robust_mean_cov_splashpoints}
    \end{subfigure}
    \caption{Terminal state mean and covariances of (a) certainty-equivalence design, (b) CE mean design + robust covariance design, and (c) robust mean + covariance designs.}
    \label{fig:splashpoints}
\end{figure*}

\subsection{DD-CS Robustness Analysis}

\begin{figure*}[!htb]
    \centering
    \hspace*{-0.25cm}
    \includegraphics[scale=0.45, trim={0cm 0cm 0cm 0cm}, clip]{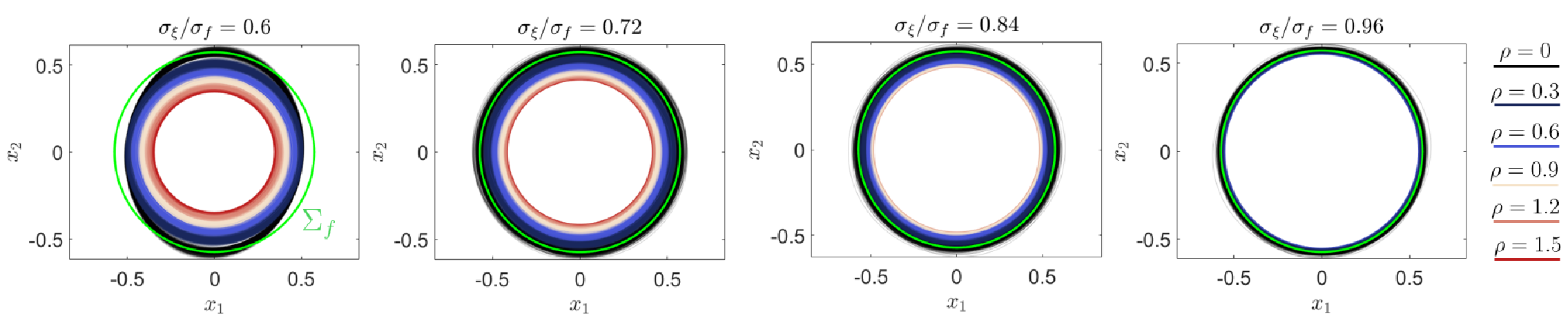}
    \caption{Terminal $3\sigma$ covariance ellipses for varying levels of robustness $\rho$ and desired precision to noise ratio $\sigma_{f} / \sigma_{\vct\xi}$.}
    \label{fig:cov_robustness}
\end{figure*}

Next, we proceed with the DD-CS analysis in the presence of noisy data, and subsequently with the synthesis of both the mean and covariance control designs.
To this end, and similarly to the DD-MS analysis, we first begin with a study on the effect of robustness level $\rho$ on the resulting terminal covariances.
For reference, given the disturbance matrix $D = 0.1 I_{2}$, confidence level $\delta = 0.1$, and sampling horizon $T = 30$, the associated MLE noise error uncertainty set is $\DeltaSet^* = \{\|\Delta\Xi_{0,T}\| \leq \rho\}$, with $\rho = 0.3263$.
Figure~\ref{fig:cov_robustness} shows the terminal covariances from the R-DD-CS design for varying levels of $\rho\in\{0, 0.3, 0.6, 0.9, 1.2, 1.5\}$, as well as for varying noise-to-precision ratio (NPR) $\sigma_{\vct\xi} / \sigma_{f} \in \{0.6, 0.72, 0.84, 0.96\}$,
evaluated on the true dynamics model $\{A, B, D\}$, where we assume $\Sigma_{f} = \sigma_f^2 I_{2}$ and $\Sigma_{\vct\xi} = \sigma_{\vct\xi}^2 I_{2}$.
As the level of robustness to noise estimation errors increases, the terminal covariances become smaller when simulated on the true model because the optimal feedback gains anticipate more uncertainty than there is in reality.
\begin{table}[!htb]
    \caption{Percentage of feasible solutions to R-DD-CS for varying noise estimation error bounds $\delta$ and disturbance variance $\sigma_{\vct\xi}^{2}$.}
    \label{table:feasibility_DDCS}
    \centering
    \begin{tabular}{c@{\hspace{3pt}}|@{\hspace{-3pt}}c@{\hspace{-8pt}}c@{\hspace{-8pt}}c@{\hspace{-8pt}}c}
        \toprule
        \midrule
        & \multicolumn{4}{c}{Noise variance $\sigma_{\vct\xi}^2$} \\
        \midrule
        $\rho$ & $0.1^2$ & $0.12^2$ & $0.14^2$ & $0.16^2$ \\
        \midrule
        0.0 & \ApplyGradient{100} & \ApplyGradient{100} & \ApplyGradient{100} & \ApplyGradient{100} \\ 
        0.3 & \ApplyGradient{100} & \ApplyGradient{100} & \ApplyGradient{100} & \ApplyGradient{100} \\
        0.6 & \ApplyGradient{100} & \ApplyGradient{100} & \ApplyGradient{100} & \ApplyGradient{4.00} \\
        0.9 & \ApplyGradient{100} & \ApplyGradient{99.0} & \ApplyGradient{70.8} & \ApplyGradient{0.00} \\
        1.2 & \ApplyGradient{87.0} & \ApplyGradient{54.2} & \ApplyGradient{1.00} & \ApplyGradient{0.00} \\
        1.5 & \ApplyGradient{22.8} & \ApplyGradient{1.2} & \ApplyGradient{0.00} & \ApplyGradient{0.00} \\
        \midrule
        \bottomrule
    \end{tabular}
\end{table}
Note also that with no robustness (i.e., $\rho = 0$), the terminal covariances (black) do \textit{not}, in general, satisfy the constraints (green) due to the non-zero estimation errors.
Hence, robustness against estimation errors is essential for feasible covariance designs.

It is interesting to note that as the NPR increases, not only does the terminal covariance becomes larger, as expected, but also the R-DD-CS is more likely to become infeasible.
According to \cite{CS_george_journal}, there is a theoretical lower bound on the achievable terminal covariance, given by $\Sigma_{N} \succeq D_{N-1} D_{N-1}^\intercal$. 
As the noise covariance $\sigma_{\vct\xi}$ increases and approaches $\sigma_f$, this lower bound becomes more constraining. 
At the same time, increasing the robustness level $\rho$ requires the covariance steering algorithm to aim for smaller values of $\Sigma_{N}$ to ensure $\Sigma_{N} \preceq \Sigma_{f}$ holds for all bounded errors within $\rho$. 
However, when $\rho$ becomes too large relative to the gap between $\sigma_\xi$ and $\sigma_f$, the convex program \eqref{eq:convexProblem} with robust constraints \eqref{eq:robust_covariance_prop_constraints}
becomes infeasible as it cannot reduce $\Sigma_{N}$ below the theoretical lower bound while meeting the robustness constraints. Table~\ref{table:feasibility_DDCS} empirically verifies this relationship, showing the percentage of feasible solutions for each NPR across 500 random trials.
\begin{figure*}[!htb]
    \centering
    \hspace*{0.3cm}
    \begin{subfigure}{0.35\columnwidth}
        \centering
        \hspace*{-0.8cm}
        \includegraphics[scale=0.35, trim={0cm 0cm 0cm 0cm}, clip]{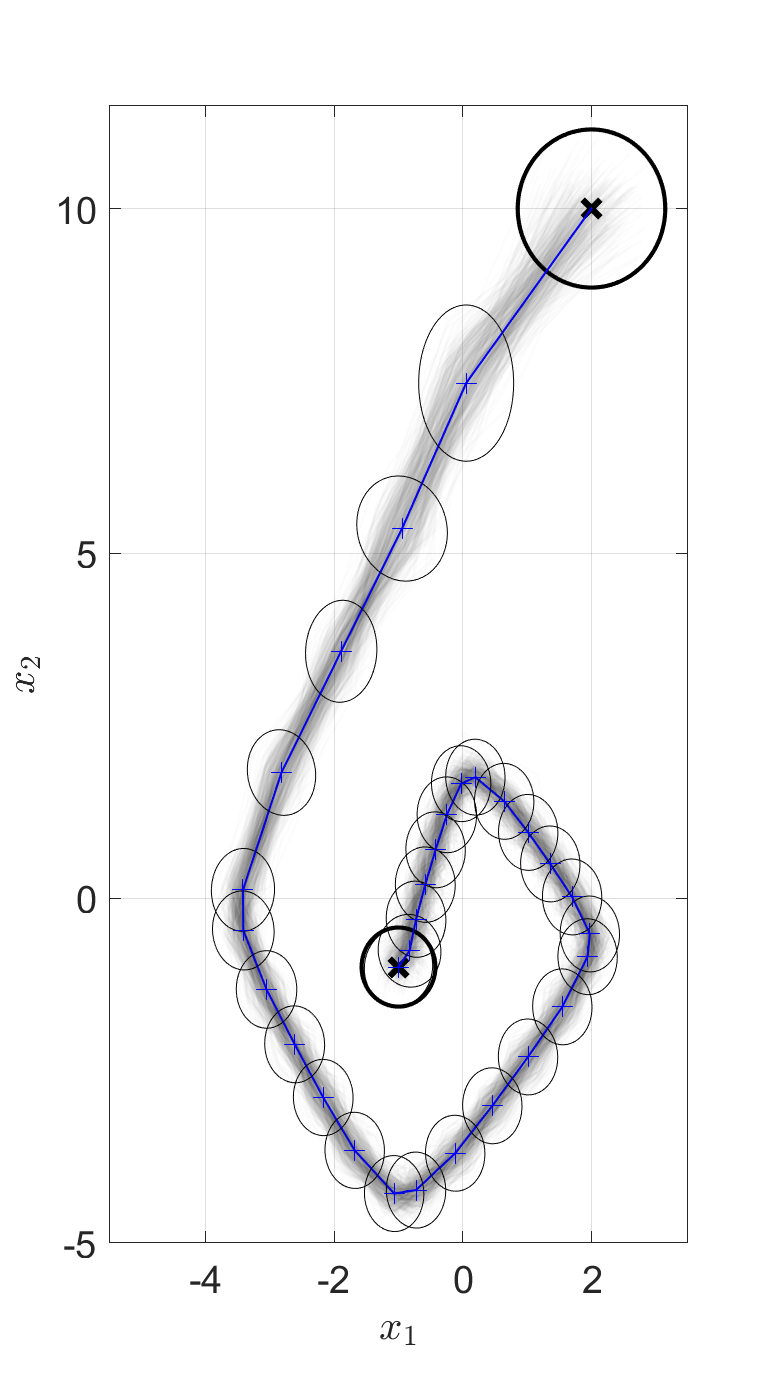}
        \caption{Model-based DS.}
        \label{fig:optimal_trajs_MBCS}
    \end{subfigure}
    \hspace{1cm} 
    \begin{subfigure}{0.35\columnwidth}
        \centering
        \hspace*{-0.8cm}
        \includegraphics[scale=0.35, trim={0cm 0cm 0cm 0cm}, clip]{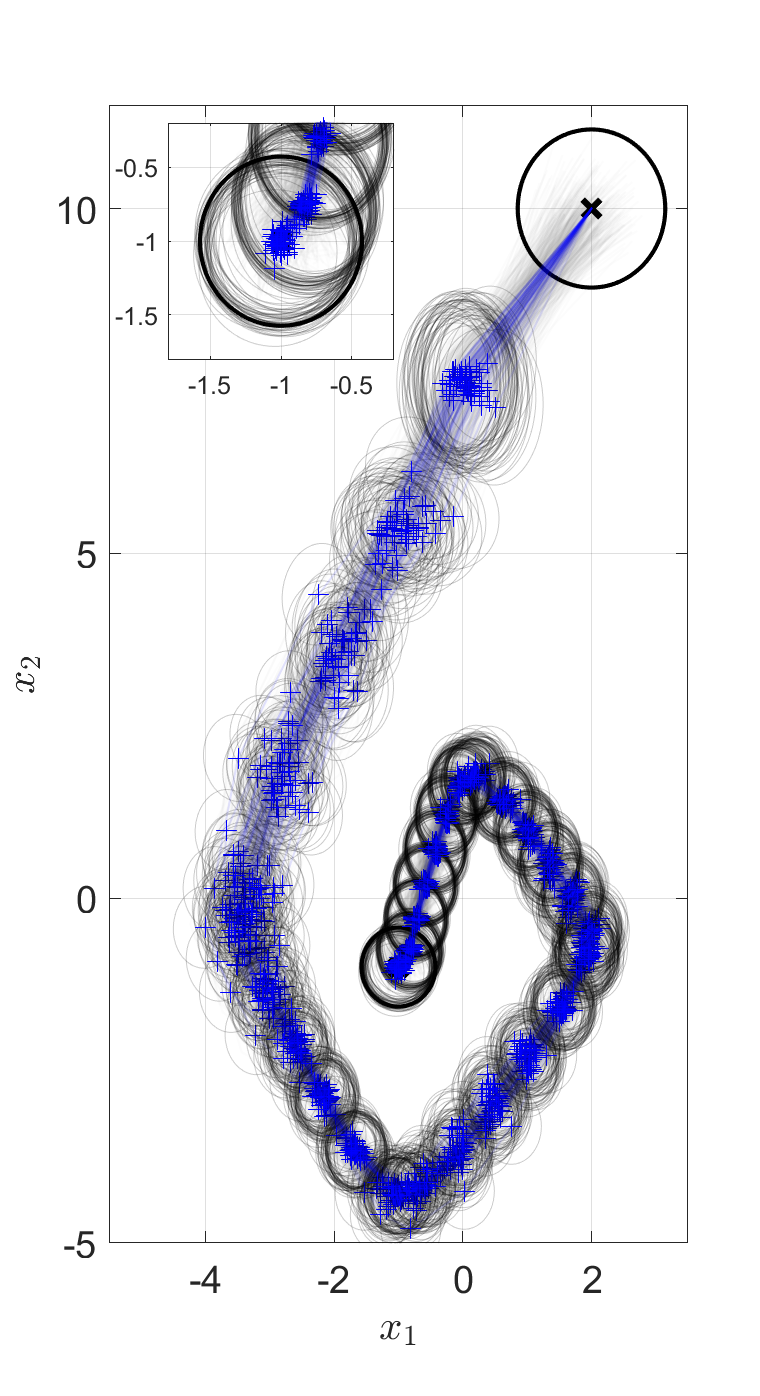}
        \caption{Data-driven DS.}
        \label{fig:optimal_trajs_DDCS}
    \end{subfigure}
    \hspace{1cm} 
    \begin{subfigure}{0.35\columnwidth}
        \centering
        \hspace*{-0.8cm}
        \includegraphics[scale=0.35, trim={0cm 0cm 0cm 0cm}, clip]{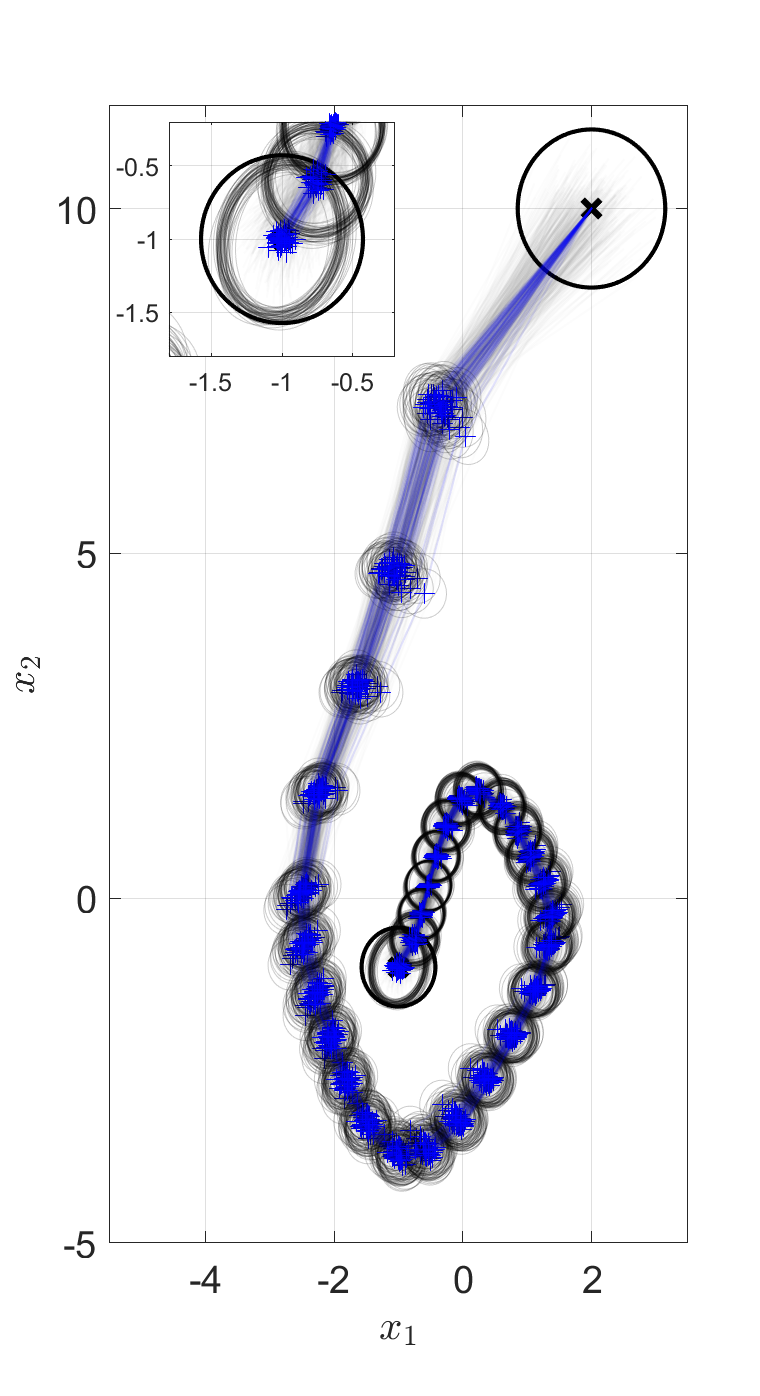}
        \caption{Robust DD-DS.}
        \label{fig:optimal_trajs_RDDCS}
    \end{subfigure}
        \hspace{0.5cm} 
    \begin{subfigure}{0.6\columnwidth}
        \centering
        \includegraphics[scale=0.35, trim={0cm 0cm 0cm 0cm}, clip]{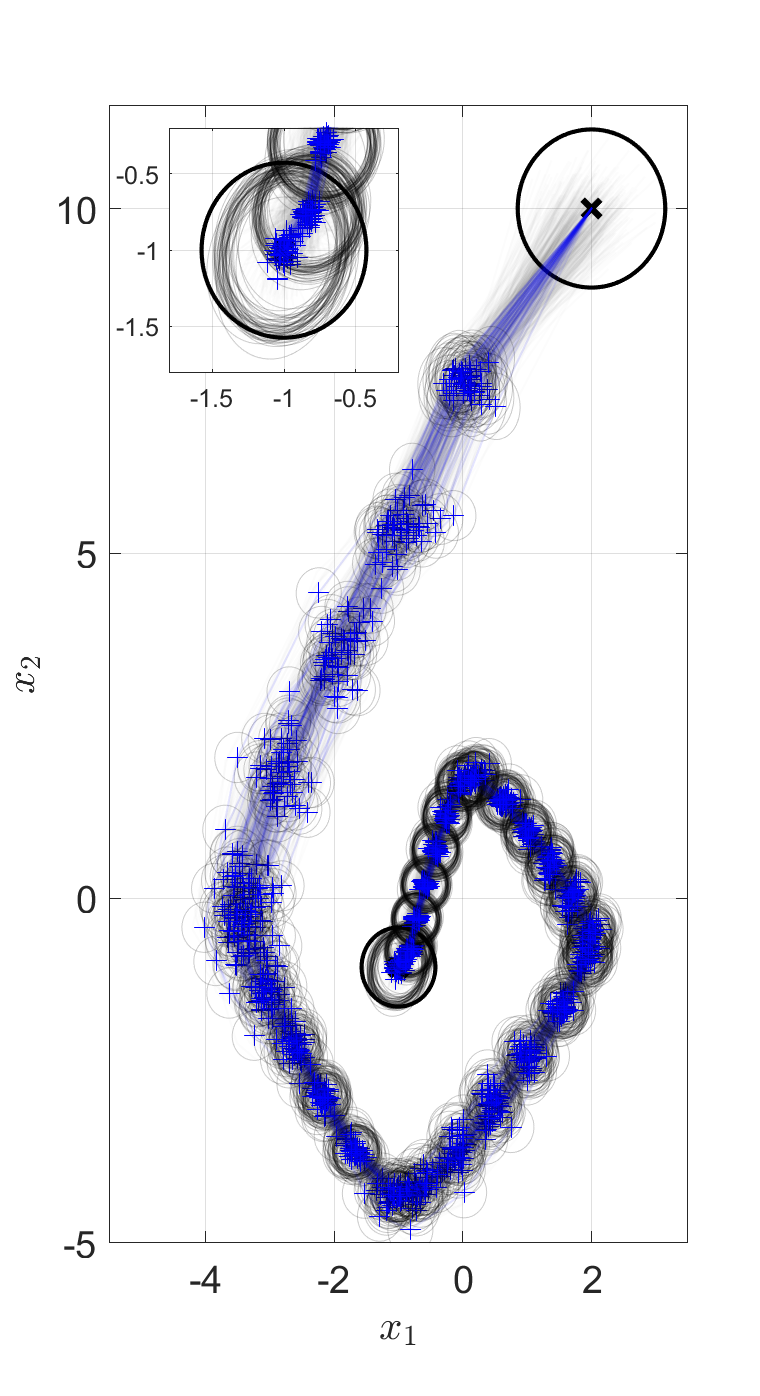}
        \caption{Parametric uncertainty DD-DS.}
        \label{fig:optimal_trajs_PUDDCS}
    \end{subfigure}
    \caption{Comparison of optimal closed-loop trajectories between (a) model-based, and (b-d) data-driven control design. The solid lines represent the mean trajectories and the ellipses represent the 3$\sigma$ covariance ellipses for each randomized trial.}
    \label{fig:optimal_trajs}
\end{figure*}

\subsection{Closed-Loop Density Steering}

Lastly, we combine the DD-MS and DD-CS designs and look at the resulting optimal trajectories.
We choose a planning horizon of $N = 30$ and ran a set of 50 random trials for data collection, where, for each trial, we plot 10 Monte Carlo trajectories from randomly sampling the additive noise.
Of all the designs, the R-DD-DS design performs the best in terms of achieving the closest terminal distribution to the desired one in the presence of noisy data.
The PU-DD-DS design, which does \textit{not} estimate the noise realizations but instead incorporates the distributional knowledge as multiplicative uncertainty, satisfies the terminal covariances for each trial, but is not robust against mean estimation errors, similar to vanilla DD-MS.
Figuring out a way to robustly satisfy the terminal mean constraints in this parametric uncertainty framework is an interesting problem for future work.

\section{Discussion and Open Problems}

The numerical study in Section~\ref{sec:sims} highlights the principal strengths and trade-offs of the proposed DUST framework.
On the strengths side, the uncertainty quantification used in the robust formulations is both principled and tight: the empirical distribution of the joint noise estimation error norm $\|\Delta\Xi_{0,T}\|$ is essentially invariant to the horizon $T$ (once persistence of excitation fixes the subspace dimension), while the subspace-based bound from \eqref{eq:unc_set_subspace}
matches the empirical quantiles across confidence levels, in sharp contrast to the earlier, overly loose construction~\cite{DD_CS_noise_conservative} whose conservatism worsens with $T$.
This is evidenced by the invariance of $\|\Delta\Xi_{0,T}\|$ across $T$ (Fig.~\ref{fig:noise_estimation_errors_norm}) and in Table~\ref{table:unc_set_upper_bounds}, where the ``Tight" subspace decomposition nearly coincides with the true quantile, whereas the original ``Loose" bound deteriorates markedly.
These uncertainty sets enable tractable robust mean and covariance steering optimization programs (e.g., SDPs) that are solved to global optimality with standard solvers delivering improved performance and constraint satisfaction compared with the certainty-equivalent and model-based baselines used in this study.

The main limitations of the proposed approach are two-fold: (i) modeling assumptions under which the proposed guarantees are defined, namely, linear time-invariant dynamics and Gaussian disturbances, and (ii) feasibility of the reformulated steering problems, namely, there is no guarantee that for the given uncertainty error bound $\rho$ (in robust CS), or similarly for the computed model-mismatch error bound $\alpha$ (in robust MS), the convex programs will have a solution.

We remark that, using the statistical properties of the noise for prediction and estimation is not unique to this work.
Indeed, in the context of DeePC, recent methods such as the signal matrix model \cite{DD_SMM,DD_MLE} and Wasserstein estimation \cite{DD_Wasserstein_estimator} have been proposed to find a relationship between past inputs/outputs and future outputs.
Additionally, \cite{DD_confidence_regions} extended these estimators to generate confidence sets, similar to our work, which can be used in the control design.
Our work differs from these methods in the sense that we estimate the \textit{noise realizations} (not just the statistics of the noise process) arising from the noisy data, from which we generate confidence sets for robust control.

The framework developed in this work assumes normally distributed additive noise as well as Gaussian boundary conditions for the state distribution.
As such, we envision the theory to serve as a baseline for future work incorporating increasing layers of complexity to the problem.
One natural extension is to incorporate probabilistic (or chance) constraints on the state and the input along the planning horizon, which typically represent physical limitations or safety considerations, such as thrust saturation or no-fly zones (NFZs).
For example, assuming that the allowable state constraint space is modeled through the polytope $\mathcal{X} \triangleq \{x : \max_{j\in\dbracket{J}} \alpha_j^\intercal x + \beta_j \leq 0\}$, these chance constraints, under Gaussian noise and linear dynamics, can be sufficiently enforced through the following inequalities~\cite{CS_george_journal}
\begin{align}~\label{eq:chance_constraints]}
    &\alpha_j^\intercal \mu_k + \Phi^{-1}(1 - \delta_{j, k}) \sqrt{\alpha_j^\intercal \Sigma_k \alpha_j} + \beta_j \leq 0, \nonumber \\
    &\hspace{3.95cm}\forall j\in\dbracket{J}, \ \forall k\in\dbracket{N},
\end{align}
where $\Phi(\cdot)$ is the standard normal CDF and $\delta_{j,k} > 0$ are the individual risk allocations.
Notice from \eqref{eq:chance_constraints]} that the underlying unknown model $\{A, B, D\}$ does not show up in the reformulated constraints.
Furthermore, since $\{\mu_k, \Sigma_k\}$ are decision variables in our formulation, enforcing these chance constraints will not require much modification compared to that of the model-based design.

It is often the case that we do not know the exact distributional form of the additive noise.
Most works (including the current one) assume, for simplicity, normally distributed disturbances.
This too may be a limiting assumption depending on the context.
There has been a surge in recent works on distributionally robust (DR) motion planning and trajectory optimization that is fueled by this very point \cite{capture_propagate_control,taskesen2023distributionally,DeePC_dist_robust}.
To this end, another natural extension of the DD-DS framework is to the class of problems where there is distributional uncertainty in the disturbances belonging, for example, to a Wasserstein ambiguity set $\mathbb{B}_{\varepsilon}(\hat{\P})$ of radius $\varepsilon$ centered around the nominal distribution $\hat\P$ that may either be chosen as a Gaussian or is empirically estimated from data.
In this regard, one would not only steer the mean and covariance of the center distribution of the state (as we have done here), but, additionally, steer the radius of the ambiguity set through feedback control \cite{CS_dist_robust_wasserstein}.
The synthesis of DR optimization techniques with direct DD control methods stemming from notions of PE data is a fruitful avenue for future work.

On a more conceptual note, while we parameterize the feedback gains in \eqref{eq:WFL_gains} in terms of the collected data using the new decision variables $G_k$, this required the intermediate step of estimating the noise realization $\hat\Xi_{0,T}$ (e.g., via MLE) which was essential to generate an uncertainty set $\{\|\Delta\Xi_{0,T}\| \leq \rho\}$ for terminal constraint satisfaction with high probability.
We leave it as an open problem to design feedback controllers that satisfy the terminal constraints, which use \textit{only} noisy input/state data without any intermediate noise estimation.
	
The PU-DD-DS design is attractive because it does not rely on the necessity for any type of noise estimation or error bounds, and directly uses the known statistics of $\vct\xi_k^{(d)} \sim \mathcal{N}(0,\Sigma_{\vct\xi})$ for control design.
In this Bayesian viewpoint the parameters are treated as random variables rather than known, fixed quantities \cite{Hackenberger2019}.
However, the resulting mean trajectories are not robust to the noisy data, as we assumed a CE design.
Hence, a possible extension, as mentioned previously, is to robustify this parametric uncertainty design.
	
The robust fundamental lemma, outlined in Section~\ref{subsec:RFL}, is a great theoretical tool to bound estimation errors from CE indirect designs based on generalized notions of persistence of excitation.
However, it is not very practical because the parameters needed for these upper bounds are functions of the underlying (unknown) model.
An interesting question is how to formulate tractable upper bounds based on the RFL (for example, using RMT) for later use in robust DD-DS.
 
Willems' fundamental lemma has inspired much of the work on direct data-driven control, including the current work.
%
%
However, the original WFL as stated, is only valid for deterministic dynamics.
The authors in \cite{DD_SPC_PCE1} state a stochastic FL, however, this is limited to the context of polynomial chaos expansions of random variables.
We leave it for future investigation to derive a more general moment-based FL characterizing the space of the state mean and covariance trajectories of a stochastic LTI system under additive Gaussian noise.


\section{Conclusion}

We have presented a novel framework for data-driven stochastic optimal control for unknown linear systems with distributional boundary conditions, referred to as data-driven density steering (DD-DS). 
The proposed framework provides a comprehensive approach to the design of optimal controllers that steer the state distribution of an uncertain linear system to a desired terminal Gaussian distribution, using only input-state data collected from the actual system.
By parameterizing the feedback gains directly in terms of the collected data, we have reformulated the data-driven distribution steering problem as an uncertain convex problem in terms of the unknown noise realizations.
Using techniques from behavioral systems theory and statistical learning, we were able to develop tight, tractable uncertainty sets for the
{errors of these}
estimated noise realizations, which were subsequently used to formulate and solve robust data-driven extensions for the mean steering (DD-MS) and covariance steering (DD-CS) of the state, which guarantee high-probability constraint satisfaction under bounded estimation errors.
Additionally, an alternative parametric uncertainty formulation (PU-DD-DS) was developed that treats model uncertainties probabilistically rather than deterministically.
Extensive numerical studies demonstrated the efficacy of the proposed methods compared to certainty-equivalence and model-based approaches.
%

\begin{ack}
This work has been supported by NASA University Leadership Initiative award 80NSSC20M0163.
The article solely reflects the opinions and conclusions of its authors and not any NASA entity.
\end{ack}

\appendix

\section*{Appendices}

\section{DD-CS Relaxation}
\label{AppA}

For ease of readability, we fix the notation in this Appendix as follows.
Let $X \triangleq X_{1,T} - \Xi_{0,T} \in \Re^{n\times T}, U\triangleq U_{0,T} \in \Re^{m\times T}$, and $X_0 \triangleq X_{0,T} \in \Re^{n\times T}$.
We assume that $X_0$ has full row rank $n$ (almost surely guaranteed under persistence of excitation).
It follows that $\Sigma_k$ 
can be represented as $\Sigma_k = X_0 S_k$ for some $S_k \in \Re^{T\times n}$.
Let $\tilde{R}_k \triangleq U^\intercal R_k U \succeq 0$.
The LMI relaxation $[\Sigma_k, S_k^\intercal; S_k, Y_k] \succeq 0$ in \eqref{eq:convexProblem_ineqConstraint} is equivalent (by the Schur complement) to $Y_k \succeq S_k \Sigma_k^{-1} S_k^\intercal$.

The relaxed optimization problem~\eqref{eq:convexProblem}
is re-written below as
\begin{subequations}\label{eq:cov_relaxation_A}
    \begin{align}
        \min_{\Sigma_k,S_k,Y_k}\quad
        & J_{\Sigma} = \sum_{k=0}^{N-1}\!\Big(\mathrm{tr}(Q_k\Sigma_k)+\mathrm{tr}(\tilde R_k Y_k)\Big) \\
        \text{such that,} &\ \text{for all } k \in \dbracket{N-1}_{0}, \\
        & \Sigma_k \succeq 0,\; S_k\in \mathbb{R}^{T\times n},\; Y_k\succeq 0, \\
        & \mathcal M_k \triangleq
          \begin{bmatrix}\Sigma_k & S_k^\top\\ S_k & Y_k\end{bmatrix} \succeq 0, \\
        & \mathcal N_k \triangleq
          \begin{bmatrix}\Sigma_{k+1}-\Sigma_{\xi} & X S_k\\ S_k^\top X^\top & \Sigma_k\end{bmatrix} \succeq 0, \label{Sk:prop}\\
        & \Sigma_k = X_0 S_k, \\
        & \Sigma_0=\Sigma_i \succ 0,\quad \Sigma_N=\Sigma_f\succ 0.
    \end{align}
\end{subequations}

Let $P_{k}$ denote the \textit{true} state covariance, and
let the state-feedback gains be recovered from $(\Sigma_k, S_k)$ via
\begin{equation*}
    G_k \triangleq S_k \Sigma_k^{-1}, \quad 
    \begin{bmatrix}
        K_k \\
        I_{n}
    \end{bmatrix} = 
    \begin{bmatrix}
        U \\
        X_0
    \end{bmatrix} G_k.
\end{equation*}
It follows that the true closed-loop covariance obeys
\begin{equation*}
    P_{k+1} = (A + BK_k) P_k (A + BK_k)^\intercal + \Sigma_{\vct\xi}, \quad P_0 = \Sigma_{i}.
\end{equation*}
\begin{thm}~\label{thm:cov_certificates}
Assume that the optimization problem \eqref{eq:cov_relaxation_A} is feasible.
If $\Sigma_i,\Sigma_{\vct\xi} \in \mathbb{S}^n_{++}$, then $P_k \preceq \Sigma_k$ for all $k$, and, in particular, $P_{N} \preceq \Sigma_f$.
\end{thm}

To prove Theorem~\ref{thm:cov_certificates}, we first need the following lemmas.

\begin{lem}~\label{lem:induction_lemma}
    Let $A_k \triangleq A + BK_k$ and define the monotone affine-congruence map $\mathcal{T}_k(X) \triangleq A_k X A_k^\intercal + \Sigma_{\vct\xi}$.
Let the sequence $\{P_k\}$ generated by the recursion
    \begin{equation*}
        P_{k+1} = \mathcal{T}_k(P_k), \quad P_0 = \Sigma_i,
    \end{equation*}
    and let the sequence $\{\Sigma_k\}$ generated by the recursion
    \begin{equation*}
        \Sigma_{k+1} \succeq \mathcal{T}_k(\Sigma_k), \quad \Sigma_0 = \Sigma_i.
    \end{equation*}
    Then, $P_k \preceq \Sigma_k$ for all $k$.
\end{lem}
\begin{pf}
    We prove the lemma by induction: 
    By hypothesis, $P_0 = \Sigma_0$.
Assume that $P_k \preceq \Sigma_k$.
Then, by congruence monotonicity%
    \footnote{
        Congruence monotonicity states that if $0\preceq P \preceq\Sigma$ and $M$ is any matrix, then $MPM^\intercal \preceq M\Sigma M^\intercal$.
    }
    of $\mathcal{T}_k$, it follows that
    \begin{equation*}
        P_{k+1} = \mathcal{T}_k(P_k) \preceq \mathcal{T}_k(\Sigma_k) \preceq \Sigma_{k+1}.
    \end{equation*}
    Hence, $P_{k+1} \preceq \Sigma_{k+1}$, thus
    completing the induction.
    \qedblack
\end{pf}

\begin{lem}~\label{fact:gram_factorization_PD}
    Let $r \in \mathbb{N}$, and let $M_j \in \mathbb{R}^{n\times p_j}$ for $j = 1,\ldots, r$.
    Define $\Sigma = \sum_{j=1}^{r} M_j M_j^\intercal$ and $G = [M_1 \ M_2 \cdots M_r] \in \mathbb{R}^{n\times p}$, with $p = \sum_{j=1}^{r} p_j$.
    Then, $\Sigma \succ 0$ if an only if $\mathrm{rank}( G ) = n$.
\end{lem}

\begin{pf}
It follows immediately from the fact that, for all $y\in\mathbb{R}^{n}$, we have
\begin{equation*}
    y^\intercal \Sigma y = \sum_{j=1}^{r} \|M_j^\intercal y\|^{2} = \|G^\intercal y\|^{2} \geq 0,
\end{equation*}
and the fact that $\mathrm{rank}(G) = n$ if and only if $\mathrm{ker}(G^\intercal) = \{0\}$.
\qedblack
\end{pf}

\begin{lem}~\label{lem:always_PD_cov}
    Let the closed-loop system matrix be $A_k \triangleq A + BK_k$, and assume $\Sigma_0 \succ 0$.
    Define $\Phi(k,\ell) \triangleq A_{k-1} \cdots A_{\ell}$ for $k > \ell$ and $\Phi(\ell,\ell) = I$.
    Suppose at least one of the following conditions holds:
    \begin{enumerate}
        \item[$i)$] $\Sigma_{\vct\xi} \succ 0$
        
        \item[$ii)$] $\Sigma_{\vct\xi} \succeq 0$ and $A_k$ is non-singular for all $k\in\dbracket{N}$
        
        \item[$iii)$] For all $k \in\dbracket{N}$,
\begin{equation}~\label{eq:sufficient_condition_PD_cov}
            \mathrm{rank}\Big[ \Phi(k,0)\Sigma_0^{1/2} \ \Phi(k,1)\Sigma_{\vct\xi}^{1/2} \ \cdots \ \Phi(k,k) \Sigma_{\vct\xi}^{1/2} \Big] = n
\end{equation}
    \end{enumerate}
    Then, $\Sigma_k \succ 0$ for all $k\in\dbracket{N}$.
\end{lem}
\begin{pf}
If $i)$ holds, then $\Sigma_{\vct\xi} \succ 0$ implies $\Sigma_{k+1} = A_k \Sigma_k A_k^\intercal + \Sigma_{\vct\xi} \succ 0$ for all $k \geq 0$.

If $ii)$ holds, we proceed by induction: By hypothesis, $A_k$ is non-singular and
the claim is true for $k=0$ since $\Sigma_0 \succ 0$.
Assume that $\Sigma_k \succ 0$.
Then, $A_k \Sigma_k A_k^\intercal \succ 0$.
Adding $\Sigma_{\vct\xi} \succeq 0$ preserves positive definiteness, hence $\Sigma_{k+1} \succ 0$, thus completing the induction.%

The state covariance propagation \eqref{eq:covDynamics} can be equivalently expressed as
\begin{equation}~\label{eq:cov_dynamics_STM}
    \Sigma_{k} = \Phi(k,0) \Sigma_0 \Phi(k,0)^\intercal + \sum_{i=0}^{k-1} \Phi(k,i+1) \Sigma_{\vct\xi} \Phi(k,i+1)^\intercal.
\end{equation}
Using \eqref{eq:cov_dynamics_STM}, write $\Sigma_k = G_kG_k^\intercal$, where
\begin{equation}~\label{eq:gram_decomposition}
    G_k \triangleq [\Phi(k,0)\Sigma_0^{1/2} \ \Phi(k,1)\Sigma_{\vct\xi}^{1/2} \ \cdots \ \Phi(k,k)\Sigma_{\vct\xi}^{1/2}].
\end{equation}

In case $iii)$ holds, the result follows immediately from \eqref{eq:gram_decomposition} and Lemma~\ref{fact:gram_factorization_PD}.
\qedblack

\end{pf}

We are now ready to provide the proof of Theorem~\ref{thm:cov_certificates}.

\begin{pf}(of Theorem~\ref{thm:cov_certificates}).
    We will show that, for each time index $k$, the actual covariance of the closed-loop system $P_{k}$ is upper-bounded by $\Sigma_k$ where $\Sigma_k$ is given by the solution of the optimization problem~\eqref{eq:cov_relaxation_A}.
    
    By the Schur complement of the covariance propagation constraint \eqref{Sk:prop}, we have that
    \begin{equation}~\label{eq:cov_prop_relaxed_Schur}
        \Sigma_{k+1} - \Sigma_{\vct\xi} \succeq X S_k \Sigma_{k}^{-1} S_k^\intercal X^\intercal.
    \end{equation}
    Since $\Sigma_i, \Sigma_{\vct\xi} \succ 0$, it follows from Lemma~\ref{lem:always_PD_cov} that $\Sigma_k \succ 0$ for all $k \in \dbracket{N}$.
    Using $G_k = S_k \Sigma_k^{-1}$ 
    along with the identity $XG_k = (X_{1,T} - \Xi_{0,T}) G_k = [B \ A] [K_k; I_{n}] = A + BK_k = A_k$, 
    equation~\eqref{eq:cov_prop_relaxed_Schur}
    is equivalent to
    \begin{equation*}
        \Sigma_{k+1} \succeq \Sigma_{\vct\xi} + A_k \Sigma_{k} A_k^\intercal = \mathcal{T}_k(\Sigma_k).
    \end{equation*}
    From Lemma~\ref{lem:induction_lemma} in follows that  $P_k \preceq \Sigma_k$ for all $k$ and, in particular, $P_{N} \preceq \Sigma_{N} = \Sigma_{f}$.
    \qedblack

\end{pf}

\section{Proof of Theorem~\ref{thm:MLE_solution}}
\label{AppB}

Substituting the multivariable normal statistics PDF into \eqref{eq:ML_problem_cost} yields the objective function
\begin{subequations}
    \begin{align*}
    \hspace*{-3mm}
        \mathcal{J}_{\mathrm{ML}} &= \sum_{k=0}^{T-1} \bigg(-\frac{n}{2}\log(2\pi) - \frac{1}{2}\log\det\Sigma_{\vct\xi} - \frac{1}{2}\xi_k^\intercal \Sigma_{\vct\xi}^{-1} \xi_k\bigg) \nonumber \\
        & = -\frac{T}{2}\ \log\det\Sigma_{\vct\xi} - \frac{1}{2} \mathrm{tr} 
        \big( \Xi_{0,T}^\intercal \Sigma_{\vct\xi}^{-1}\Xi_{0,T} \big).
    \end{align*}
\end{subequations}
Thus, the ML problem becomes
\begin{subequations}~\label{eq:ML_problem_nonconvex}
    \begin{align}
        &\min_{\Xi_{0,T}, \Sigma_{\vct\xi}} \ &&\left(\frac{T}{2}\log\det\Sigma_{\vct\xi} + \frac{1}{2}\mathrm{tr} \big( \Xi_{0,T}^\intercal \Sigma_{\vct\xi}^{-1} \Xi_{0,T}\big) \right) \\
        &\quad\quad  &&(X_{1,T} - \Xi_{0,T})(I_{T} - \S^\dagger \S) = 0.
    \end{align}
\end{subequations}
From the Lagrangian of \eqref{eq:ML_problem_nonconvex}
\begin{multline*}
    \mathcal{L} = \frac{T}{2}\log\det\Sigma_{\vct\xi} + \frac{1}{2}\mathrm{tr}
    \big( \Xi_{0,T}^\intercal \Sigma_{\vct\xi}^{-1}\Xi_{0,T} \big) \\
    + \mathrm{tr} \big( \Lambda^\intercal (X_{1,T} - \Xi_{0,T})(I_{T} - \S^\dagger \S) \big),
\end{multline*}
the first-order necessary conditions yield
\begin{align}
    \frac{\partial\mathcal{L}}{\partial\Xi_{0,T}} &= \Sigma_{\vct\xi}^{-1} \Xi_{0,T} - \Lambda (I_{T} - \S^\dagger \S) = 0,
\end{align}
and
\begin{align} \label{eqn:NC2}
    \frac{\partial\mathcal{L}}{\partial\Sigma_{\vct\xi}} &= \frac{T}{2}\Sigma_{\vct\xi}^{-1} - \frac{1}{2}\Sigma_{\vct\xi}^{-1}\Xi_{0,T}\Xi_{0,T}^\intercal\Sigma_{\vct\xi}^{-1} = 0. 
\end{align}     
Hence,
\begin{equation} \label{Xistar}
    \Xi_{0,T}^\star = \Sigma_{\vct\xi}^* \Lambda (I_{T} - \S^\dagger \S),
\end{equation}
where we use the fact that $I_{T} - \S^\dagger \S$ is symmetric.
Combining equation \eqref{Xistar} with the equality constraint \eqref{eq:consistency_equation} yields
\begin{align*}
    &[X_{1,T} - \Sigma_{\vct\xi} \Lambda (I_{T} - \S^\dagger \S)](I_{T} - \S^\dagger \S) = 0 \\
    \Leftrightarrow \ &X_{1,T} (I_{T} - \S^\dagger \S) - \Sigma_{\vct\xi} \Lambda (I_{T} - \S^\dagger \S) = 0 \\
    \Leftrightarrow \ &\Sigma_{\vct\xi} \Lambda (I_{T} - \S^\dagger \S) = X_{1,T} (I_{T} - \S^\dagger \S) \\
    \Leftrightarrow \ &\Xi_{0,T}^\star = X_{1,T} (I_T - \S^\dagger \S),
\end{align*}
where in the first equivalence we use the fact that $I_{T} - \S^\dagger \S$ is idempotent.
    
The necessary condition~\ref{eqn:NC2} yields
\begin{align}
    \Sigma_{\vct\xi}^* &= \frac{1}{T}\Xi_{0,T}^\star (\Xi_{0,T}^\star)^\intercal.
\end{align}
Lastly, plugging in $\Xi_{0,T}^\star$ from \eqref{Xistar} achieves the desired result.
\qedblack

\section{Proof of Lemma~\ref{lem:error_estimate_distribution}}
\label{AppC}

For the unconstrained ML problem of estimating the normally distributed parameters
$\xi=[\xi_0^\intercal,\ldots,\xi_{T-1}^\intercal]^\intercal$,
the Fisher information matrix (FIM) is
$\mathcal{I}=I_T\otimes\Sigma_{\vct\xi}^{-1}$ and hence
$\mathcal{I}^{-1}=I_T\otimes\Sigma_{\vct\xi}$.
Let the vectorized equality constraints be
$C(\xi)=(\Gamma\otimes I_n)\xi-\lambda=0$,
with Jacobian $J=\Gamma\otimes I_n$, 
where
$\Gamma=I_T-\S^\dagger \S$ is an orthogonal projector.

The asymptotic covariance of the constrained MLE 
is
\begin{equation}\label{eq:CMLE-cov-pinv}
\Sigma_{\Delta}
    = \mathcal{I}^{-1}
    - \mathcal{I}^{-1} J^\intercal\!\big(J\mathcal{I}^{-1}J^\intercal\big)^{\dagger} J \mathcal{I}^{-1}.
\end{equation}
Compute the expression in parenthesis in \eqref{eq:CMLE-cov-pinv}
\begin{align*}
    J\mathcal{I}^{-1}J^\intercal
    &= (\Gamma\otimes I_n)(I_T\otimes \Sigma_{\vct\xi})(\Gamma\otimes I_n) \\
    &= (\Gamma\Gamma\otimes \Sigma_{\vct\xi}) = \Gamma\otimes \Sigma_{\vct\xi}.
\end{align*}
Using the facts that $\Gamma$ is an orthogonal projector ($\Gamma^{\dagger}=\Gamma$) and $\Sigma_{\vct\xi}\succ0$ (hence, $\Sigma_{\vct\xi}^{\dagger}=\Sigma_{\vct\xi}^{-1}$), and using the fact that $(A\otimes B)^{\dagger}=A^{\dagger}\otimes B^{\dagger}$, we obtain
\begin{equation*}
    \big(J\mathcal{I}^{-1}J^\intercal\big)^{\dagger}
    = (\Gamma\otimes \Sigma_{\vct\xi})^{\dagger}
    = \Gamma\otimes \Sigma_{\vct\xi}^{-1}.
\end{equation*}
Substituting into \eqref{eq:CMLE-cov-pinv} and simplifying with mixed-product rules,
\begin{align*}
    \mathcal{I}^{-1} &J^\intercal \big(J\mathcal{I}^{-1}J^\intercal\big)^{\dagger} J \mathcal{I}^{-1} \\
    &= (I_T\otimes \Sigma_{\vct\xi})(\Gamma\otimes I_n)\,(\Gamma\otimes \Sigma_{\vct\xi}^{-1})\,(\Gamma\otimes I_n)\,(I_T\otimes \Sigma_{\vct\xi}) \\
    &= (\Gamma\otimes \Sigma_{\vct\xi})\,(\Gamma\otimes \Sigma_{\vct\xi}^{-1})\,(\Gamma\otimes \Sigma_{\vct\xi})
    = \Gamma\otimes \Sigma_{\vct\xi}.
\end{align*}
Therefore, the error covariance simplifies to
\begin{equation*}
    \Sigma_{\Delta}
    = (I_T\otimes \Sigma_{\vct\xi}) - (\Gamma\otimes \Sigma_{\vct\xi})
    = (I_T-\Gamma)\otimes \Sigma_{\vct\xi}
    = \S^\dagger \S \otimes \Sigma_{\vct\xi}.
\end{equation*}
\qedblack

\color{black}

\section{Proof of Proposition~\ref{prop:uncertainty_set_general}}
\label{AppD}

    Since $\Delta\vct\xi\sim\mathcal{N}(0,\Sigma_{\Delta})$,
    it is known that the uncertainty set
    \begin{equation}~\label{eq:uncertainty_set_xi}
        \DeltaSet_{\xi} = \big\{\Delta\xi\in\mathrm{range}(\Sigma_{\Delta}) : \Delta\xi^\intercal\Sigma_{\Delta}^{\dagger}\Delta\xi \leq Q_{\vct\rchi^2_{nT}}(1-\delta)\big\},
    \end{equation}
    contains the $(1-\delta)$-quantile of the distribution of $\Delta\vct\xi$ \cite{probability_theory},
    that is, $\P(\Delta\vct\xi \in \DeltaSet_{\xi}) \geq 1-\delta$.%
    To turn \eqref{eq:uncertainty_set_xi} into an uncertainty set for $\Delta\vct\Xi_{0,T} = \mathrm{vec}^{\textcolor{blue}{-1}}(\Delta\vct\xi)$, note that 
    \begin{equation*}
        \lambda_{\min}^{+}(\Sigma_{\Delta}^{\dagger})\|\Delta\xi\|^2 \leq \Delta\xi^\intercal\Sigma_{\Delta}^{\dagger}\Delta\xi\leq\lambda_{\max}(\Sigma_{\Delta}^{\dagger})\|\Delta\xi\|^2,
    \end{equation*}
    for all $\Delta\xi\in\mathrm{range}(\Sigma_{\Delta})$, where $\lambda_{\min}^{+}$ denotes the minimum \textit{non-zero} eigenvalue.
    Hence, $\Delta\xi^\intercal\Sigma_{\Delta}^{\dagger}\Delta\xi \leq Q_{\vct\rchi^2_{nT}}(1-\delta)$ implies that $\lambda_{\min}^{+}(\Sigma_{\Delta}^{\dagger})\|\Delta\xi\|^2 \leq Q_{\vct\rchi^2_{nT}}(1-\delta)$.
    Also, from the definition of the Frobenius norm, it follows that
    \begin{equation*}
        \|\Delta\Xi_{0,T}\|_{\rm F}^{2} = \mathrm{tr}(\Delta\Xi_{0,T}^\intercal\Delta\Xi_{0,T}) = \|\Delta\xi\|^2.
    \end{equation*}
    Thus, the uncertainty set \eqref{eq:uncertainty_set_xi} can be 
    over-approximated by the set
    \begin{equation*}
        \DeltaSet_{\Xi} = \left\{\Delta\Xi_{0,T} : \|\Delta\Xi_{0,T}\|_{\rm F} \leq \sqrt{\frac{Q_{\vct\rchi^2_{nT}}(1-\delta)}{\lambda_{\min}^{+}(\Sigma_{\Delta}^{\dagger})}} \right\},
    \end{equation*}
    in the sense that
    $\Delta\xi \in \DeltaSet_{\xi}$ implies $\Delta \Xi_{0,T} \in \DeltaSet_{\Xi}$.
    Since $\DeltaSet_{\xi}$ is a $(1-\delta)$ confidence set for the distribution $\Delta\vct\xi$, it then follows that
    \begin{equation*}
        \P(\Delta\vct\Xi_{0,T} \in \DeltaSet_{\Xi}) \geq \P(\Delta\vct\xi \in \DeltaSet_{\xi}) \geq 1-\delta,
    \end{equation*}
   and  hence $\DeltaSet_{\Xi}$ is a $(1-\delta)$ confidence set for the distribution of $\Delta\vct\Xi_{0,T}$.
    Letting $\rho^2 = Q_{\vct\rchi^2_{nT}}(1-\delta)/\lambda_{\min}^{+}(\Sigma_{\Delta}^{\dagger})$, and noting that $\|\Delta\Xi_{0,T}\| \leq \|\Delta\Xi_{0,T}\|_{\rm F}$, we achieve the desired result.
    \qedblack

\section{Proof of Corollary~\ref{cor:uncertainty_set_MLE}}
\label{AppE}

    First, we begin with a result on the minimum non-zero eigenvalue of singular matrices.

    \begin{lem}~\label{prop:min_nonzero_eig}
        Let $\Sigma \in\Sb_{+}^{q}$ be a symmetric positive semi-definite matrix of rank $0 < r \leq q$.
        For $\epsilon > 0$, let $\Sigma^{\epsilon} \triangleq \Sigma + \epsilon I_{q}$.
        Then,
        \begin{equation}
            \lim_{\epsilon \downarrow 0} \lambda_{\min}\big((\Sigma^{\epsilon})^{-1}\big) = \lambda_{\max}^{-1}(\Sigma) = \lambda_{\min}^{+}(\Sigma^{\dagger}).
        \end{equation}
    \end{lem}

    \begin{pf}
        Consider the spectral decomposition
        \begin{equation}~\label{eq:singular_cov_decomposition}
            \Sigma = U
            \begin{bmatrix}
                \Lambda_r & 0 \\
                0 & 0
            \end{bmatrix} U^\intercal,
        \end{equation}
        where $\Lambda_r = \mathrm{diag}(\lambda_1,\ldots,\lambda_r)$, with $\lambda_1 \geq \cdots \geq \lambda_r > 0$ in decreasing order.
        From the decomposition \eqref{eq:singular_cov_decomposition}, it follows that $\Sigma^\epsilon = U\Lambda_r^\epsilon U^\intercal$, where
        \begin{equation*}
            \Lambda_r^{\epsilon} = \mathrm{diag}(\lambda_1 + \epsilon, \ldots, \lambda_r + \epsilon, \underbrace{\epsilon, \ldots, \epsilon}_{q-r}).
        \end{equation*}
Hence,
        \begin{equation*}
            (\Sigma^\epsilon)^{-1} = U \mathrm{diag}\left(\frac{1}{\lambda_1+\epsilon},\ldots,\frac{1}{\lambda_r+\epsilon},\underbrace{\frac{1}{\epsilon},\ldots,\frac{1}{\epsilon}}_{q-r}\right)U^\intercal.
        \end{equation*}
        Therefore,
        \begin{equation}~\label{eq:limit_min_eig}
            \lim_{\epsilon\downarrow 0} \lambda_{\min}\big((\Sigma^\epsilon)^{-1}\big) = \lim_{\epsilon\downarrow 0}\frac{1}{\lambda_1 + \epsilon} = \frac{1}{\lambda_1} = \frac{1}{\lambda_{\max}(\Sigma)}.
        \end{equation}
        From \eqref{eq:singular_cov_decomposition} and the definition of the pseudoinverse, we have that
        \begin{equation*}
            \Sigma^\dagger = U
            \begin{bmatrix}
                \Lambda_r^{-1} & 0 \\
                0 & 0
            \end{bmatrix} U^\intercal.
        \end{equation*}
        Hence, the non-zero eigenvalues of $\Sigma^\dagger$ are precisely $\{1/\lambda_i\}_{i=1}^{r}$.
        Therefore, the minimum (non-zero) eigenvalue of the pseudoinverse coincides with \eqref{eq:limit_min_eig}, which concludes the proof of the lemma.
        \qedblack
    \end{pf}

We can now proceed with the proof of Corollary~\ref{cor:uncertainty_set_MLE}.
    From Lemma~\ref{lem:error_estimate_distribution}, the covariance of the noise estimation error from the MLE scheme is $\Sigma_{\Delta} = \S^\dagger \S\otimes \Sigma_{\vct\xi}$.
    Let $\Sigma_{\Delta}^{\epsilon} \triangleq (\S^\dagger \S + \epsilon I_{T})\otimes \Sigma_{\vct\xi}$, for some $\epsilon > 0$.
    From the properties of the Kronecker product \cite{kronecker_product},
    it follows that the eigenvalues of the matrix $(\Sigma_{\Delta}^{\epsilon})^{-1}$ are given by
    \begin{align*}
        &\mathrm{spec}(\Sigma_{\Delta}^{\epsilon})^{-1} = \bigg\{\frac{1}{\lambda_i\mu_j}, \lambda_i \in \mathrm{spec}(\Sigma_{\vct\xi}), \\
        &\hspace{4cm}\mu_j \in \mathrm{spec}(S^\dagger S + \epsilon I_{T})\bigg\}.
    \end{align*}
    It then follows that
    \begin{align*}
        \lambda_{\min}(\Sigma_{\Delta}^{\epsilon})^{-1} &= \frac{1}{\lambda_{\max}(\Sigma_{\vct\xi})\mu_{\max}(S^\dagger S + \epsilon I_{T})} \\
        &= \|\Sigma_{\vct\xi}^{1/2}\|^{-2}(1 + \epsilon)^{-1},
    \end{align*}
    where we have used the fact that $\lambda_{\max}(\Sigma_{\vct\xi}) = \sigma_{\max}(\Sigma_{\vct\xi}) = \|\Sigma_{\vct\xi}^{1/2}\|^2$, and that $\mathrm{spec}(\S^\dagger \S + \epsilon) = \{\epsilon, 1 + \epsilon\}$, since $\S^\dagger \S$ is a projection matrix.
    Taking the limit as $\epsilon \rightarrow 0$, we get
    \begin{equation*}
        \lim_{\epsilon\rightarrow 0} \lambda_{\min}(\Sigma_{\Delta}^{\epsilon})^{-1} = \|\Sigma_{\vct\xi}^{1/2}\|^{-2},
    \end{equation*}
    and the result follows immediately from Lemma~\ref{prop:min_nonzero_eig}.
    \qedblack

\section{Finite-Sample Uncertainty Set Construction}
\label{AppF}

Using the MLE solution \eqref{eq:MLE_noise_soln}, the estimation error becomes
\begin{equation}~\label{eq:exact_estimation_error_matrix}
    \Delta\Xi_{0,T} \triangleq \Xi_{0,T} - \hat{\Xi}_{0,T} = \Xi_{0,T} \scriptP, \quad \scriptP \triangleq \scriptS^\dagger \scriptS,
\end{equation}
where $\scriptP \in \Re^{T\times T} \succeq 0$ is the orthogonal projector onto the row-space of $\scriptS$.
The matrix $\scriptP$ is symmetric, idempotent, and since the input/state data is persistently exciting (PE) it follows that $r = \mathrm{rank}(\scriptP) = n + m$.
Since $\vct\Xi_{0,T} = [\vct\xi_0, \ldots, \vct\xi_{T-1}]$, where each individual disturbance satisfies $\vct\xi_k\in\Re^{n}\sim\mathcal{N}(0,\Sigma_{\vct\xi})$, it follows that
\begin{equation}
    \mathrm{vec}(\Delta\vct\Xi_{0,T})  \sim \mathcal{N}(0, \scriptP \otimes \Sigma_{\vct\xi}),
\end{equation}
which follows from $\mathrm{vec}(\Delta\vct\Xi_{0,T}) = \mathrm{vec}(\vct\Xi_{0,T}\scriptP) = (\scriptP^\intercal \otimes I) \mathrm{vec}(\vct\Xi_{0,T})$ and $\mathrm{vec}(\vct\Xi_{0,T}) \sim \mathcal{N}(0, I_{T} \otimes \Sigma_{\vct\xi})$.
Since $\scriptP$ is symmetric and idempotent, the covariance {of $\mathrm{vec}(\Delta\vct\Xi_{0,T})$}
simplifies to $\scriptP \otimes \Sigma_{\vct\xi}$.
This is {exactly} the covariance 
obtained via the constrained-MLE Fisher information calculation in Lemma~\ref{lem:error_estimate_distribution}.

To derive the non-asymptotic distribution for the norm of the estimation error, 
note first that 
we can equivalently express the disturbance matrix as $\vct\Xi_{0,T} = \Sigma_{\vct\xi}^{1/2} \vct G$, where $\vct G \in \Re^{n\times T}$ is a standard Gaussian random \textit{matrix} (each entry $\vct G_{ij}$ is a standard normal random \textit{variable}).
By decomposing the projection matrix as $\scriptP = V_r V_r^\intercal$, where $V_r\in\Re^{T\times r}$ defines an orthonormal basis on $\Re^{r}$, and using \eqref{eq:exact_estimation_error_matrix}, yields the equivalent expression of the estimation error as 
\begin{equation}~\label{eq:random_matrix_uncertainty}
    \Delta\vct\Xi_{0,T} = \Sigma_{\vct\xi}^{1/2} \vct G V_r V_r^\intercal.
\end{equation}
Using \eqref{eq:random_matrix_uncertainty} and the unitary invariance of the Frobenius norm, it follows that
\begin{equation}~\label{eq:normed_estimation_error_matrix}
    \|\Delta\vct\Xi_{0,T}\|_{\mathrm{F}} = \|\Sigma_{\vct\xi}^{1/2} \vct G_r\|_{\mathrm{F}}, \quad \vct G_r \triangleq \vct G V_r \in \Re^{n\times r}.
\end{equation}
The above identity is finite-sample and depends \textit{only} on $r$ (the relevant subspace dimension) and $\Sigma_{\vct\xi}$.
Importantly, it does not involve the data collection horizon $T$.

To translate the normed estimation error distribution in \eqref{eq:normed_estimation_error_matrix} to a high-probability uncertainty set, first notice that
\begin{equation}
    \|\Delta\vct\Xi_{0,T}\|_{\mathrm{F}}^{2} = \mathrm{tr} ( \vct G_r^\intercal \Sigma_{\vct\xi} \vct G_r)  \leq \lambda_{\max}(\Sigma_{\vct\xi}) \|\vct G_r\|_{\mathrm{F}}^{2} \quad \mathrm{a.s.}
\end{equation}

\begin{lem}~\label{lem:random_matrix_independence}
    Let $\vct G \in \Re^{n\times T}$ have i.i.d. entries $\vct G_{ij} \sim \mathcal{N}(0,1)$.
    Let $V_r \in \Re^{T\times r}$ have orthonormal columns, i.e., $V_r^\intercal V_r = I_{r}$.
    Then, all entries of $\vct G_r \triangleq \vct G V_r \in \mathbb{R}^{n\times r}$ are i.i.d. $\mathcal{N}(0,1)$.
\end{lem}

\begin{pf}
    Write the $i$th row of $\vct G$ as the vector $\vct g_i \in \Re^{T}$.
    Because $\vct G$ has i.i.d. $\mathcal{N}(0,1)$ entries, the rows $\{\vct g_i\}_{i=1}^{n}$ are independent and $\vct g_i \sim \mathcal{N}(0, I_{T})$ for each $i$.
    Define $\vct \eta_i \triangleq V_r^\intercal \vct g_i \in \Re^{r}$.
    The mean and covariance of $\vct \eta_i$ are given by
    \begin{align*}
        \E[\vct \eta_i] &= V_r^\intercal \E[\vct g_i] = 0, \\
        \mathrm{Cov}[\vct \eta_i] &= V_r^\intercal \mathrm{Cov}(\vct g_i) V_r = V_r^\intercal I_{T} V_r = I_{r}.
    \end{align*}
    Hence, $\vct \eta_i \sim \mathcal{N}(0, I_r)$ for each $i=1,\ldots,n$.
    %
    %
  It follows that the entries of the matrix $\vct G_r = [\vct \eta_1^\intercal; \ldots; \vct \eta_n^\intercal]$ are i.i.d. $\mathcal{N}(0,1)$.
    \qedblack
\end{pf}
Since $\vct G_r$ is a standard Gaussian random matrix from Lemma~\ref{lem:random_matrix_independence}, it follows that $\|\vct G_r\|_{\mathrm{F}}^{2}$ is the sum of $nr$ independent
$\vct\rchi_{1}^{2}$ random variables, hence $\|\vct G_r\|_{\mathrm{F}}^{2} = \vct\rchi_{nr}^{2}$.
\color{black}
For any random variables $\vct X, \vct Y$ such that $\vct X \leq \vct Y$ a.s., their respective CDFs satisfy $F_{\vct X}(t) \geq F_{\vct Y}(t)$ for all $t$~\cite{stochastic_order}.
Hence, for all $\alpha \in (0,1)$, the respective $\alpha$-quantiles obey $Q_{\vct X}(\alpha) \leq Q_{\vct Y}(\alpha)$.
Applying this fact to the present setting with $\vct X = \|\Delta\vct\Xi_{0,T}\|_{\mathrm{F}}^{2}$ and $\vct Y = \lambda_{\max}(\Sigma_{\vct\xi})\vct\rchi_{nr}^{2}$, we obtain
\begin{equation}
    Q_{\vct X}(1-\delta) \leq \lambda_{\max}(\Sigma_{\vct\xi})Q_{\vct\rchi^{2}_{nr}}(1-\delta) \triangleq t^\star.
\end{equation}
Finally, based on the definition of the quantile function, it follows that $\P(\vct Y \leq t^\star) = 1-\delta$, and, from the fact that $\vct X \leq \vct Y$ a.s., we have the event inclusion $\{\vct Y \leq t^\star\} \subseteq \{\vct X \leq t^\star\}$.
Putting these two facts together, it follows readily that
$
    \P(\vct X\leq t^\star) \geq \P(\vct Y \leq t^\star) = 1-\delta,
$
and hence,
$
 \P \big(\|\Delta\vct\Xi_{0,T}\|_{\mathrm{F}}^{2} \leq \lambda_{\max}(\Sigma_{\vct\xi}) Q_{\vct\rchi_{nr}^{2}}(1-\delta) \big) \geq 1-\delta.
$
\section{Proof of Theorem~\ref{thm:robust_control_SLS}}
\label{AppG}

    From the matrix decompositions $\Lambda = Z[\Delta A \ \Delta B] = [\Lambda^{0}; \Lambda^{\tilde{w}}]$, it follows that $\Lambda^{0} = 0$ and $\bar\Phi\Lambda = \bar\Phi^{0}\Lambda^{0} + \bar\Phi^{\tilde{w}}\Lambda^{\tilde{w}} = \bar\Phi^{\tilde{w}}\Lambda^{\tilde{w}}$, where $\bar\Phi$ is similarly decomposed as $\bar\Phi = [\bar\Phi_{x}; \bar\Phi_{u}] = [\bar\Phi^{0} \ \ \bar\Phi^{\tilde{w}}]$.
    As a result, we can equivalently write elementwise the state and control constraints~\eqref{eq:R-DD-MS_constraints} as
    \begin{align}
        F_{j}^\intercal
        \begin{bmatrix}
            \mu \\
            v
        \end{bmatrix} &= F_j^\intercal \bar{\Phi} \tilde{w} + F_j^\intercal \bar{\Phi} \Lambda (I - \bar{\Phi} \Lambda)^{-1} \bar{\Phi} \tilde{w} \nonumber \\
        &= F_j^\intercal [\bar{\Phi}^{0} \ \bar{\Phi}^{\tilde{w}}]
        \begin{bmatrix}
            \mu_0 \\
            0
        \end{bmatrix} \nonumber \\
        &\hspace{20pt}+ F_j^\intercal \bar{\Phi}^{\tilde{w}} \Lambda^{\tilde{w}} (I - \bar{\Phi}^{\tilde{w}}\Lambda^{\tilde{w}})^{-1} [\bar{\Phi}^{0} \ \bar{\Phi}^{\tilde{w}}]
        \begin{bmatrix}
            \mu_0 \\
            0
        \end{bmatrix} \nonumber \\
        &= F_j^\intercal \bar\Phi^{0} \mu_0 + F_j^\intercal \bar\Phi^{\tilde{w}}\Lambda^{\tilde{w}}(I - \bar\Phi^{\tilde{w}}\Lambda^{\tilde{w}})^{-1}\bar\Phi^{0}\mu_0 \nonumber \\
        &\leq b_j, \ \forall j\in\dbracket{J}, \ \forall \|\Delta A\|\leq\varepsilon_{A}, \ \forall \|\Delta B\| \leq \varepsilon_{B}, \label{eq:robust_constraints}
    \end{align}
    %
    Next, note that since $Z\Delta A$ and $Z \Delta B$ are strictly block lower triangular matrices, we have $(\bar\Phi^{\tilde{w}}\Lambda^{\tilde{w}})^{N+1} = 0$ and hence $(I - \bar\Phi^{\tilde{w}}\Lambda^{\tilde{w}})^{-1} = \sum_{k=0}^{N}(\bar\Phi^{\tilde{w}}\Lambda^{\tilde{w}})^{k}$, which follows from the matrix Neumann expansion~\cite{matrix_analysis}.
    It follows that the second term in the constraints \eqref{eq:robust_constraints} can be upper bounded as follows
    \begin{align}
        F_j^\intercal \bar\Phi^{\tilde{w}}&\Lambda^{\tilde{w}}(I - \bar\Phi^{\tilde{w}}\Lambda^{\tilde{w}})^{-1}\bar\Phi^{0}\mu_0 = F_j^\intercal \sum_{k=1}^{N} (\bar\Phi^{\tilde{w}}\Lambda^{\tilde{w}})^{k}\bar\Phi^{0} \mu_0 \nonumber \\
        &= F_j^\intercal \bar\Phi^{\tilde{w}}\sum_{k=0}^{N-1}(\Lambda^{\tilde{w}}\bar\Phi^{\tilde{w}})^{k}\Lambda^{\tilde{w}}\bar\Phi^{0} \mu_0 \nonumber \\ 
        &\leq \|F_j^\intercal \bar\Phi^{\tilde{w}}\| \left\|\sum_{k=0}^{N-1}(\Lambda^{\tilde{w}}\bar\Phi^{\tilde{w}})^{k}\Lambda^{\tilde{w}}\bar\Phi^{0} \mu_0\right\| \nonumber \\ 
        &\leq \|F_j^\intercal \bar\Phi^{\tilde{w}}\| \left\|\sum_{k=0}^{N-1} (\Lambda^{\tilde{w}}\bar\Phi^{\tilde{w}})^{k}\right\| \|\Lambda^{\tilde{w}}\bar\Phi^{0}\mu_0\| \nonumber \\
        &\leq \|F_j^\intercal \bar\Phi^{\tilde{w}}\| \sum_{k=0}^{N-1} \|\Lambda^{\tilde{w}}\bar\Phi^{\tilde{w}}\|^{k} \|\Lambda^{\tilde{w}}\bar\Phi^{0}\mu_0\|. \label{eq:robust_constraints_upper_bound_intermediate}
    \end{align}
   
    Fix $\theta \in (0,1)$ and suppose there exists a pair of scalars $\tau,\gamma > 0$ such that there is a feasible solution $\bar\Phi$ to the pair of constraints
    \begin{equation}~\label{eq:hyperparam_constraints}
        \left\|
        \begin{bmatrix}
            \frac{\varepsilon_A}{\theta}\bar{\Phi}_{x}^{\tilde{w}} \\ 
            \frac{\varepsilon_B}{1-\theta}\bar{\Phi}_{u}^{\tilde{w}}
        \end{bmatrix}
        \right\| \leq \tau, \quad
        \left\|
        \begin{bmatrix}
            \frac{\varepsilon_A}{\theta} \bar{\Phi}_{x}^{0} \\
            \frac{\varepsilon_B}{1-\theta} \bar{\Phi}_{u}^{0}
        \end{bmatrix} \mu_0
        \right\| \leq \gamma.
    \end{equation}
    We claim that the constraints in \eqref{eq:hyperparam_constraints} imply, uniformly for all $\|\Delta A\| \leq \varepsilon_A$ and $\|\Delta B\| \leq \varepsilon_B$, that
    \begin{equation}~\label{eq:hyperparam_constraints_intermediate}
        \|\Lambda^{\tilde{w}} \bar{\Phi}^{\tilde{w}}\| \leq \tau, \quad \|\Lambda^{\tilde{w}} \bar{\Phi}^{0} \mu_0\| \leq \gamma.
    \end{equation}
    To see this, take the first constraint in \eqref{eq:hyperparam_constraints_intermediate} and note the following chain of inequalities
    \begin{align*}
        \|\Lambda^{\tilde{w}}\bar{\Phi}^{\tilde{w}}\|
        &= \left\|\big[\Lambda_A^{\tilde{w}}\ \ \Lambda_B^{\tilde{w}}\big]
        \begin{bmatrix}\bar{\Phi}_x^{\tilde{w}}\\ \bar{\Phi}_u^{\tilde{w}}\end{bmatrix}\right\| \\
        &\le \Big\|\big[\tfrac{\theta}{\varepsilon_A}\Lambda_A^{\tilde{w}}\ \ \tfrac{1-\theta}{\varepsilon_B}\Lambda_B^{\tilde{w}}\big]\Big\|
        \left\|
        \begin{bmatrix}
        \tfrac{\varepsilon_A}{\theta}\bar{\Phi}_{x}^{\tilde{w}} \\[2pt]
        \tfrac{\varepsilon_B}{1-\theta}\bar{\Phi}_{u}^{\tilde{w}}
        \end{bmatrix}
        \right\| \\
        &\le \sqrt{\Big(\tfrac{\theta}{\varepsilon_A}\|\Lambda_A^{\tilde{w}}\|\Big)^2
        + \Big(\tfrac{1-\theta}{\varepsilon_B}\|\Lambda_B^{\tilde{w}}\|\Big)^2}\;
        \left\|
        \begin{bmatrix}
        \tfrac{\varepsilon_A}{\theta}\bar{\Phi}_{x}^{\tilde{w}} \\[2pt]
        \tfrac{\varepsilon_B}{1-\theta}\bar{\Phi}_{u}^{\tilde{w}}
        \end{bmatrix}
        \right\| \\
        &\le \sqrt{\theta^2+(1-\theta)^2}\;\tau \ \le\ \tau,
    \end{align*}
    where the first inequality follows from the submultiplicativity of the spectral matrix norm, and
    the second inequality holds because, for any matrices $A_1 ,A_2$ (with the same number of rows) and scalars $a_1,a_2$, we have that $\|[a_1 A_1 \ a_2 A_2]\|^{2} = \lambda_{\max}(a_1^2 A_1 A_1^\intercal + a_2^2 A_2 A_2^\intercal) \leq a_1^2 \|A_1\|^2 + a_2^2\|A_2\|^{2}$.
    Finally, the last inequality holds because $\sqrt{\theta^2 + (1-\theta)^2} \leq 1$ for all $\theta \in (0,1)$.
    A similar upper bound can be obtained in terms of $\|\Lambda^{\tilde{w}}\bar\Phi^{0} \mu_0\|$. %

    Combining \eqref{eq:hyperparam_constraints_intermediate} with \eqref{eq:robust_constraints_upper_bound_intermediate}, we obtain
    \begin{align*}
     F_{j}^\intercal
        \begin{bmatrix}
            \mu \\
            v
        \end{bmatrix} &=
         F_j^\intercal \bar{\Phi}^{0} \mu_0 + F_j^\intercal \bar\Phi^{\tilde{w}}\Lambda^{\tilde{w}}(I - \bar\Phi^{\tilde{w}}\Lambda^{\tilde{w}})^{-1}\bar\Phi^{0}\mu_0 \nonumber \\
         &\leq F_j^\intercal \bar{\Phi}^{0} \mu_0 + \|F_j^\intercal \bar{\Phi}^{\tilde{w}}\| \sum_{k=0}^{N-1} \tau^{k} \gamma  \nonumber \\
         &= F_j^\intercal \bar{\Phi}^{0} \mu_0 + \|F_j^\intercal \bar{\Phi}^{\tilde{w}}\| \frac{1-\tau^{N}}{1-\tau}\gamma \leq b_j, \label{eq:hyperparam_constraint1_final}
    \end{align*}
    where the last inequality is exactly the convex constraint enforced in \eqref{eq:R-DD-MS_SLS_convex}.%

    In summary, satisfaction of the constraints \eqref{eq:hyperparam_constraints} is sufficient for satisfaction of the constraints $F_j^\intercal[\mu; v] \leq b_j$, for all $\|\Delta A\| \leq \varepsilon_{A}$ and $\|\Delta B\| \leq \varepsilon_{B}$.
    This implies that the original constraints \eqref{eq:robust_constraints} are satisfied.
    \qedblack


\bibliographystyle{elsarticle-num}
\bibliography{refs}

\end{document}